\documentclass[reprint,
superscriptaddress,
 amsmath,amssymb,
 aps,
]{revtex4-2}

\usepackage{graphicx}
\usepackage{dcolumn}
\usepackage{bm}
\usepackage{hyperref}

\usepackage{xcolor}
\begin{document}

\preprint{APS/123-QED}

\title{Phonon-mediated strong coupling between a three-dimensional topological insulator and a two-dimensional antiferromagnetic material}

\author{D. Quang To}%
\affiliation{Department of Materials Science and Engineering, University of Delaware, Newark, DE 19716, USA}%

\author{Weipeng Wu}
\affiliation{Department of Physics and Astronomy, University of Delaware, Newark, DE 19716, USA}%

\author{Subhash Bhatt}
\affiliation{Department of Physics and Astronomy, University of Delaware, Newark, DE 19716, USA}%

\author{Yongchen Liu}
\affiliation{Department of Materials Science and Engineering, University of Delaware, Newark, DE 19716, USA}%

\author{Anderson Janotti}%
\affiliation{Department of Materials Science and Engineering, University of Delaware, Newark, DE 19716, USA}%

\author{Joshua M.O. Zide}%
\affiliation{Department of Materials Science and Engineering, University of Delaware, Newark, DE 19716, USA}%

\author{Mark J.H. Ku}
\affiliation{Department of Materials Science and Engineering, University of Delaware, Newark, DE 19716, USA}%
\affiliation{Department of Physics and Astronomy, University of Delaware, Newark, DE 19716, USA}%

\author{John Q. Xiao}%
\affiliation{Department of Physics and Astronomy, University of Delaware, Newark, DE 19716, USA}%

\author{M. Benjamin Jungfleisch}%
\affiliation{Department of Physics and Astronomy, University of Delaware, Newark, DE 19716, USA}%

\author{Stephanie Law}%
 \email{slaw@udel.edu}
\affiliation{Department of Materials Science and Engineering, University of Delaware, Newark, DE 19716, USA}%

\author{Matthew F. Doty}%
 \email{doty@udel.edu}
\affiliation{Department of Materials Science and Engineering, University of Delaware, Newark, DE 19716, USA}%

\date{\today}

\begin{abstract}
Van der Waals antiferromagnetic and topological insulator materials provide powerful platforms for modern optical, electronic, and spintronic devices applications. The interaction between an antiferromagnet (AFM) and a topological insulator (TI), if sufficiently strong, could offer emergent hybrid material properties that enable new functionality exceeding what is possible in any individual material constituent. In this work, we study strong coupling between THz excitations in a three dimensional (3D) topological insulator and a quasi-two dimensional (2D) antiferromagnetic material resulting in a new hybridized mode, namely a surface Dirac plasmon-phonon-magnon polariton. We find that the interaction between a surface Dirac plasmon polariton in the 3D TI and a magnon polariton in the 2D AFM is mediated by the phonon coupling in the 3D TI material. The coupling of phonons with an electromagnetic wave propagating in the 3D TI enhances the permittivity of the TI thin film in a way that results in a strong correlation between the dispersion of Dirac plasmon polaritons on the surfaces of the TI with the thickness of the TI. As a result, the dispersion of surface Dirac plasmon polaritons in the TI can be tuned toward resonance with the magnon polariton in the AFM material by varying the TI's thickness, thereby enhancing the strength of the coupling between the excitations in the two materials. The strength of this coupling, which results in the surface Dirac plasmon-phonon-magnon polariton, can be parameterized by the amplitude of the avoided-crossing splitting between the two polariton branches at the magnon resonance frequency. We numerically study the dependence of the strong coupling on a variety of structural parameters of the (3D)TI/(2D)AFM hybrid material. Our results reveal that the strength of the coupling depends primarily on the anisotropy constant of the 2D AFM material, as well as on its thickness, and reaches a maximum when the AFM layer is sufficiently thick to be considered a half-infinite slab. Finally, we show that the extremely large anisotropy constant reported for certain 2D van der Waals antiferromagnetic materials results in a coupling strength that should be experimentally observable even in the presence of realistic scattering losses.       
\end{abstract}

\maketitle

\section{Introduction}
THz technologies have grown rapidly in the past few decades, with important applications including medical diagnostics, security screening, bio-agent detection, telecommunication, high-speed electronics, and industrial quality control \cite{Siegel2003,Pawar2013,Walowski2016,Zaytsev2019,Amini2021}. The THz electromagnetic spectrum lies between that of microwave and infrared radiation and the ``THz gap" describes the frequency range between 0.1 and 30~THz where there is a lack of essential components for device applications. In addition to developing suitable, efficient sources and detectors capable of operating at these frequencies\cite{Burford2017,Dang2020a, Papaioannou2021,Wu2021,Seifert2022}, it is also important to find material platforms that can guide and transduce THz signals and information within integrated circuits. Fortunately, the tremendous progresses in materials science and engineering in recent years has resulted in the synthesis of numerous new classes of materials with unprecedented properties with the potential to overcome some of these obstacles and open new avenues for the development of THz devices. For instance, three dimensional topological insulators (3D TIs) such as Sb$_2$Te$_3$, Bi$_2$Te$_3$ or Bi$_2$Se$_3$, which host two-dimensional surface Dirac plasmons with energy in the THz regime, could be utilized to guide THz signals within integrated circuits \cite{Di2013,Stauber2017,Ginley2018,Di2020}. Similarly, two-dimensional van der Waals antiferromagnetic (2D AFM) materials like FePS$_3$, NiPS$_3$, MnBi$_2$Te$_3$, or CrI$_3$, which host magnons in the same THz energy range, could be employed to transfer THz frequency information without energy dissipation due to the absence of charge current \cite{Jungfleisch2018,Gibertini2019,Huang2020, Zhang2020,Yang2021,Jiang2021, Zhang2021,Belvin2021}. However, to date the generation of THz magnons in AFM materials is still not well controlled, with common techniques relying on conversion from a thermal source \cite{Han2019,Fulara2019,Wang2019,Liu2021}. Moreover, the magnon in an AFM material is insensitive to small external magnetic fields because of a vanishing macroscopic magnetic moment. Those material properties make it difficult to utilize magnons in AFMs within devices. Finding ways to generate, control, and detect magnons in an AFM material-based heterostructure is therefore one essential step toward improved devices. In that context, a strong interaction between the electric and magnetic degree of freedoms in a TI/AFM heterostructure, which results in a hybridization between the magnetic and plasmonic resonances of the two constituents, may provide an effective alternative for the excitation, manipulation, and detection of the magnon via optical control of the dispersion of surface plasmons in the TI. Moreover, the hybridization of magnons with photons \cite{Yuan2017,Golovchanskiy2021,Xiao2021,Henriques2022} or phonons \cite{Liu2021,Zhang2021} could lead to emergent properties that offer even more device opportunities. 

\subsection{Hybridized states and strong coupling}
Hybridized states are established when two distinct excitations interact with sufficient strength to create a new mode whose character and dispersion relation cannot be understood by considering either excitation alone \cite{Torma2014,FornDiaz2019,Frisk2019}. A good example is the formation of a surface plasmon polariton, which is a hybridized state formed from an electromagnetic wave (photon) and charges oscillating at a metallic sample surface (plasmon). The emergence of such a hybridized state is typically observed through an anti-crossing (avoided crossing) in the dispersion relation. The strength of the interaction can be parameterized by the amplitude of the avoided-crossing splitting between the two polariton branches. By analogy to cavity quantum electrodynamics, we define strong coupling to be the regime in which the observed mode splitting $\delta$ becomes comparable to the line width of the involved excitation, making the cooperativity factor $C=\frac{\delta^{2}}{4\Gamma_{1} \Gamma_{2}} \geq 1$ \cite{Sivarajah2019}, where $\Gamma_{1}$ and $\Gamma_{2}$ are the line widths of the isolated excitations that comprise the hybridized states. These line widths originate in the loss (dissipation) for each excitation.

The two excitations whose hybridization we consider here are the Dirac plasmon-phonon polariton (DPPP) on the surface of a 3D TI and a magnon polariton (MP) in a 2D AFM.  The DPPP on the surface of 3D TI is itself a hybridized state, as described above, and such polaritons have been studied extensively \cite{Pitarke2006,Zhang2012,Torma2014,Stauber2013,Qi2014,Deshko2016, Wang2020}. Magnon polaritons (MPs), which are the collective excitations of electronic spins in a magnetic material (i.e. spin waves), have also been studied extensively in numerous material platforms \cite{Almeida1988,Dumelow1997,Sloan2019,Macedo2019,Vasconcelos2020,Hao2021}. To date there have been just a few reports on the interaction between the surface DPPP and the MP in heterostructures composed of a 3D TI and an AFM, and these have been limited to 3D antiferromagnetic materials such as NiO, FeF$_2$, or MnF$_2$ \cite{Bludov2019,Pikalov2021,To2022b}. The computationally-predicted anticrossing splitting in the systems studied to date is too small to be observed experimentally. In other words, these previous reports suggest that it will not be possible to create hybridized states or reach the strong coupling regime in such systems with presently available materials. 

In this paper we show, via numerical simulation, that three changes to TI / AFM hybrid materials can overcome this limitation and allow for entry into the regime in which strong coupling should be experimentally observable. First, we study hybrid 3D TI / AFM materials in which the AFM is a 2D van der Waals material such as FePS$_3$. FePS$_3$ has an anisotropy energy with magnitude between $2.66~meV$ and $3.6~meV$ \cite{Wildes2012,Lan2016,Olsen2021,Liu2021}, up to three orders of magnitude larger than that of a typical 3D antiferromagnetic material like MnF$_3$. This remarkably large anisotropy energy significantly increases the strength of coupling between the magnon polariton in the 2D AFM and the surface DPPP in a 3D TI. The relatively high magnon energy ($\approx$ ~3.7~THz) in FePS$_3$ \cite{McCreary2020, Liu2021} also reduces the need for an extremely high quality 3D TI such as that reported previously for a hybrid composed of a 3D TI and a traditional 3D AFM \cite{To2022b}. Second, increasing the thickness of the AFM material allows one to tune the number of magnons in the hybridized states, which in turn increases the coupling constant. Third, the coupling of an electromagnetic wave with a phonon in the bulk of a 3D TI allows one to tune the energy of the DPPP by changing the thickness of the TI. This provides a tool for tuning the DPPP toward resonance with the magnon polariton in the AFM material, thereby enhancing the strength and visibility of the coupling between the excitations in the two materials. 

\subsection{Conceptual Model}
The reason that tuning the DPPP into resonance with the MP results in stronger and more easily observable coupling can be understood conceptually from a $2\times 2$ matrix Hamiltonian:
\begin{equation}
    \hat{H}=\left[ {\begin{array}{cccc}
   E_{DPPP}(k, d_{TI}) & V_{int} \\
   V_{int} & E_{MP}(k) \\
  \end{array} } \right]
\label{2x2}
\end{equation}
where $E_{DPPP}(k,d_{TI})$ is the energy of the DPPP in the TI, which depends on the wave vector $k$ and the TI thickness $d_{TI}$, $E_{MP}$ is the energy of the magnon polariton in the AFM, and $V_{int}$ is the strength of the coupling between the DPPP and the MP. The energies of the hybridized state that arises due to coupling are found from the eigenvalues of this matrix. The eigenstates are the hybridized modes with both DPPP and MP characterized, i.e the superposition $\Psi_{Hybrid} = \Psi_{TI}+\Psi_{AFM}$ where $\Psi_{TI}$ and $\Psi_{AFM}$ describe the surface Dirac plasmon-phonon-polariton state in the TI and the magnon polariton state in the AFM, respectively.  

When $E_{DPPP}$ and $E_{MP}$ are significantly different, the eigenstates remain largely dominated by either the DPPP or MP modes. The perturbation induced by the coupling is small and difficult to distinguish from the normal k-dependence of the energy for the independent DPPP or MP. In other words, the two excitations are only weakly coupled. Two factors impact the strength and visibility of the coupling. First, when $d_{TI}$ is chosen so that $E_{DPPP}(k)$ and $E_{MP}(k)$ are degenerate for some value of $k$, the eigenstates at the degeneracy point have energy $E_{DPPP}(k)\pm V_{int}$ (which is equal to $E_{MP}(k)\pm V_{int}$ for this value of $k$). In other words, the eigenstates are fully hybridized polaritons with equal DPPP and MP composition. For this reason, the dependence of the DPPP energy on $d_{TI}$ provides a powerful tool for tuning the excitations into resonance and creating a fully hybridized state. Second, the magnitude of the interaction parameter $V_{int}$ controls the magnitude of the anti-crossing splitting ($\delta = 2 V_{int}$). As we will show below, the choice of a 2d AFM with large anistropy energy and an increasing thickness of the AFM material both increase the strength of the interaction between magnons and the EM wave. 

\subsection{Summary of Approach}
We investigate the interaction between the 3D TI and 2D van der Waals AFM FePS$_3$ by calculating the dispersion relations for the entire hybrid structure as a function of various structural parameters. The dispersion relation describes the dependence of the energy (or frequency) of the excitations ($E(k)$ / $\omega (k)$) on the wave vector $k$. We calculate the dispersion relations by solving Maxwell's equation for an electromagnetic wave propagating in the structure employing a so-called global scattering matrix technique that allows us to pull out information about the electric field amplitudes at any point or interface within the heterostructure. One of the important theoretical advances reported here is that we use a Heisenberg Hamiltonian model that captures the magnetic interactions in the quasi 2D AFM material to derive an analytical expression for the magnetic susceptibility tensor of FePS$_3$. This analytical expression is generalizable to any 2D AFM material in the family XPS$_3$ (X=Mn, Fe, Co, Ni). This magnetic susceptibility tensor, which is the input for our global scattering matrix method, is distinct from that of bulk (3D) AFM materials because one has to consider interactions between the spin moments of magnetic atoms up to the third next-nearest-neighbor. The output of the global scattering method then provides the dispersion relation of the surface plasmon-phonon-magnon polariton. We analyze the computed dispersion relations to understand the impact of various structural parameters on the strength and visibility of the coupling.   

The paper is organized as follow. In Sect.~\ref{method}, we present the methods and models employed in this article to investigate the interaction between the 3D TI layer and the 2D AFM material. We first introduce the optical response functions of TIs and AFMs to the electric and magnetic components of an electromagnetic wave propagating within each constituent material. We then describe the global scattering matrix method we employ to solve Maxwell's equations within the TI / AFM heterostructure. In Sect.~\ref{results} we discuss the calculated dispersion relations for the surface Dirac plasmon-phonon-magnon polaritons. We explore the dependence of these dispersion relations on various material properties and, in particular, explore the material and device properties required to obtain strong coupling between the 3D TI and the 2D AFM heterostructure. The roles of the material parameters in tuning the strength of this coupling provide important guidance as to how the strong coupling regime can be reached experimentally. Finally, we provide conclusions and perspectives for this work in Sect.~\ref{conc}.

\section{Theory and model}
\label{method}
We employ a semi-classical approach to explore the structural parameters and materials properties that allow us to achieve strong coupling between THz excitations in a 3D TI and a 2D AFM. We use a global scattering matrix to find a solution to Maxwell's equations for an electromagnetic (EM) wave propagating in the considered structure subject to standard boundary conditions at interfaces. The EM wave will excite both surface Dirac plasmon phonon polaritons (DPPPs) in the TI and magnon polaritons (MPs) in the AFM via its electric and magnetic field components. Those excitations will interact with each other, resulting in a new hybridization between plasmon-phononic and magnetic resonance, namely the creation of a surface Dirac plasmon-phonon-magnon polaritons (SDPP-MPs) which leads to a change in the dispersion relationship $\omega (k)$. For that reason, an analysis of the dispersion relationship for these hybridized modes will allow us to explore the physical origins that underlie the interactions. From the output of this technique we plot the imaginary part of the reflection coefficient, which describes the amplitudes of the evanescent waves propagating along the surface of the TI layer as a function of in-plane wave vector and the frequency of EM wave. Local maxima of the imaginary part of the reflection coefficient represent the existence of the modes and thus this type of plot effectively reveals the dispersion relation. The inputs for this method are the optical response function and thickness of the corresponding material constituents of the system. In the following, we introduce the optical frequency-dependent formulas for a TI and an AFM and then describe the global scattering matrix we use in this paper. 

\subsection{Optical response function}
We consider two potential 3D TI materials, Bi$_{2}$Se$_{3}$ and Sb$_{2}$Te$_{3}$, that host two dimensional spin-polarized Dirac plasmon on the surface. The behavior of these Dirac plasmon is analogous to that in graphene and the Dirac plasmon system on the surface of a pristine 3D TI layer can be treated as a conducting electron sheet with optical conductivity given by
\begin{align}
    \sigma_{\rm TI}&=\frac{e^{2}E_{F}}{4\pi \hbar^{2}}\frac{i}{\omega+i\tau^{-1}}.
    \label{optiTI}
\end{align}
where $E_{F} \approx 260~meV$ is the Fermi energy of surface states, $\tau \approx 0.06~ps$ is the relaxation time \cite{Wang2020}, and $e$ is the electron charge. 

We note that a TI thin film can acquire a nonzero local magnetic moment due to proximity with an AFM material when the two materials are put in contact. However, this effect is normally weak and can be neglected, especially in the case of an AFM material \cite{Zhu2018}. In addition, the hybridized states at the interface between a TI and another material (e.g. the AFM in this work) may change the carrier density at the interface, as predicted by density functional theory for the case of a TI/III-V semiconductor interface\cite{To2022a}. In the case of a structure composed of two van der Waals materials, this effect is expected to be small and can be ignored. We therefore assume the same optical conductivity expression for the conducting surface of the TI and the interface between the TI and the AFM. In other words, in the following $\sigma_{0} \equiv \sigma_{1} \equiv \sigma$ as given by Eq. \ref{optiTI} (where $\sigma_{0}$ and $\sigma_{1}$ are respective the optical conductivity of the Dirac plasmon on the surface of the TI and at the interface between the TI and the AFM). 

Remarkably, interactions between the Dirac plasmon mode and the lattice vibrations, i.e. phonons, in a bulk TI significantly alter the dispersion of the surface Dirac plasmon polariton in the TI, resulting in the formation of a Dirac plasmon phonon polariton (DPPP) mode that is different from the polariton modes of 2D materials like Graphene \cite{Stauber2013,Stauber2017}. In the case of chalcogenide materials with a rhombohedral lattice and quantum layer structure, like that of Bi$_2$Se$_3$ and Sb$_{2}$Te$_{3}$, two characteristic phonon modes are observable when the AC electric field is perpendicular to the c axis: the alpha phonon, also known as the (Eu1) mode, and the beta phonon, also known as the (Eu2) mode \cite{Richter1977}. The strong alpha phonon mode oscillation contributes to a large variation in the TI permittivity in the THz regime we consider in this work. In contrast, the contribution of the beta phonon is usually small and is negligible for the case of Sb$_2$Te$_3$. Incorporating all of these effects, the frequency-dependent permittivity of the bulk TI in the far-IR range of interest can be described by the Drude–Lorentz model \cite{Wang2020,To2022a,To2022b}:
\begin{equation}
    \varepsilon_{TI} = \varepsilon_{\infty} + \frac{S_{\alpha}^{2}}{\omega_{\alpha}^{2}-\omega^{2}-i\omega\Gamma_{\alpha}} + \frac{S_{\beta}^{2}}{\omega_{\beta}^{2}-\omega^{2}-i\omega\Gamma_{\beta}}
\label{dielectricTI}
\end{equation}
where $\varepsilon_{\infty}$ is the dielectric constant at high frequency ($\omega \rightarrow \infty$), $\omega_{x}$, $\Gamma_{x}$, and $S_{x}$ are the frequency, the scattering rate, and the strength of the Lorentz oscillator associated with the $\alpha$ ($x=\alpha$) and the $\beta$ ($x=\beta$) phonons of the TI thin film. Numerical values for all TI parameters are taken from reference \cite{Deshko2016} and are listed in Table \ref{tableTI}. All the TIs used in this work are non-magnetic materials, so their permeabilities are set to unity, $\mu_{TI}=1$.
\begin{widetext}
\center
\begin{table}[h]
\caption{The TI parameters used in this work, taken from \cite{Deshko2016}.}
\begin{tabular}{cccccccc}
\hline
\hline
Materials &$\varepsilon_{\infty}$ &S$_{\alpha}$(cm$^{-1}$) &$\omega_{\alpha}$(cm$^{-1}$) &$\Gamma_{\alpha}$(cm$^{-1}$)  &S$_{\beta}$ (cm$^{-1}$) &$\omega_{\beta}$(cm$^{-1}$) &$\Gamma_{\beta}$ (cm$^{-1}$) \\
\hline
Bi$_{2}$Se$_{3}$  &1 &675.9 &63.03 &17.5 &100 &126.94 &10 \\
Sb$_{2}$Te$_3$  &51 &1498.0 &67.3 &10 &NA &NA &NA\\
\hline
\hline
\end{tabular}
\label{tableTI}
\end{table}\vspace{-15pt}
\end{widetext}

The AFM materials we consider (FePS$_3$, MnPS$_3$, NiPS$_3$, and CoPS$_3$) belong to a family of quasi-two-dimensional van der Waals AFMs in which the magnetic lattice is a honeycomb-like structure akin to the graphene \cite{Jiang2021, Yang2021}. Because van de Waals layered structures have very weak interlayer coupling, the dielectric tensor of FePS$_3$ is frequency independent in the AFM phase and has a strong anisotropy between the in-plane and out-of-plane dielectric constants of the bulk materials, which can be written as
\begin{equation}
    \varepsilon_{AFM} = \begin{pmatrix}
        \varepsilon^{xx} &0 &0\\
        0 &\varepsilon^{yy} &0 \\
        0 &0 &\varepsilon^{zz}
    \end{pmatrix} 
\end{equation}
where  $\varepsilon^{xx}=\varepsilon^{yy}=\varepsilon^{\parallel}=25$ and $\varepsilon^{zz}=\varepsilon^{\perp}=5$ \cite{Ghosh2022}. Below the Neel temperature of $T_{N}=123~K$ \cite{Joy1992}, the magnetic moment of FePS$_3$ is out of plane along the c-direction (z-direction). We assume that the samples are below their Neel temperatures in the calculations we conduct here. The permeability of FePS$_3$ in the absence of an external magnetic field therefore can be expressed as 

\begin{equation}
    \mu = \begin{bmatrix}
    \mu^{xx} &0 &0 \\
    0 & \mu^{yy} &0 \\
    0 &0 &1 
    \end{bmatrix}
    \label{mu}
\end{equation}
where $\mu^{xx}=\mu^{yy} = 1 + 4\pi \frac{2 \gamma^{2} H_{a} M_{0} }{\Omega_{0}^{2} - \left(\omega^{2} +i/\tau_{mag} \right) }$, and $\mu^{zz} =1$. See Appendix \ref{apd1} for the detailed derivation of Eqn.~\ref{mu}. Here, $\gamma$ is the gyromagnetic ratio, $H_{a}$ is the effective anisotropy field, $M_{0}$ is the sublattice magnetization saturation, $\Omega_{0}$ is the antiferromagnetic resonance or zero-wave vector magnon frequency in the AFM material, and $\tau_{mag}$ is the magnetic relaxation time. For FePS$_3$, $M_{0} \approx 830~ G$, $H_{a} = 9840~kOe$, $\Omega_{0}=3.7~THz$, and $\Gamma_{AFM}=1/\tau_{mag}=0.035~THz$ \cite{Zhang2021}. Below we will consider how the scattering loss rate in the AFM material influences the strength of the coupling between the TI and AFM materials. Finally, the substrate MgO used in this study is a non-magnetic material so that its permeability $\mu_{MgO}=1$ and its dielectric constant is given by $\varepsilon_{MgO}=9.9$ \cite{Subramanian1989}. 

\subsection{Global scattering matrix approach}
\label{ScatteringMethod}
Now that we have obtained the optical response functions for the material constituents of our hybrid structure, we study the interaction between the TI and the AFM constituents by solving Maxwell's equations to derive the dispersion relationship for a monochromatic electromagnetic (EM) wave propagating in our optical structure. We do this using the scattering matrix formalism that has proven to be a powerful tool for investigating the electric and spin transport properties of layered structures \cite{To2019,Dang2020,To2021}. Here we adapt that robust tool to our optical structure. We note that we have previously used a recursive method \cite{To2022a,To2022b} to efficiently calculate the transmission and reflection coefficients of hybrid structures, but this recursive approach does not make it easy to pull out what happens at specific interfaces within the structure. The ability to isolate and understand what happens at interfaces within the structure, or in subsets of the structure, provides important insight into the underlying physics and the ways in which the structure and composition can be used to tune the optical response. We therefore develop here a new so-called “global scattering matrix” method from which we can easily extract what happens at each interface and within each layer. We present a detailed description of the global scattering matrix formalism in Appendix~\ref{globalmatrix}. The most important outcome of this formalism for the work presented here is that we can compute the optical response of the entire structure and the constituent parts from a global scattering matrix constructed based on interfacial scattering and propagation matrices that capture what happens at each interface and within each layer of the structure. The inputs to these interfacial scattering and propagation matrices are the materials parameters of the system and the optical response functions of each layer. 

Starting from the optical response functions derived in the previous sections, we employ the global scattering matrix formalism to compute the reflection coefficients for our hybrid material system. The imaginary part of the reflection coefficient, $Im(r)$, is proportional to the losses in the system \cite{Woessner2015,Kumar2015,Bezares2017,Epstein2020,Wang2020,To2022a}. The presence of loss in the reflectance spectrum indicates that the incident EM wave has generated an excitation that is carrying energy away laterally, i.~e., propagating in the x- or y-direction rather than transmitting or reflecting in the +z or -z directions, respectively. The frequency dependence of such loss thus generates the dispersion curves for the hybridized excitations in the coupled system, which is the aim of this study. In the next section we consider how this dispersion relation depends on structural and material properties, which allows us to probe the physics underlying the formation of hybridized excitations.

\section{Result and discussion}
\label{results}

\begin{figure}[h!]
\centering
    \includegraphics[width=.4\textwidth]{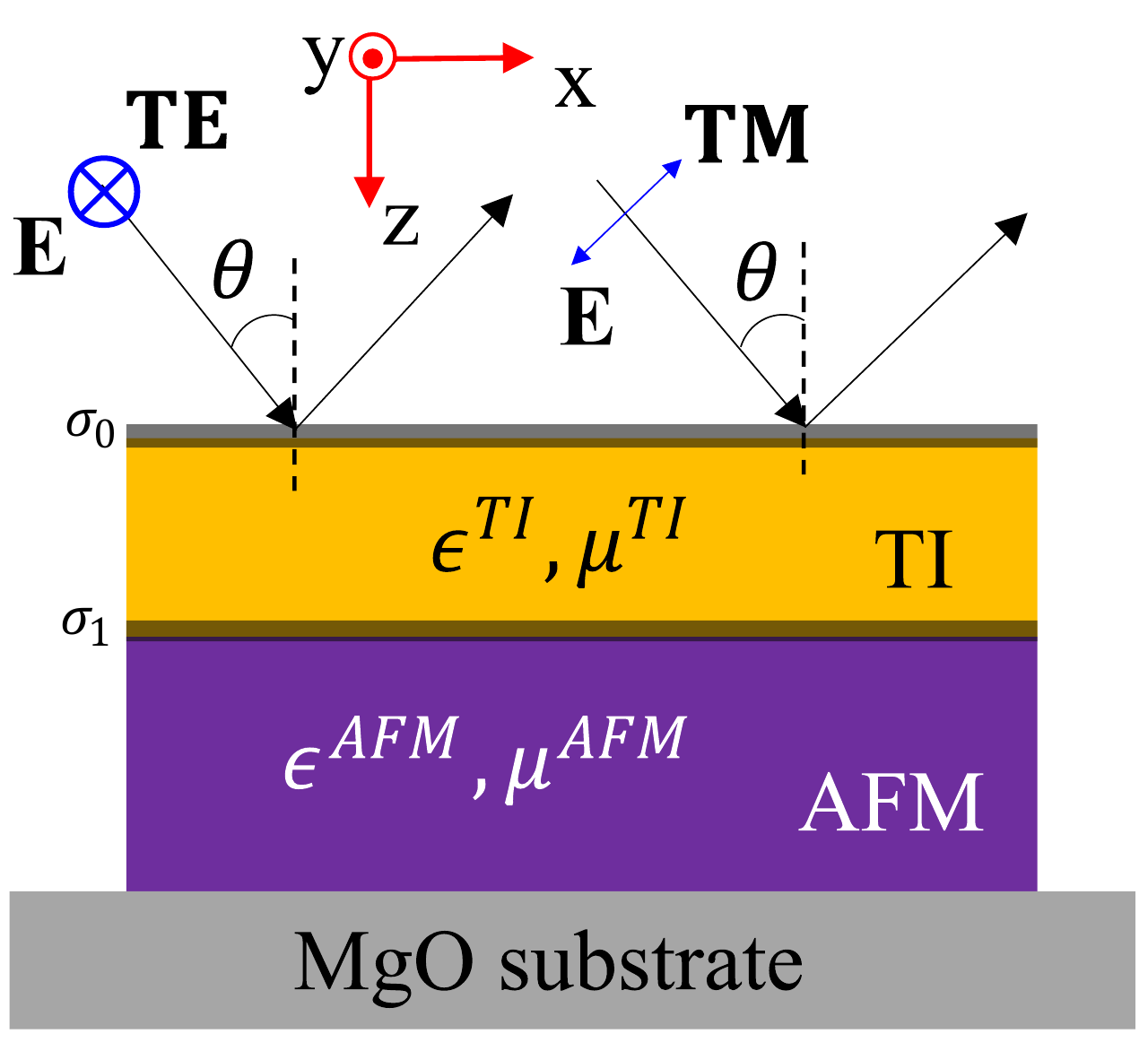}
 \caption{The TI/AFM bilayer structure on an MgO substrate investigated here. The  optical response functions in each material are the permittivity $\varepsilon_{TI/AFM}$ and permeability $\mu_{TI/AFM}$. An EM wave with both TE- and TM-polarization is incident on the TI from above with angle of incidence $\theta$. However only TM-polarized light will excite both electric and magnetic degrees of freedoms in the structure, namely surface Dirac-plasmon-phonon polaritons in the TI and magnon polaritions in the AFM.}
  \label{structre}
\end{figure}

The structure under investigation in this paper is shown in Fig.~\ref{structre}: an AFM material (FePS$_3$) on a substrate (MgO) is capped with a TI thin film. In this model, an electromagnetic wave with both TE- and TM-polarized components is incident on the top TI layer. As a result of the electromagnetic interaction with the electric and magnetic field components of the EM wave, surface Dirac plasmon polaritons in the TI thin film and magnon polaritons in the AFM material will be excited at certain resonant frequencies. The excited surface Dirac plasmon polaritons can then interact with the phonon in the bulk of the TI and also couple to the magnon polaritons in the AFM layer. We note that the TE-polarized EM wave cannot excite the surface Dirac plasmon polaritons in the TI \cite{To2022a}. Consequently we consider only TM-polarized incident EM waves in the our analysis. For convenience, we denote the Cartesian coordinates as in Fig.~\ref{structre}: the z-axis is along the growth direction of the structure, the heterostructure has finite width $W$ in the x direction, and the heterostructure is infinite in the y-direction. We set the direction of propagation of the EM wave to be parallel to the x-z plane so that the magnetic field of TM-polarized EM waves is along the y axis. Throughout our analysis the color plots in the following figures represent the amplitude of the imaginary part of the Fresnel reflection coefficient $Im(r)$ of the entire structure. The maxima of the function $Im(r)$ reveals the dispersion relationship for the coupled modes. We first discuss the emergence and characteristics of coupled surface Dirac plasmon-phonon-magnon modes and then consider how the strength of the coupling depends on structural and material parameters. 

We first note that in the long-wavelength limit $\left(k_{x}d_{TI} \ll 1 \right)$, the analytical expression for the surface Dirac plasmon mode in the TI thin film was derived in \cite{Stauber2017, Ginley2018}

\begin{equation}
    \omega_{TI_{+}}^{2} = \frac{v_{F}\sqrt{2\pi n_{2D}}e^{2}}{\varepsilon_{0}h}\frac{k_{x}}{\varepsilon_{top} + \varepsilon_{bot}+k_{x}d_{TI}\varepsilon_{TI}}
    \label{TIdis}
\end{equation}
and 
\begin{equation}
    \omega_{TI_{-}}^{2} = \frac{2 \varepsilon_{0}\varepsilon_{TI}hv_{F}+e^{2}\sqrt{2\pi n_{D}}d_{TI}}{\sqrt{4\varepsilon_{0}^{2}\varepsilon_{TI}^{2}h^{2}v_{F}^{2}+2\varepsilon_{0}\varepsilon_{TI}e^{2}\sqrt{2 \pi n_{D}}d_{TI}}}k_{x}^{2}
\end{equation}
where the subscripts $TI_{+}$ and $TI_{-}$ stand for the optical and acoustic mode, respectively. Here $v_{F}$ is the Fermi velocity for the Dirac plasmon in the TI; $n_{2D}$ is the sheet carrier concentration of the entire TI thin film, including the contribution from both surfaces; $\varepsilon_{top}$, $\varepsilon_{bot}$ and $\varepsilon_{TI}$ are the permittivity of the top and bottom dielectric media and the TI, respectively; $k_{x}$ is the in-plane wave vector; and $d_{TI}$ is the thickness of the TI layer. In this work, we focus on studying the optical mode of the surface Dirac plasmon in the TI; only this mode can be excited in a traditional optical experiment because the acoustic mode does not have any contribution in the optical dipole matrix element \cite{Ginley2018}. In the following parts we will use relation \ref{TIdis} as a reference for our further analysis of the hybridized modes. 

\subsection{Surface Dirac plasmon-phonon-magnon polariton: Signature of strong coupling}

\begin{figure}[h!]
\centering
    \includegraphics[width=.45\textwidth]{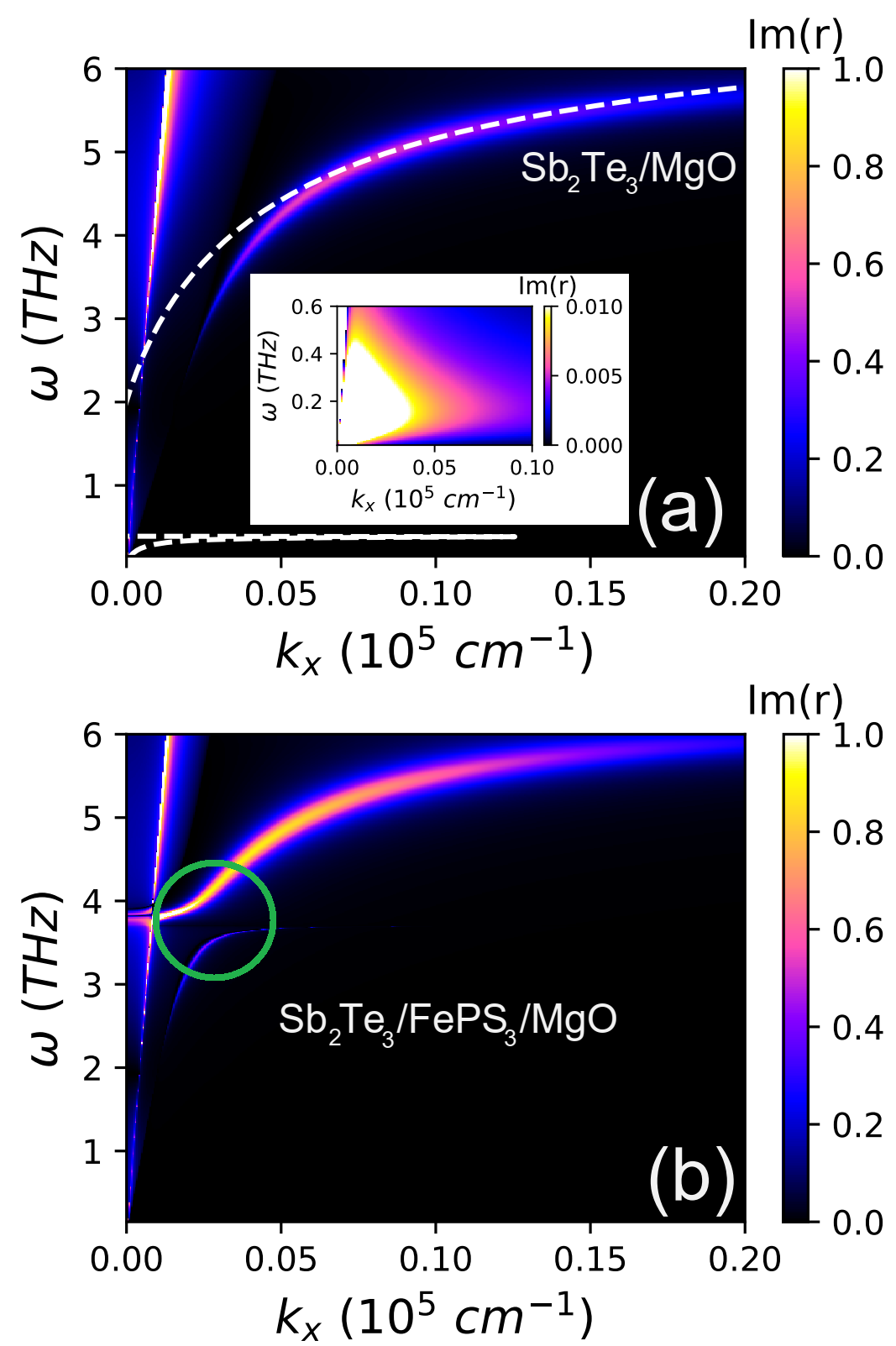}
 \caption{(a) The dispersion relation of the surface Dirac plasmon-phonon polariton in a bare Sb$_2$Te$_3$ thin film on the half-infinite MgO substrate. The dashed white line provides, for reference, an analytical calculation of the dispersion of the surface Dirac plasmon mode in a pristine Sb$_2$Te$_3$ layer on a MgO substrate, as described in the text. (b) The surface Dirac plasmon-phonon-magnon polariton in the Sb$_2$Te$_3$/FePS$_3$ structure. Both dispersion relations are plotted as a function of in-plane wave vector k$_x$ and frequency $\omega$. These calculations were both performed with the thickness of the TI thin film $d_{TI}=500~nm$ and the FePS$_3$ layer in figure (b) is sufficiently thick to be considered a semi-infinite layer.}
  \label{SbTe_FePS3_500}
\end{figure}

We will start by treating the AFM as a semi-infinite slab (i.e. infinitely thick) so that we can focus on the physics of the TI/AFM interface and the effect of the TI parameters on the resulting emergent hybridized state. We apply the global scattering matrix technique described in Sect.~\ref{ScatteringMethod} to two different configurations of the structure shown in Fig.~\ref{structre}: 1) a Sb$_2$Te$_3$ layer with thickness $d_{TI}=500~nm$ on a half-infinite bare MgO substrate and 2) the same Sb$_2$Te$_3$ layer with thickness $d_{TI}=500~nm$ on a half-infinite FePS$_3$ material [the thickness of the FePS$_3$ is very large in comparison to that of the Sb$_2$Te$_3$ layer so that, in these calculations, $d_{AFM} \approx 10 d_{TI}$) ]. The color plot in Fig.~\ref{SbTe_FePS3_500} displays the imaginary part of the Fresnel reflection coefficient $Im(r)$ calculated for the entire structure as a function of the the frequency $\omega$ and the in-plane wave vector $k_{x}$.

In Fig. \ref{SbTe_FePS3_500}(a) we plot the dispersion relation for the surface Dirac plasmon-phonon polariton (SDPPP) in a bare Sb$_2$Te$_3$ layer on the half-infinite MgO substrate. The dispersion of the SDPPP appears in the color plot in the range between $k_{x}=0.02 \times 10^{5}~cm^{-1}$ and $0.2\times 10^{5}~cm^{-1}$. The steeper line in the color plot between $k_{x}=0$ and $0.02\times 10^{5}~cm^{-1}$, in both this and subsequent figures, is the dispersion of the photon in vacuum $\omega = ck$. This photon dispersion is not important to the focus of this work and we normally neglect it without further notification. The dashed white curve is an analytical calculation of the dispersion of the surface Dirac plasmon mode in a pristine Sb$_2$Te$_3$ layer on a half-infinite MgO substrate obtained by using Eq. \ref{TIdis}. One can see that the dispersion of the SDPPP represented in the color plot in Fig. \ref{SbTe_FePS3_500}(a) is comparable to the analytical curve, with very good agreement for polariton branches above 2~THz. We note that beside the upper surface Dirac plasmon-phonon polariton branch with frequency above 2 THz, which can be observed clearly in the Fig. \ref{SbTe_FePS3_500}(a) color plot, there is also a mode at around 0.2~THz shown in the inset. This lower polariton mode can be seen clearly from the dashed white analytical curve around 0.2~THz (the horizontal dashed white line) in the Fig. \ref{SbTe_FePS3_500}(a), but its intensity is two orders of magnitude less than the intensity of the modes above 2~THz. This lower intensity is due to a large scattering loss rate of the surface Dirac plasmon in the Sb$_2$Te$_3$ material at room temperature. The surface Dirac plasmon, with high loss, dominates the modes at low frequency and consequently this lower frequency mode is barely visible in our color plot. In contrast, for the higher frequency mode (above 2~THz), the interaction with the $\alpha$ phonon plays an important role and makes the surface Dirac plasmon-phonon polariton mode become visible. Overall, Fig. \ref{SbTe_FePS3_500}(a) simply verifies that the global scatting matrix approach (color plot) agrees with the analytical dispersion (dashed white line) when applied to a sample in which interactions with the AFM material are suppressed. We will next turn on interactions with the AFM. Because the energy of magnons in the AFMs considered here is far higher than the low-energy Dirac plasmon polariton mode, the interaction between the magnon polarition in the AFM and the TI mode below 2~THz is small and can be ignored. 

\begin{figure}[h!]
\centering
    \includegraphics[width=.45\textwidth]{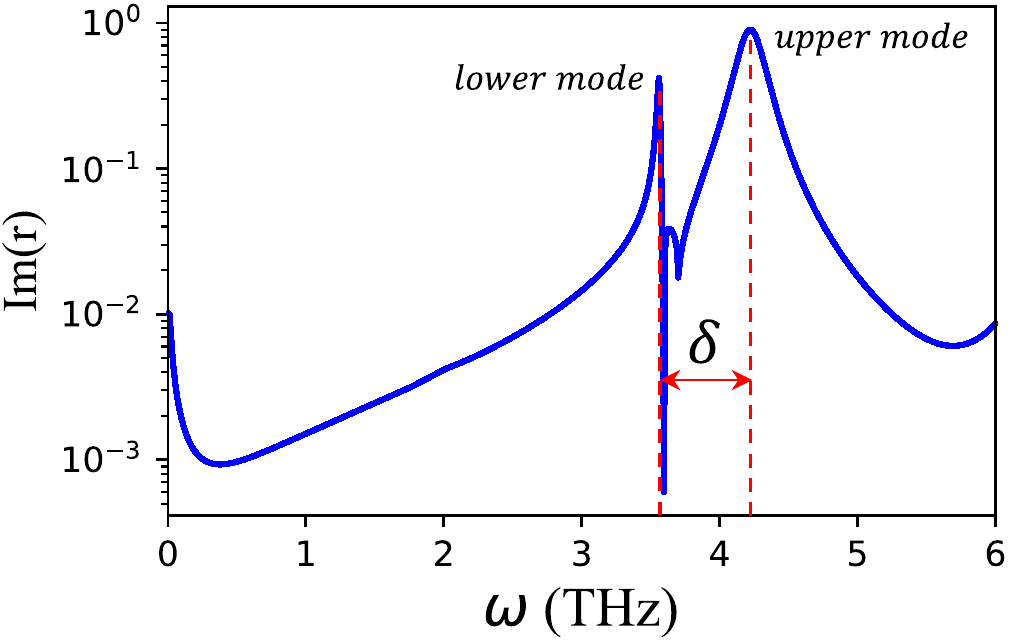}
 \caption{The imaginary part of the reflectivity Im(r) on a logarithmic (log.) scale as a function of frequency $\omega$ calculated for a Sb$_2$Te$_3$/FePS$_3$ structure with the thickness of TI thin film $d_{TI}=500~nm$ and the FePS$_3$ layer sufficiently thick to be considered a semi-infinite layer. The result is calculated for an plane wave vector $k_{x} = 0.03 \times 10^{5}~ cm^{-1}$ around the resonance point for the surface Dirac plasmon-phonon-magnon polariton.}
  \label{Imr}
\end{figure}

In Fig. \ref{SbTe_FePS3_500}(b) the Sb$_2$Te$_3$ is put on top of a very thick FePS$_3$ layer. We observe a significant change in the spectrum of the dispersion relation around $\omega \approx 3.7~THz$ owing to the interaction between the surface Dirac plasmon-phonon polariton (SDPPP) in the Sb$_2$Te$_3$ layer and the magnon polariton (MP) in the FePS$_3$. The coupling between the SDPPP and MP results in an anti-crossing highlighted by the green circle in Fig. \ref{SbTe_FePS3_500}(b). This interaction and anti-crossing lead to the formation of an upper and a lower mode that are evident through the reduction of the amplitude of $Im(r)$ around $\omega=3.7~THz$ and $k_{x}\approx 0.3 \times 10^{5}~cm^{-1}$ in the color plot. The magnitude of the splitting between the two modes that occurs at $3.7~THz$ due to the coupling between the SDPPP and MP can be evaluated by plotting the function $Im(r)$ vs. frequency $\omega$ at a fixed $k_{x}\approx 0.3 \times 10^{5}~cm^{-1}$ (resonance point) as shown in Fig.~\ref{Imr}. In this plot, the peaks at around $\omega\approx 3.5~THz$ and  $\omega\approx 4.2~THz$ indicate, respectively, the lower and upper modes in the color plot of Fig. \ref{SbTe_FePS3_500}(b). The separation between the two peaks denoted by $\delta$ is the splitting between the two modes at the resonance point, which is twice the strength of the coupling between the two excitations in our system. The splitting $\delta \approx 0.65~THz$ extracted from Fig.~\ref{Imr} for the interaction between SDPPP and MP should be experimentally detectable because it is comparable to the line width of the isolated mode in the system. This interaction is entering the strong coupling regime if the cooperativity factor $C = \frac{\delta^{2}}{4\Gamma_{TI}\Gamma_{AFM}}$ is greater than 1, where $\Gamma_{TI}$ and $\Gamma_{AFM}$ are, respectively, the scattering loss rates of the Dirac Plasmon phonon polariton in the TI and the magnon polariton in the AFM. The full width half maximum line width that represents the scattering loss rate for the surface Dirac plasmon phonon polariton in the TI is $\Gamma_{TI} \approx 3~THz$ \cite{To2022a}. The line width of the magnon polariton in the FePS$_3$ is $\Gamma_{AFM}=0.035~THz$ \cite{Zhang2021}. Inputting these values results in a cooperativity factor $C \approx 1$, which indicates the formation of a hybridized state that is approaching the strong coupling regime.  

\subsection{Dependence of the coupling strength on the TI thickness: the role of the phonon in the TI}
\begin{figure}[h!]
\centering
    \includegraphics[width=.45\textwidth]{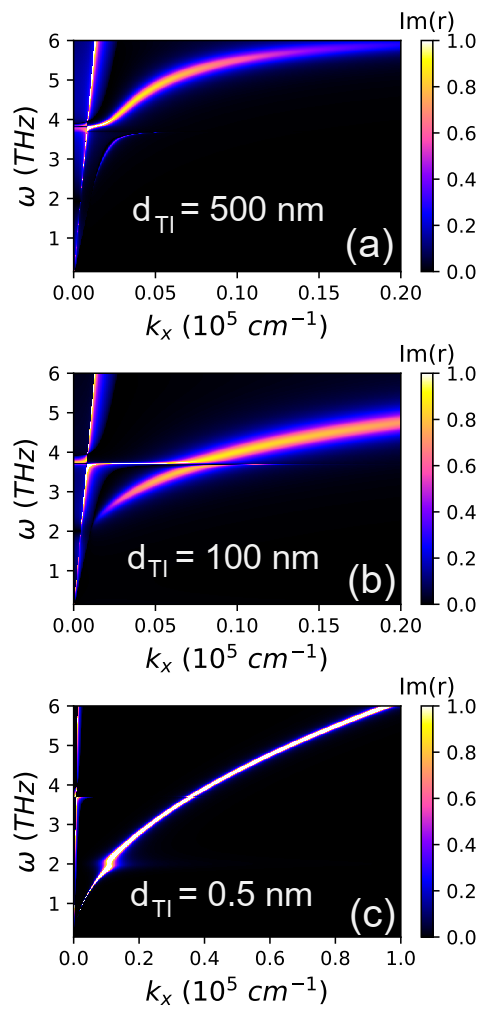}
 \caption{The imaginary part of the reflectivity Im(r) as a function of frequency $\omega$ calculated for a Sb$_2$Te$_3$/FePS$_3$ structure with TI thickness (a) $d_{TI}=500~nm$, (b) $d_{TI}=100~nm$, and (c) and $d_{TI}=0.5~nm$ on top of a semi-infinite FePS$_3$ layer. }
  \label{SPMP_TI_Graphene}
\end{figure}

\begin{figure}[h!]
\centering
    \includegraphics[width=.45\textwidth]{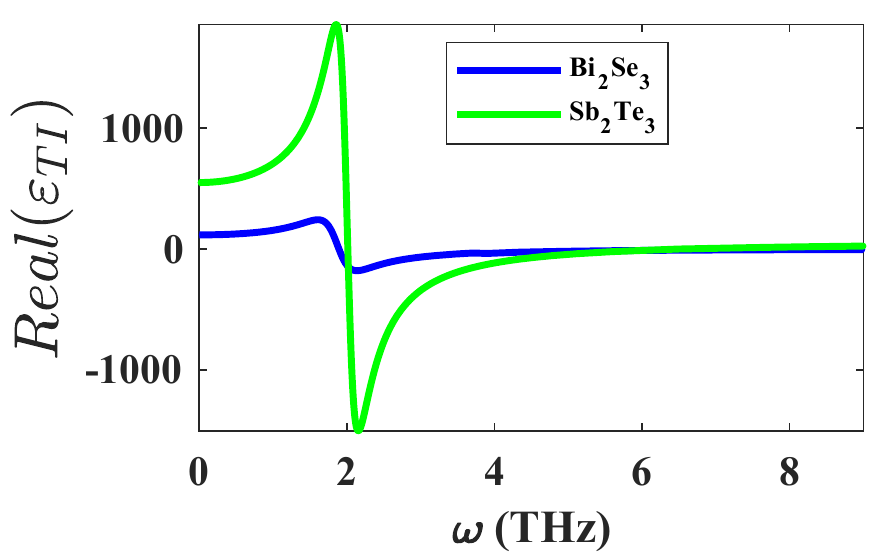}
 \caption{Dielectric function (real part) of Bi$_2$Se$_3$ (blue) and Sb$_2$Te$_3$ (green) as a function of frequency plotted using Eq. \ref{dielectricTI}.}
  \label{dielec}
\end{figure}

Our primary aim in this study is to explore the material and structural parameters that enable us to reach the strong coupling regime for the interaction between THz excitations in a TI/AFM structure. We will now investigate the impact of TI structural parameters on the strength of the coupling between the surface Dirac plasmon phonon polaritons (SDPPPs) and magnon polaritons (MPs) in our system. In this section we maintain the very large thickness of the AFM, i.e. the AFM is always a half-infinite medium while the TI's thickness is varied to understand how $d_{TI}$ influences the strength of the coupling. In Fig.~\ref{SPMP_TI_Graphene} we plot the dispersion relation of hybridized surface Dirac plasmon phonon magnon polaritons (SDPP-MPs) for different thicknesses of the TI layer (a) $d_{TI}=500~nm$, (b) $d_{TI}=200~nm$, and (c) $d_{TI}=0.5~nm$. We note that $d_{TI}=0.5~nm$ is about the thickness of a single quintuple layer of Sb$_2$Te$_3$, which is the minimum practical thickness. One observes from those plots that the dispersion of SDPP-MPs redshifts, i.e. shifts toward the low frequency regime, as the thickness of the Sb$_2$Te$_3$ layer is reduced. This arises as a result of the interaction between the $\alpha$ phonon and the surface Dirac plasmon polaritons in the TI thin film, which makes the dispersion of the surface Dirac plasmon polaritons become thickness-dependent. Indeed, due to a strong coupling between the EM wave and the $\alpha$ phonon in the TI, the real part of the dielectric constant of the TI at low frequency possesses a transition from positive to negative sign when the frequency $\omega$ of EM wave increases from zero and crosses 2~THz for both Bi$_2$Se$_3$ and Sb$_2$Te$_3$ TI materials, as shown in Fig.~\ref{dielec}. When the $\omega$ keeps increasing, the dielectric constant becomes positive again and converges to the $\varepsilon_{\infty}$. For the Sb$_2$Te$_3$ considered here, the dielectric constant is negative in the range between $2~THz$ and $6~THz$, which is why the SDPP-MP mode above 2~THz redshifts as the TI thickness decreases. This dependence can also be seen in the analytical expression for the surface Dirac plasmon mode in Eq.~\ref{TIdis} where the thickness of the TI and its dielectric constant appear simultaneously in the denominator. Physically, this redshift occurs because the surface Dirac plasmon polariton modes in the TI are coupled modes of the two surfaces. The energy of that coupled modes depends on the coupling constant, which is proportional to both the dielectric constant and the thickness of the TI. 

A direct consequence of the dependence of the SDPP-MPs on the thickness of the TI thin film is that the strength of the coupling between the surface Dirac plasmon phonon polariton (SDPPP) and the magnon polariton (MP), which is measured by the magnitude of the splitting between the upper and lower mode at $3.7~THz$, reduces as the thickness of TI decreases. This reduction occurs because the SDPPP shifts away from the resonance with the MP, thus reducing the contribution of the magnon to the hybridized mode and reducing the coupling strength \cite{To2022b}. We note that Fig.~~\ref{SPMP_TI_Graphene}(c) effectively describes the dispersion relation of a surface Dirac-plasmon-magnon-polariton in a Graphene-like/AFM system. This is because the thickness of the TI is vanishingly-small in this case, creating a degeneracy of the two surfaces of the TI and creating a Graphene-like system with extremely small coupling strength compared to that of the Sb$_2$Te$_3$ materials with finite thickness (e.g. $d_{TI}=500~nm$). The analysis here reveals the important role of the phonon in the TI as a mediator of the interaction between the surface Dirac Plasmon-phonon polariton in the TI and the magnon polariton in the AFM. 

\begin{figure}[h!]
\centering
    \includegraphics[width=.45\textwidth]{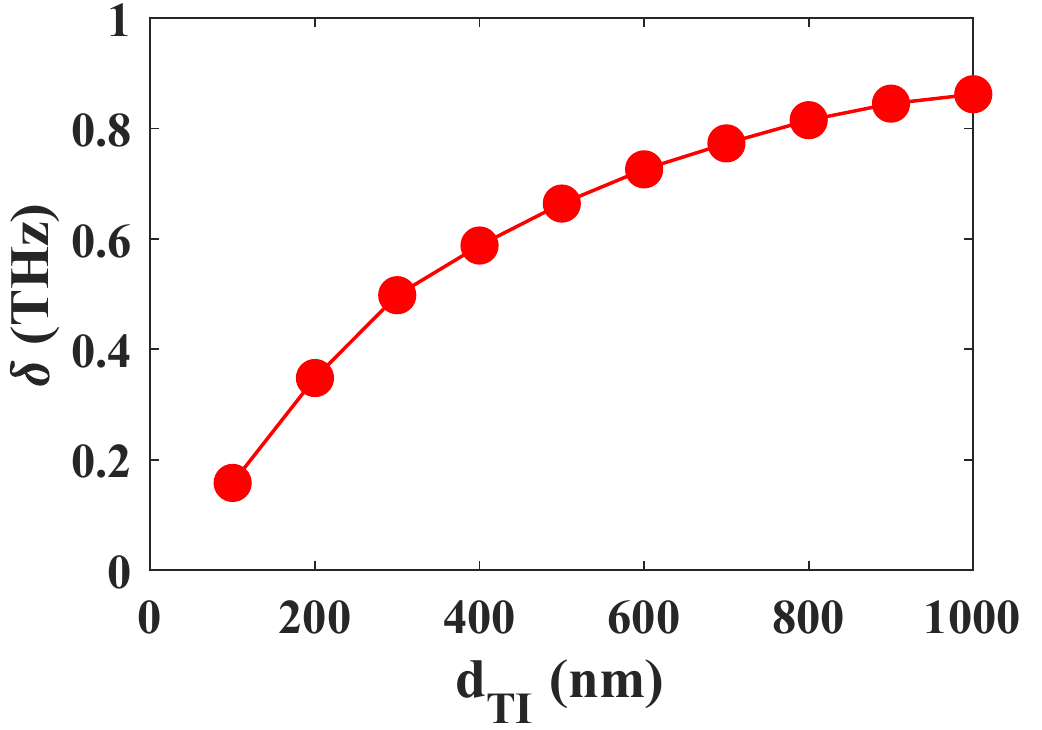}
 \caption{Splitting $\delta$ indicating the strength of the coupling between the surface Dirac plasmon phonon polariton in the Sb$_2$Te$_3$ and the magnon polariton the the FePS$_3$ as a function of the Sb$_2$Te$_3$ thickness. This calculation is done with the assumption that the FePS$_3$ layer is very thick and can be considered as a half-infinite medium.}
  \label{Delta_TIthickness}
\end{figure}

Using the TI's thickness to tune the coupling strength between the surface Dirac Plasmon-phonon polariton in the TI and the magnon polariton in the AFM provides a significant advantage relative to what could be achieved using graphene instead of a TI. Specifically, one can enhance the interaction and reach the strong coupling regime by varying the TI's thickness whereas the coupling strength for a graphene/AFM structure is fixed. Our analysis also indicates that pursuing a TI with larger negative dielectric constant in the frequency regime in which the hybridized mode is formed would reduce the time required to grow the TI sample: a larger coupling strength could be achieved with a thinner TI material. Specifically Sb$_2$Te$_3$ is a much better candidate than Bi$_2$Se$_3$ for this application because the stronger interaction with the $\alpha$ phonon in Sb$_2$Te$_3$ leads to larger magnitude of the real part of the permittivity, as can be seen in Fig.\ref{dielec}. Finally, to get a more complete picture of the TI thickness-dependent coupling strength we plot in Fig.\ref{Delta_TIthickness} the splitting $\delta$ vs the Sb$_2$Te$_3$ thickness $d_{TI}$. The splitting $\delta$ simply rises monotonically without saturation upon increasing $d_{TI}$ across this range of sample thicknesses, from $\delta \approx 0.18~THz$ at $d_{TI}=100~nm$ up to $\delta \approx 0.9~THz$ when $d_{TI}=1000~nm$. This calculation shows that $d_{TI}\geq 400~nm$ would give a splitting $\geq 0.6~THz$ that should be experimentally observable and get us into the strong coupling regime for the interaction between THz excitations in the Sb$_2$Te$_3$/FePS$_3$ structure. 

\begin{figure}[h!]
\centering
    \includegraphics[width=.45\textwidth]{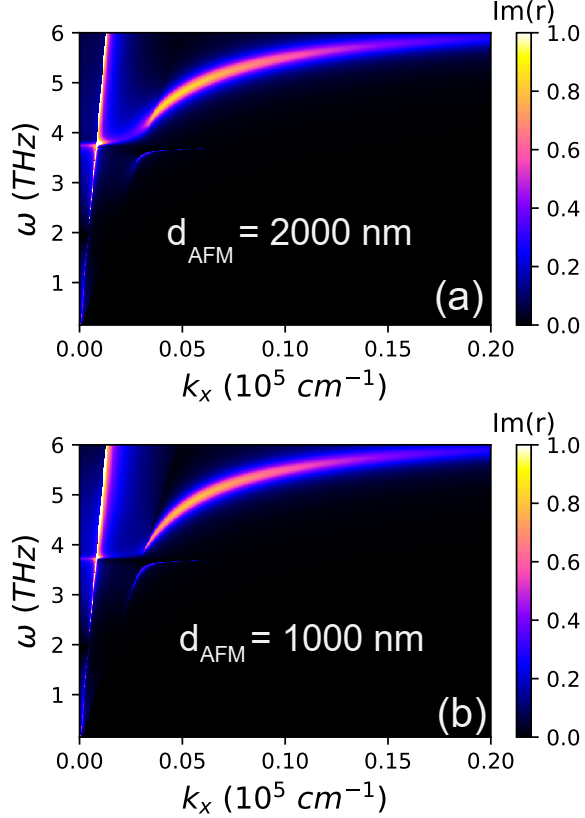}
 \caption{The imaginary part of reflectivity Im(r) as a function of frequency $\omega$ calculated for Sb$_2$Te$_3$/FePS$_3$ structure with the thickness of TI thin film $d_{TI}=500~nm$ and the thickness of the FePS$_3$ layer (a) $d_{AFM}=2000~nm$ and (b) $d_{AFM}=1000~nm$. }
  \label{SPMP_TI_AFMthickness}
\end{figure}

\begin{figure}[h!]
\centering
    \includegraphics[width=.45\textwidth]{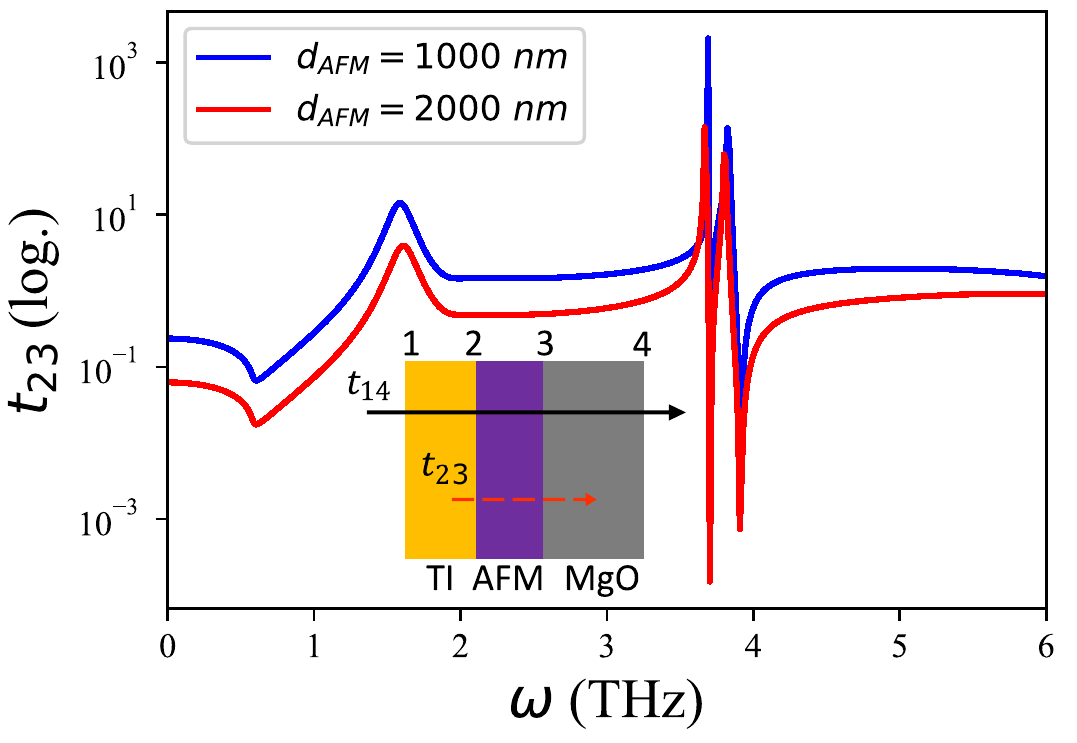}
 \caption{Transmission coefficient $t_{23}$ as a function of frequency $\omega$ at fixed wave vector $k_{x}=0.03\times 10^{5}~cm^{-1}$ for $d_{AFM}=500~nm$ (blue) and  $d_{AFM}=200~nm$ (red). The inset represents the TI/AFM structure and indicates how the transmission coefficient is calculated for different paths.}
  \label{Transmission}
\end{figure}

\subsection{Dependence of the coupling strength on 2D AFM structure parameters and material quality}

We now consider the influence of the AFM material properties and structural parameters on the interaction between the surface Dirac plasmon phonon polariton (SDPPP) and the magnon polariton (MP) in the TI/AFM structure. To do this, we replace the semi-infinite AFM slab with a slab of finite thickness on a semi-infinite MgO substate. The dispersion relations shown in Fig.\ref{SPMP_TI_AFMthickness} are calculated by applying the global scattering matrix method with a fixed Sb$_2$Te$_3$ thickness of $d_{TI}=500~nm$ for different thickness of the FePS$_3$ layer (a) $d_{AFM}=2000~nm$ and (b) $d_{AFM}=1000~nm$.

\begin{figure}[h!]
\centering
    \includegraphics[width=.45\textwidth]{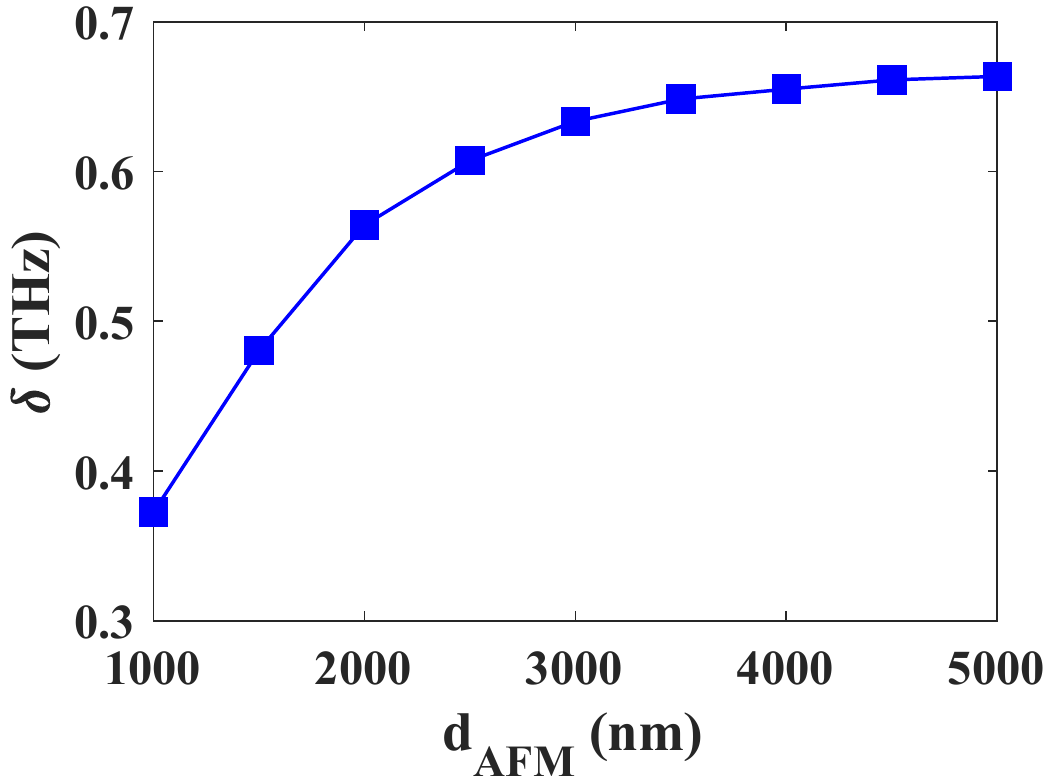}
 \caption{Splitting $\delta$ indicating the strength of the coupling between the surface Dirac plasmon phonon polariton in the Sb$_2$Te$_3$ and the magnon polariton in the FePS$_3$ as a function of the FePS$_3$ thickness. This calculation is done for fixed Sb$_2$Te$_3$ thickness $d_{TI}=500~nm$.}
  \label{Delta_AFMthickness}
\end{figure}

We previously saw that decreasing the thickness of the TI redshifted the SDPPP mode, which in turn altered the strength of the surface Dirac plasmon phonon magnon polaritons (SDPP-MP) coupling. Varying the AFM thickness does not modify the dispersion of SDPP-MP in the same way. There is no shift in either the MP or SDPPP mode with AFM thickness. However, the coupling strength, as measured by the splitting, increases with increasing AFM thickness. To understand what is happening in this case, we plot in Fig.~\ref{Transmission} the transmission coefficient $t_{23}$, on a logarithmic scale, for the EM wave travelling between the 2nd and 3rd interfaces. These interfaces are, respectively, (2nd) the interface between the TI and the AFM and (3rd) the interface between the AFM and the MgO substrate, as indicated in the inset of Fig.\ref{Transmission}. Please refer to Appendix \ref{globalmatrix} for a detailed description of how we calculated this transmission coefficient from the global scattering matrix technique. Fig.\ref{Transmission} shows the result for $d_{AFM}=1000~nm$ (blue curve) and $d_{AFM}=2000~nm$ (red curve) while keeping $d_{TI}=500~nm$ fixed. One can see that the transmission coefficient $t_{23}$ decreases over the entire range of frequencies upon increasing the thickness of the AFM layer from 1000~nm to 2000~nm. This shows that the thinner FePS$_3$ layer is more transparent to the EM wave. One can think of this in analogy to an optical absorption: there is a fixed interaction probability (cross-section) and consequently the probability of interaction between the EM wave and the magnetic degree of freedom in the AFM layer (MP) increases with AFM thickness. Essentially, a thinner FePS$_3$ results in smaller amplitude of the magnon polariton mode and thus a smaller interaction between the surface Dirac plasmon phonon polariton in the TI and the magnon polarition in the AFM layer because fewer magnons participate. 

We plot the splitting $\delta$ as a function of AFM thickness in Fig.~\ref{Delta_AFMthickness}. One observes that the splitting $\delta$ increases rapidly from 0.38~THz at $d_{AFM}=1000~nm$ to 0.6~THz at $d_{AFM}=2500~nm$. The splitting begins to saturate at $d_{AFM}=3000~nm$ with $\delta \approx 0.64~THz$. The saturation of the splitting occurs because of a competition between two effects. The number of magnons generated continues to increase with increasing AFM thickness. However, the surface electromagnetic wave associated with the SDPPP decays exponentially with z, which means that the cross-section for interaction between the EM wave and the local spin moment also decreases exponentially with z. In other words, magnons generated sufficiently far from the TI/AFM interface do not contribute to the formation of hybridized states and the splitting saturates at $\delta \approx 0.64~THz$ when $d_{AFM}=3000~nm$. Fig.~\ref{Delta_AFMthickness} tells us that the FePS$_3$ layer should be thicker than 3000~nm in order to obtain a coupling strength close to the saturation, but that increasing the AFM thickness above this value is unlikely to be useful.

\begin{figure}[h!]
\centering
    \includegraphics[width=.45\textwidth]{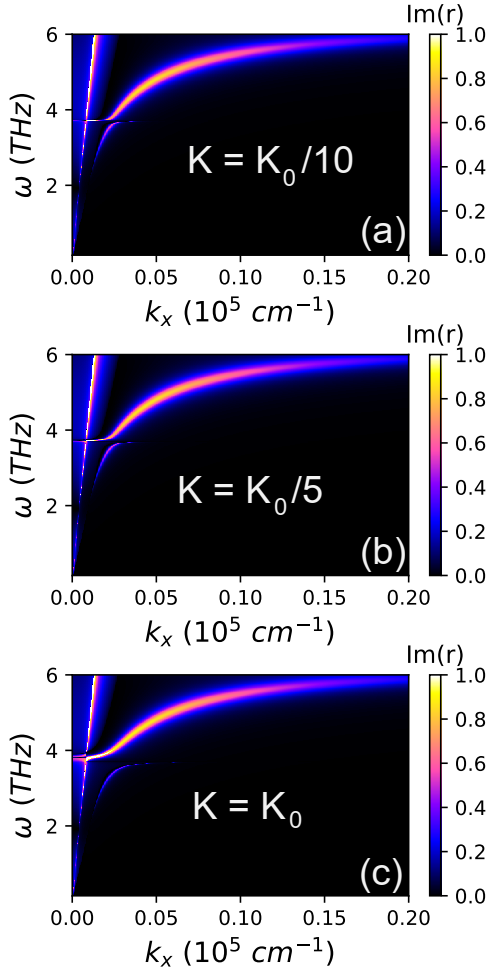}
 \caption{Dispersion relation of surface Dirac plasmon phonon magnon polariton (SDPP-MP) in Sb$_2$Te$_3$/FePS$_3$ bilayer structure with thickness of Sb$_2$Te$_3$ $d_{TI}=500~nm$ and half-infinite FePS$_3$ layer for different value of anisotropy constant $K=\gamma^{2}H_{a}M_{0}$ (a) $K=\frac{1}{10}K_{0}$, (b) $K= \frac{1}{5}K_{0}$ and (c) $K=K_{0}$ respectively. Here $K_{0}$ is the primary value of anisotropy constant in FePS$_3$.}
  \label{DPPMP_vs_anisotropy_E}
\end{figure}

We next consider the impact of the anisotropy constant of the AFM material constituent of the TI/AFM heterostructure. The anisotropy constant is defined by $K=\gamma^{2}H_{a}M_{0}$, where $\gamma$, $H_{a}$, and $M_{0}$ are, respectively, the gyromagnetic ratio, effective anisotropy field, and magnetization of the AFM spin sublattice. In Fig.~\ref{DPPMP_vs_anisotropy_E} we plot the dispersion of the SDPP-MP for $d_{TI}=500~nm$ and $d_{AFM}=5000~nm$ for different values of the anisotropy constant of the AFM material: (a) $K=\frac{1}{10}K_{0}$, (b) $K=\frac{1}{5}K_{0}$, and (c) $K=K_{0}$, where $K_{0}$ is the value of anisotropy constant for FePS$_3$ used in our previous calculations. We find that the strength of the TI/AFM coupling is proportional to the magnitude of this parameter K. In other words, a larger value of the anisotropy constant results in stronger coupling and a larger $\delta$, meaning a larger and more easily detectable splitting between the SDPP-MP hybrid modes. 

We now explain the physical origin of the increased coupling strength with increasing K shown in Fig \ref{DPPMP_vs_anisotropy_E}. The magnitude of the anisotropy constant K determines the magnetic dipole of the AFM material. A larger magnetic dipole leads to a stronger interaction between the magnetic component of the EM wave propagating in the system and the local spin moment in the AFM. A stronger interaction between the magnetic component of the EM wave and the local spin moment means that the EM wave excites magnon polaritons containing a larger number of magnons. The increased number of magnon polaritons results in a stronger interaction between the magnon states in the AFM and the Dirac plasmon phonon states in the TI, resulting in a larger contribution of magnons to the formation of Dirac plasmon phonon magnon hybrid modes. Because the anisotropy constant is primarily determined by the anisotropy energy and spin sublattice magnetization saturation of an AFM material, this  suggests that any AFM material with anisotropy energy comparable to that of FePS$_3$ (of order one meV) may be a promising alternative candidate for realizing strong coupling between a surface-plasmon-phonon polariton in a TI and magnon polaritons in an AFM. Possible alternative AFM material that are promising include: L1$_2$ IrMn$_3$ ($\Delta=6.81~meV$) \cite{Szunyogh2009}, Na$_4$IrO$_4$ ($\Delta=5.4~meV$) \cite{Wang2017}, and Cr–trihalide Janus monolayers with applied strain up to 5$\%$ (giving $\Delta=3.77~meV$ for Cl$_3$-Cr$_2$-I$_3$ monolayer) \cite{Albaridy2020}.

\begin{figure}[h!]
\centering
    \includegraphics[width=.45\textwidth]{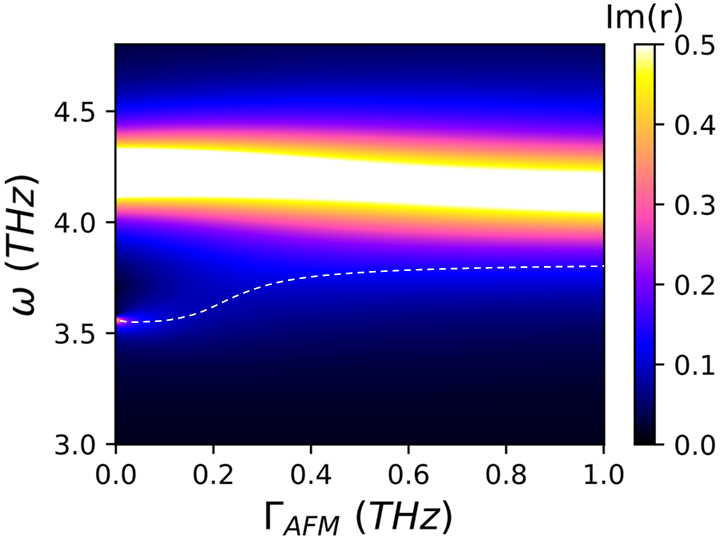}
 \caption{Mode energies of a SDPP-MP in a Sb$_2$Te$_3$/FePS$_3$ bilayer structure with Sb$_2$Te$_3$ thickness $d_{TI}=500~nm$ and a half-infinite FePS$_3$ as a function of the scattering loss rate in the AFM. The dashed white line represents the evolution of lower mode v.s. $\Gamma_{AFM}$. This calculation is performed at fixed in plane wave vector $k_{x}=0.03 \times 10^{5}~cm^{-1}$, which is at the anti-crossing point.}
  \label{lossrate}
\end{figure}

Finally, in the calculations presented thus far we have assumed that the scattering loss rate in the FePS$_3$ AFM is $\Gamma_{AFM}=0.035~THz$, which is a value taken from Ref \cite{Zhang2021}. This scattering rate parameter depends largely on crystalline and interface quality, which are specific to individual samples. We therefore consider the effect of changing scattering loss rates in the AFM material on the strength of the coupling between the TI and AFM. In Fig.~\ref{lossrate}, we plot the mode energies of SDPP-MPs in the TI/AFM structure shown in Fig.\ref{structre} using $d_{TI}=500~nm$ and a very thick (half-infinite) AFM layer. We plot the mode energies near $\omega=3.7~THz$ as a function of the scattering loss rate in the AFM material for a fixed in-plane wave vector $k_{x}=0.03 \times 10^{5}~cm^{-1}$. In other words, we focus on the anti-crossing point in the dispersion spectrum. When the scattering loss rate of the AFM material is low (left side of Fig.~\ref{lossrate}), we observe two distinct modes at $3.5~THz$ and $4.2~THz$. This is the signature of the interaction between the surface DPPPs in the TI and the MPs in the AFM layer that results in the anti-crossing splitting. The two distinct modes disappear when the scattering loss rate exceeds 0.2~THz. The loss of distinct modes (collapse of the anti-crossing) occurs when the loss rate in the AFM exceeds the coupling strength. $\Gamma_{AFM}=0.2~THz$ therefore provides a benchmark for the AFM quality required to experimentally realize observable strong coupling between a TI and an AFM. We note that the scattering loss rates of AFM materials are typically in the GHz range, which is well below this threshold.  

\section{Conclusion}
We have studied strong coupling between surface Dirac plasmon-phonon-polaritons in a TI thin film and magnon polaritons in an AFM material using a numerical semi-classical approach. Our results show that spectral signatures of strong coupling, specifically hybridized surface Dirac plasmon-phonon-magnon polaritons with cooperativity factor $C>1$, can emerge in a Sb$_2$Te$_3$ / FePS$_3$ heterostructure when (a) the thickness of the AFM material (FePS$_3$) is sufficiently large (about $\approx 3000~nm$), (b) the thickness of the TI thin film (Sb$_2$Te$_3$) is about 500~nm, and (c) the quality of the AFM material is sufficiently high that the scattering loss rate does not exceed 0.1~THz. All of these structural and materials parameters should be experimentally realizable. Equally importantly, our analysis as a function of various structural parameters allows us to understand the physical interactions that underly the coupling. For example, our analysis reveals the important role of phonons in the TI as a mediator of the interaction between the TI and AFM. Because of the important role played by phonons, and in particular the ability to tune the energy of the surface Dirac plasmon phonon polariton mode with the thickness of the TI, TIs have a significant advantage over 2D materials such as graphene for achieving strong interactions between surface Dirac plasmons and magnon polaritons. Finally, our calculations suggest that any 2D van der Waals and other types of AFM materials with a large anisotropy constant could be a viable choice for realizing strong coupling in a TI / AFM hybrid material. 
\label{conc}

\begin{acknowledgments}
This research was primarily supported by NSF through the University of Delaware Materials Research Science and Engineering Center, DMR-2011824.
\end{acknowledgments}

\appendix

\section{Magnetic susceptibility of XPS$_{3}$ (X = Mn, Fe, Co, Ni)}
\label{apd1}

\begin{figure}[h!]
\centering
    \includegraphics[width=.4\textwidth]{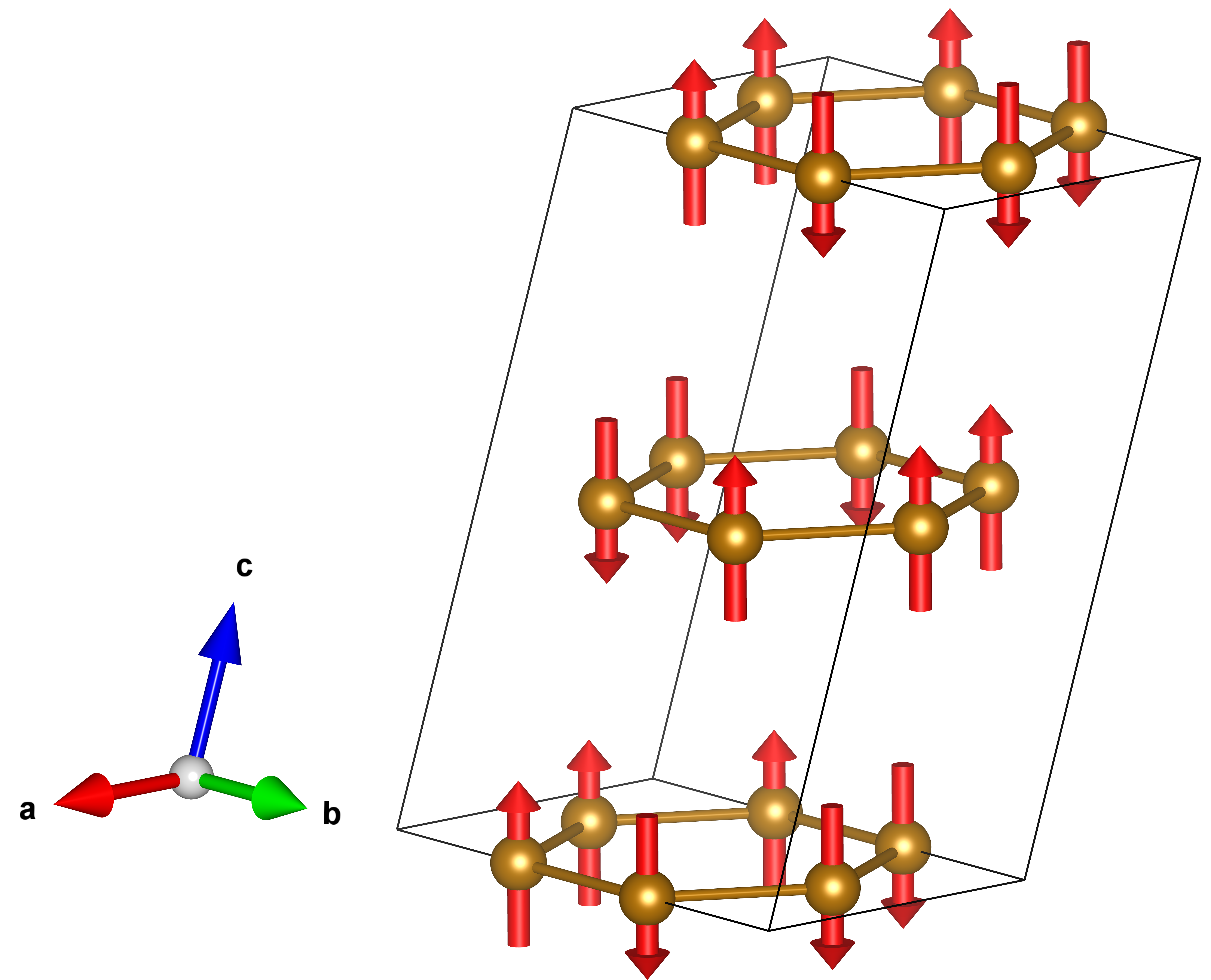}
 \caption{The layered magnetic lattice of FePS$_3$ formed by Fe atoms. The arrows indicate direction of spin moment with Zigzag AFM phases investigated in this work. This figure is plotted by using VESTA software \cite{Momma2008}}
  \label{FePS3MAG}
\end{figure}

In this appendix, we derive the frequency dependent magnetic susceptibility for 2D antiferromagnetic materials in the family XPS$_{3}$ (X = Mn, Fe, Co, Ni), which includes the FePS$_3$ studied in the main text. These materials are van der Waals magnets that form layered structures weakly bound by van der Waals forces. Figure~\ref{FePS3MAG} shows the layered magnetic structure of FePS$_3$ established by only the Fe atoms. Within each layer, the Fe atoms arrange in a honeycomb-like lattice structure with opposite spin moments. We consider in this work the FePS$_3$ magnetic structure with zigzag AFM phase, but our method presented in this section can be applied to the general case of any 2D antiferromagnetic material with different AFM phases. 

Due to the small value of the interlayer exchange interaction $J^{\prime}$ in comparison to the intralayer exchange interaction $J_{i}~(i=1,2,3)$, these AFM are, to a very good approximation, quasi-two dimensional magnets even in the bulk. The magnon dynamics in FePS$_3$ can therefore be considered by investigating a quasi-2D honeycomb structure of Fe atoms in which the magnetic interactions within the lattice are described via a Heisenberg Hamiltonian \cite{Liu2021}:
\begin{widetext}
    \begin{align}
     H &= \sum_{i,j\neq i}2J_{i,j}\boldsymbol{S}_{i} \cdot \boldsymbol{S}_{j}+ \Delta\sum_{i}\left(S_{i}^{z} \right)^{2} - \gamma \hbar \sum_{i}h_{0}^{z}S_{i}^{z} + \gamma \hbar \sum_{i} \boldsymbol{h}\cdot\boldsymbol{S}_{i}
     \label{Hmag}
\end{align}
\end{widetext}
where $\gamma$ is the gyromagnetic ratio, $\hbar$ is Planck's constant, $h_{0}^{z}$ is an external static magnetic field applyied to the lattice along the z-direction, $\boldsymbol{h}$ is a driven magnetic field, $\boldsymbol{S_{i}}$ is the spin operator, $J_{ij}$ is the exchange energy of the interaction between site $i^th$ and $j^{th}$, and  $\Delta$ is the single atom anisotropy energy. Table \ref{tableAFM} presents the spin-spin interaction parameters of the AFM materials used in this study. 
\begin{table}[h]
\caption{The spin-spin interaction parameters of the 2D AFM materials used in this work.}
\begin{tabular}{cccccccc}
\hline
\hline
Materials &J$_{1}$ (meV) &J$_{2}$ (meV) &J$_{3}$ (meV) &J$^{\prime}$  (meV) &$\Delta$ (meV)  \\
\hline
FePS$_3$ \cite{Liu2021} &1.49 &0.04 &-0.6 &-0.0073 &-3.6 \\
NiPS$_3$\cite{Olsen2021} &3.8 &-0.2 &-13.8 &N/A &-0.3\\
MnPS$_3$\cite{Olsen2021} &-1.54 &-0.14 &-0.36 &0.0019 &-0.0086 \\
\hline
\hline
\end{tabular}
\label{tableAFM}
\end{table}

\begin{figure}[h!]
\centering
    \includegraphics[width=.45\textwidth]{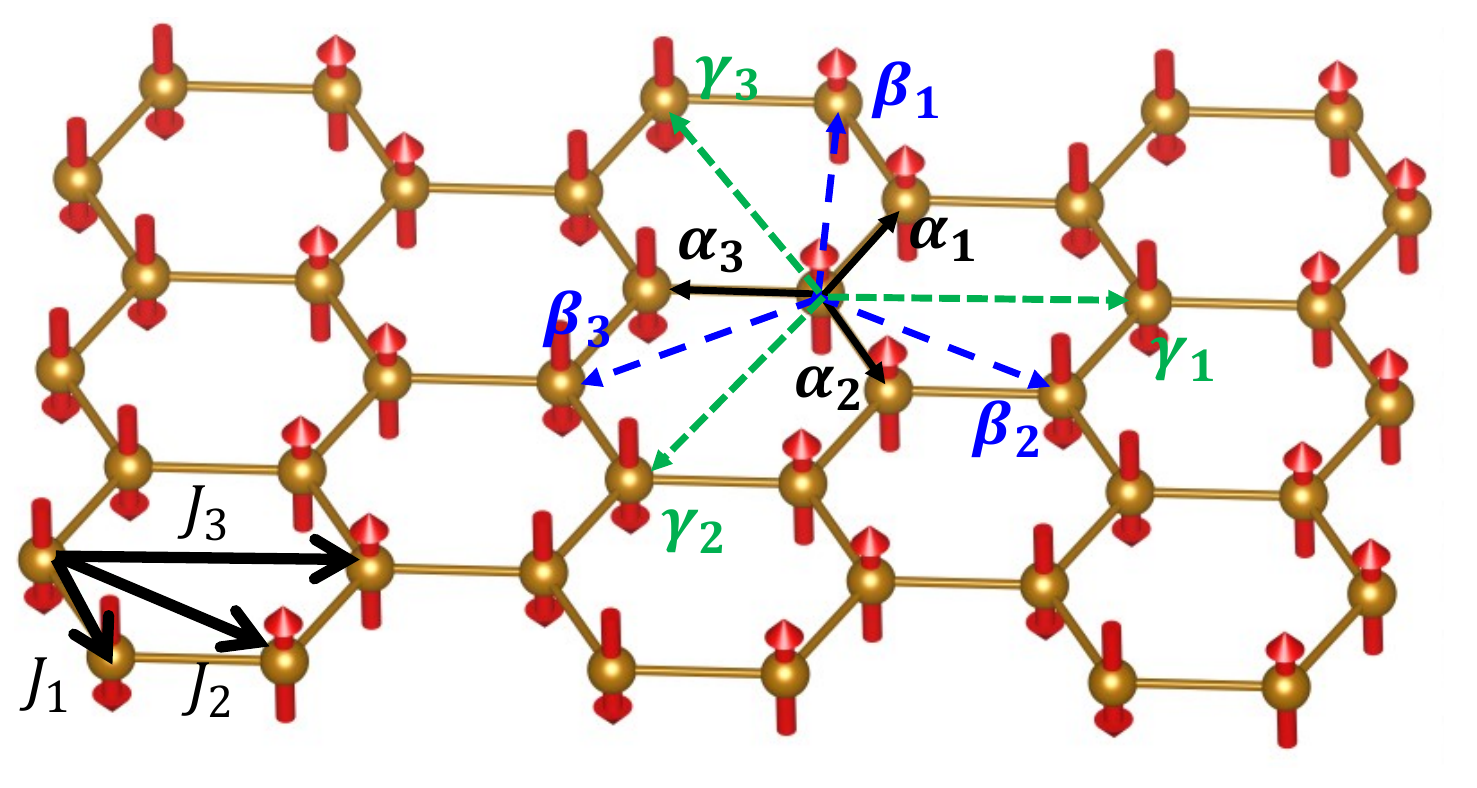}
 \caption{The quasi-2D magnetic lattice of FePS$_3$ formed by Fe atoms. The arrows indicate the direction of the spin moments with Zigzag AFM phases investigated in this work.}
  \label{MAG}
\end{figure}

Considering a uniform precession of spin moments under the driven magnetic field $\boldsymbol{h}$, we use a macrospin approximation with the uniform sublattice magnetizations in sublattice A and B, given respectively by $\boldsymbol{M}_{A,B}=\gamma \hbar N \boldsymbol{S}_{A,B}$, where N is the number of spins per unit volume and $\boldsymbol{S}_{A,B}$ is the spin in units of $\hbar$ ($S=\vert \boldsymbol{S}_{A,B}\vert=2$ in the case of Fe atom). We note that in the XPS$_{3}$ AFM family, one needs to consider the exchange interactions between two magnetic moments up to the third nearest neighbor $J_{i=1,2,3}$ associated with the vectors joining nearest $\alpha_{i=1,2,3}$, second nearest $\beta_{i=1,2,3}$, and third nearest $\gamma_{i=1,2,3}$ neighboring Fe atoms as indicated in the Fig~\ref{MAG} \cite{Li2109}. Using the Hamiltonian \ref{Hmag}, one obtains the energy per unit volume:

\begin{widetext}
    \begin{equation}
    E = \xi \left(M_{A}^{2} +M_{B}^{2} \right) + \eta \boldsymbol{M}_{A}\cdot\boldsymbol{M}_{B} + \vartheta\left[ \left(M_{A}^{z}\right)^{2}+\left(M_{B}^{z}\right)^{2} \right] - h_{0}^{z}\left(M_{A}^{z} + M_{B}^{z}\right) - \boldsymbol{h}\cdot\left(\boldsymbol{M}_{A} + \boldsymbol{M}_{B} \right)
\end{equation}
\end{widetext}
where $\xi= \frac{2\left(J_{1}+J_{2}\right)S}{\gamma \hbar M_{0}}$, $\eta=\frac{2\left(J_{1}+4J_{2}+3J_{3}\right)S}{\gamma \hbar M_{0}}$, $\vartheta=\frac{\Delta S}{ \gamma \hbar M_{0}}$, and $M_{0}$ is the magnetization of one sublattice per volume. 

Suppose a transverse magnetic field $\boldsymbol{h}=\boldsymbol{h}(t)=\left(h_{x},h_{y},0 \right)e^{-i\omega t}$ drives the spin dynamics in the lattice governed by the Landau-Lifshitz equation
\begin{equation}
    \frac{d}{dt}\boldsymbol{M}_{A,B} = \frac{g\mu_{B}}{\hbar}\boldsymbol{M}_{A,B} \times \boldsymbol{F}_{A,B}^{eff}
\end{equation}
where $\boldsymbol{F}_{A,B}^{eff}=-\boldsymbol{\nabla}_{A,B}E\left(\boldsymbol{M}_{A,B} \right)$ is the effective force acting on the A (B) spin sublattice and the magnetic moment $\boldsymbol{M}_{A,B} = m_{A,B}^{x} e^{-i \omega t}\hat{x} + m_{A,B}^{y} e^{-i \omega t}\hat{y}+ M_{A,B}^{z}\hat{z}$.

In this case one has
\begin{align}
    \frac{d}{dt}\boldsymbol{M}_{A,B}=-i\omega e^{-i\omega t} \begin{pmatrix}
        m^{x}_{A,B} \\
        m^{y}_{A,B}\\
        0
    \end{pmatrix}
\end{align}
and
   \begin{align}
    \boldsymbol{F}_{A,B}^{eff} = - \begin{pmatrix}
        2\xi m_{A,B}^{x}e^{-i \omega t}+\eta m_{B,A}^{x}e^{-i \omega t} -  h_{x} e^{-i\omega t} \\
        2\xi m_{A,B}^{y}e^{-i \omega t}+\eta m_{B,A}^{y}e^{-i \omega t} -  h_{y} e^{-i\omega t} \\
        2\left(\xi + \vartheta \right)M_{A,B}^{z}+\eta M_{B,A}^{z} -  h_{0}^{z}
    \end{pmatrix} 
\end{align} 
leading to a set of equations of transverse motion for the two-spin sublattices A and B:
    \begin{equation}
        \begin{pmatrix} 
      m_{A}^{x}\\
      m_{A}^{y}\\
      m_{B}^{x}\\
      m_{B}^{y}
    \end{pmatrix}
    = D^{-1} C \begin{pmatrix}
         h_{y} \\
        h_{x} \\
         h_{y} \\
         h_{x}
    \end{pmatrix}
    \label{eqmagnon}
    \end{equation}
where
\begin{widetext}
D = \begin{equation}
     \begin{bmatrix} 
     i \omega & -\gamma \left(2 \vartheta M_{A}^{z} + \eta M_{B}^{z} - h_{0}^{z} \right) &0 &\gamma \eta M_{A}^{z} \\
     \gamma \left(2 \vartheta M_{A}^{z} + \eta M_{B}^{z} - h_{0}^{z} \right) & i \omega &-\gamma \eta M_{A}^{z} &0 \\
     0 &\gamma \eta M_{B}^{z} & i\omega & -\gamma \left(2 \vartheta M_{B}^{z} + \eta M_{A}^{z} - h_{0}^{z} \right) \\
     -\gamma \eta M_{B}^{z} &0 &\gamma \left(2 \vartheta M_{B}^{z} + \eta M_{A}^{z} - h_{0}^{z} \right) &i\omega
    \end{bmatrix}
    \label{Dmatrix}
\end{equation}
\end{widetext}
and 
$C = diag \left(\gamma M_{A}^{z},-\gamma M_{A}^{z}, \gamma M_{B}^{z}, -\gamma M_{B}^{z}  \right)$. The determinant of matrix D (Eqn.~\ref{Dmatrix}) is given by:
\begin{widetext}
    \begin{align}
    det \vert D \vert &= \omega^{4} - 2\gamma^{2} \left[ 4\vartheta^{2} \left( M_{0}^{z} \right)^{2} -  4 \eta \vartheta\left( M_{0}^{z} \right)^{2} + \left(h_{0}^{z}\right)^{2}  \right] \omega^{2} + \gamma^{4} \left[ 4\vartheta^{2} \left( M_{0}^{z} \right)^{2} -  4 \eta \vartheta \left( M_{0}^{z} \right)^{2} - \left(h_{0}^{z}\right)^{2}  \right]^{2} \\
     &= \left[ \omega^{2} - \gamma^{2}\left(\sqrt{4\vartheta^{2} \left( M_{0}^{z} \right)^{2} -  4 \eta \vartheta \left( M_{0}^{z} \right)^{2}}  +h_{0}^{z}\right)^{2} \right]\left[ \omega^{2} - \gamma^{2}\left(\sqrt{4\vartheta^{2} \left( M_{0}^{z} \right)^{2} -  4 \eta \vartheta \left( M_{0}^{z} \right)^{2}}  -h_{0}^{z}\right)^{2} \right] \\
     &= \left[ 4\gamma^{2}\vartheta^{2} \left( M_{0}^{z} \right)^{2} -  4 \gamma^{2}\eta \vartheta \left( M_{0}^{z} \right)^{2} - \left(\omega -\gamma h_{0}^{z} \right)^{2} \right]\left[ 4\gamma^{2}\vartheta^{2} \left( M_{0}^{z} \right)^{2} -  4 \gamma^{2}\eta \vartheta \left( M_{0}^{z} \right)^{2} - \left(\omega + \gamma h_{0}^{z} \right)^{2}\right]\\
     &=\left[ \Omega_{0}^{2} - \left(\omega -\gamma h_{0}^{z} \right)^{2} \right]\left[ \Omega_{0}^{2} - \left(\omega + \gamma h_{0}^{z} \right)^{2}\right]
\end{align}
\end{widetext}
Here we have used $\Omega_{0}^{2}=4\gamma^{2}\vartheta^{2} \left( M_{0}^{z} \right)^{2} -  4 \gamma^{2}\eta \vartheta \left( M_{0}^{z} \right)^{2}$.

We now define a total magnetic moment as
\begin{equation}
    \boldsymbol{M}_{t} = \begin{pmatrix}
        m_{A}^{x}+m_{B}^{x} \\
        m_{A}^{y}+m_{B}^{y}
    \end{pmatrix} = \begin{pmatrix}
        \chi^{xx} & \chi^{xy} \\
        \chi^{yx} & \chi^{yy}
    \end{pmatrix} \begin{pmatrix}
         h_{x} \\
         h_{y} 
    \end{pmatrix}
\end{equation}
where $\begin{pmatrix}
        \chi^{xx} & \chi^{xy} \\
        \chi^{yx} & \chi^{yy}
    \end{pmatrix}$ is the magnetic susceptibility tensor.

Solving Eq.\ref{eqmagnon} within the linear approximation $M_{A}^{z}=-M_{B}^{z}=M_{0}^{z}$, one obtains the magnetic susceptibility tensor given by
\begin{equation}
    \chi^{xx}  = \frac{4\gamma^{2}\vartheta\left(M_{0}^{z} \right)^{2} \left[ \omega^{2} - \Omega_{0}^{2} + \left(\gamma h_{0}^{z}\right)^{2} \right] }{det \vert D \vert} 
    \label{sus1}
\end{equation}
\begin{equation}
      \chi^{xy} = \frac{8i\gamma^{3}\vartheta\left(M_{0}^{z} \right)^{2} h_{0}^{z} \omega }{ det \vert D \vert}  
      \label{sus2}
\end{equation}
with $\chi^{xx}=\chi^{yy}$ and $\chi^{xy}=-\chi^{yx}$.  

If we call
\begin{align}
    H_{e} &= \eta M_{0} = \frac{2\left(J_{1}+4J_{2}+3J_{3} \right) S}{\gamma \hbar} \\
    H_{a} &= 2 \vartheta M_{0} = \frac{2 \Delta S}{\gamma \hbar}
\end{align}
the, respectively, effective exchange field and effective anisotropy field, then in the case of vanishing external magnetic field $h_{0}^{z} = 0$, one obtains

\begin{align}
    \chi^{xx} &= \chi^{yy} = \frac{2 \gamma^{2} H_{a} M_{0} }{\Omega_{0}^{2} - \omega^{2} } \\
    \chi^{xy} &= \chi^{yx} = 0
\end{align}
where we have used $M_{0}^{z} \approx M_{0} $ and $\Omega_{0}^{2}=\gamma^{2} \left( H_{a}^{2} - 2 H_{e} H_{a} \right)$ is the antiferromagnetic resonance frequency or zero-wave vector magnon frequency in the antiferromagnetic material. In a system with non-vanishing scattering loss rate, one has
\begin{align}
    \chi^{xx} &= \chi^{yy} = \frac{2 \gamma^{2} H_{a} M_{0} }{\Omega_{0}^{2} - \left(\omega +i/\tau_{mag}\right)^{2} } \\
    \chi^{xy} &= \chi^{yx} = 0
\end{align}
with $\tau_{mag}$ the relaxation time of the magnon. 

The antiferromagnetic resonance frequency or zero-wave magnon frequency in the FePS$_{3}$ material $\Omega_{0}^{FePS_{3}} = 3.7 ~THz$ \cite{McCreary2020, Liu2021} and its magnetization $M_{0}^{FePS_3} \approx 830~(G)$ \cite{Wildes2020a}. In order to obtain the $H_{a}$ effective anisotropy field of FePS$_{3}$ we note that this effective anisotropy field is proportional to the magnitudes of the anisotropy energy $\Delta$, and the spin S of the antiferromagnetic material, which are respective $\Delta =3.6~meV$ taken from reference \cite{Liu2021} and $S=2$ in FePS$_{3}$. For comparison, those values in MnF$_2$ are, respectively, about 0.0024~meV and 2.5, which correspond to the effective anisotropy field $H_{a}^{MnF_2}=8.2~kOe$ \cite{Rezende2019}. We therefore estimate the value for the effective anisotropy field in FePS$_{3}$ to be about $H_{a}^{FePS_3}=9840~kOe$ and use this value in the calculations reported in the main text.  

\section{Global scattering matrix}
\label{globalmatrix}
We now present in detail the so-called global scattering matrix method used to solve Maxwell's equations to obtain the dispersion relations studied in the main text. This method is similar to the Green's function technique used to investigate scattering for a propagating wave in a multi-layered structure by an evaluation of the S-scattering matrix computed from the scattering path operator and has been successfully employed to study electric and spin transport in several system \cite{To2019,Dang2020,To2021}. Here we adopt this robust technique to the optical system studied in this article. 

\begin{center}\
\begin{figure}[h]\
 \includegraphics[width=0.45\textwidth]{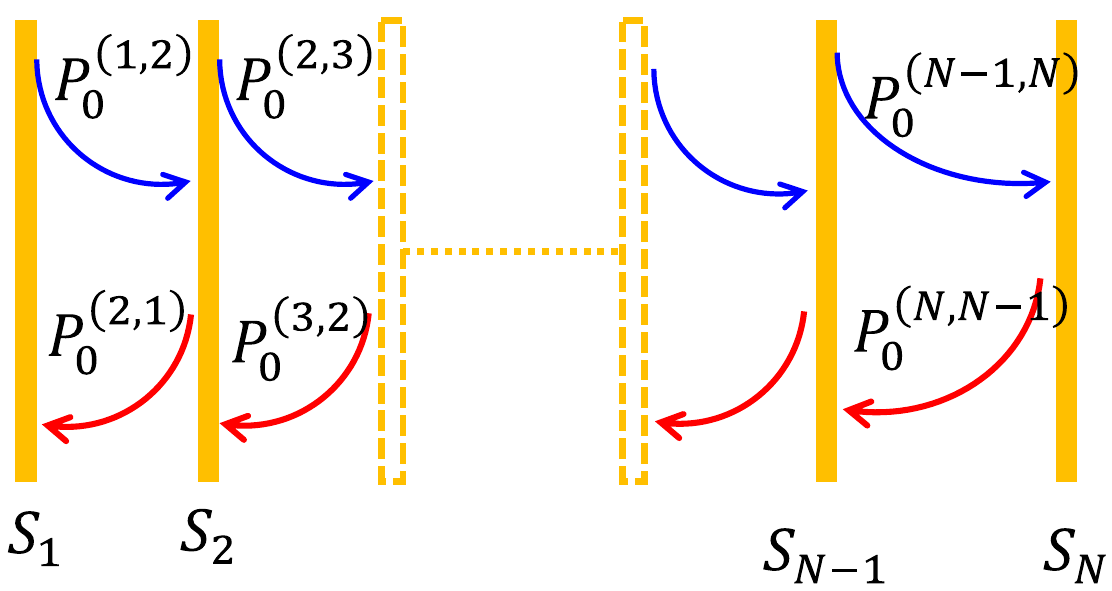}
 \caption{Schematic of a heterostructure composed of N interface with interfacial scattering matrix $S_{i}$ and propagation matrix $P_{0}^{ij}$ describing a scattering process in this structure.}
  \label{Heterostructure}
\end{figure}\
\end{center}\

Consider a heterostructure with N interface as shown in Fig \ref{Heterostructure}. We denote the z-axis as the growth direction of the structure. The dimension of the heterostructure along the y-axis is infinite and along the x-direction it is finite with a width W. Assuming that an EM wave beam is incident from the left-hand side of the structure with the direction of propagation to parallel to the x-z plane, within the m$^{th}$ layer the electric field $\boldsymbol{E}_{m}=\left(E_{x,m},E_{y,m},E_{z,m} \right)$ and the magnetic field $\boldsymbol{H}_{m}=\left(H_{x,m},H_{y,m},H_{z,m} \right)$ components of a monochromatic electromagnetic wave that is a solution of Maxwell's equations propagating along the z direction take the general form:

\begin{widetext}
\begin{align}
\boldsymbol{E}_{m} = e^{i\left(k_{x,m}x - \omega t \right)}\begin{bmatrix} e^{ik_{z,m}z} &0 &e^{-ik_{z,m}z} &0 \\
0 &e^{ik_{z,m}z} &0 &e^{-ik_{z,m}z} \\
-\frac{\varepsilon_{m}^{\parallel}k_{x,m}}{\varepsilon_{m}^{\perp}k_{z,m}}e^{ik_{z,m}z} &0 &\frac{\varepsilon_{m}^{\parallel}k_{x,m}}{\varepsilon_{m}^{\perp}k_{z,m}}e^{-ik_{z,m}z} &0
\end{bmatrix} \begin{pmatrix} A_{x,m} \\
A_{y,m} \\
B_{x,m} \\
B_{y,m}
\end{pmatrix}
\label{Electric}
\end{align}
\end{widetext}
\begin{widetext}
\begin{align}
\boldsymbol{H}_{m} =  \frac{e^{i\left(k_{x,m}x - \omega t \right)}}{\mu_{0}\mu_{m}}\begin{bmatrix} 0 &-\frac{k_{z,m}}{\omega}e^{ik_{z,m}z} &0 &\frac{k_{z,m}}{\omega}e^{-ik_{z,m}z} \\
\frac{1}{\omega k_{z,m}}\left( \frac{\varepsilon_{m}^{\parallel}}{\varepsilon_{m}^{\perp}}k_{x,m}^{2}+k_{z,m}^{2} \right) e^{ik_{z,m}z} &0 &-\frac{1}{\omega k_{z,m}}\left( \frac{\varepsilon_{m}^{\parallel}}{\varepsilon_{m}^{\perp}}k_{x,m}^{2}+k_{z,m}^{2} \right)e^{-ik_{z,m}z} \\
0 &\frac{k_{z,m}}{\omega}e^{ik_{z,m}z} &0 &\frac{k_{z,m}}{\omega}e^{-ik_{z,m}z}
\end{bmatrix} \begin{pmatrix} A_{x,m} \\
A_{y,m} \\
B_{x,m} \\
B_{y,m}
\end{pmatrix}
\label{Magnetic}
\end{align}
\end{widetext}
where $A_{(x,y),m}$ and $B_{(x,y),m}$ are the amplitudes of the x- and y- components of the forward- and backward-propagating EM waves, respectively; $\omega$ is the frequency of the EM wave; $k_{x,m}$ and $k_{z,m}$ are the x- and z-components of the wave vector of the EM wave within the $m^{th}$ layer; and x and z are the coordinates along the x- and z- directions. 

At the $m^{th}$ interface, the amplitudes of the EM wave should satisfy the standard boundary conditions \cite{Jackson1999,Zangwill_2012}:
\begin{align}
    \left.\boldsymbol{n} \times \left( \boldsymbol{E}_{m+1} -\boldsymbol{E}_{m} \right) \right\vert_{m} = 0 \label{boundary1} \\
    \left.\boldsymbol{n} \times \left( \boldsymbol{H}_{m+1} -\boldsymbol{H}_{m} \right) \right\vert_{m} = \boldsymbol{J}_{m} \label{boundary2}
\end{align}
where
\begin{equation}
    \boldsymbol{n} = \begin{pmatrix}
    0\\
    0\\
    1
    \end{pmatrix}, ~ ~\boldsymbol{J}_{m} = \sigma_{m}\boldsymbol{E}_{m+1}, ~ ~ \sigma_{m} = \begin{pmatrix}
    \sigma_{m}^{xx} &\sigma_{m}^{xy}\\
    \sigma_{m}^{yx} &\sigma_{m}^{yy}
    \end{pmatrix}
\end{equation}
Here $\sigma_{m}$ is the optical conductivity tensor of the corresponding two-dimensional carrier gas at the $m^{th}$-interface. Substituting Eqs. \ref{Electric} and \ref{Magnetic} into Eqs. \ref{boundary1} and \ref{boundary2}, one obtains
\begin{equation}
    \begin{pmatrix}
    A_{x,m} \\
    A_{y,m} \\
    B_{x,m} \\
    B_{y,m}
    \end{pmatrix} = I_{m}\begin{pmatrix}
    A_{x,m+1} \\
    A_{y,m+1} \\
    B_{x,m+1} \\
    B_{y,m+1}
    \end{pmatrix}
\end{equation}
where $I_{m}$ is an interface matrix that relates the amplitudes of the EM wave in the adjacent $m^{th}$ and $(m+1)^{th}$ layers. If we define:
\begin{equation}
    U = \begin{pmatrix}
    1 &0 &1 &0 \\
    0 &1 &0 &1
    \end{pmatrix}, ~ ~ ~ ~ V = \begin{pmatrix}
    1 &0 &0 \\
    0 &1 &0
    \end{pmatrix}
\end{equation}
then the interface matrix $I_{m}$ will read:
\begin{equation}
    I_{m} =\begin{pmatrix}
        I_{m}^{11} & I_{m}^{12} \\
        I_{m}^{21} & I_{m}^{22}
    \end{pmatrix} = \begin{pmatrix}
    U \\
    L_{m}
    \end{pmatrix}^{-1} \begin{pmatrix}
    U \\
    R_{m}
    \end{pmatrix}
    \label{interfacematrix}
\end{equation}
where $I_{m}^{ij}$ ($i,j=1,2$) are $2 \times 2$ matrices, 
\begin{widetext}
    \begin{equation}
    L_{m}= \frac{V}{\mu_{0}\mu_{m}} \begin{pmatrix}
    0 &-\frac{k_{z,m}}{\omega} &0 &\frac{k_{z,m}}{\omega} \\
    \frac{\left(  \varepsilon^{\parallel}_{m}k_{x,m}^{2}+\varepsilon^{\perp}_{m}k_{z,m}^{2} \right)}{\varepsilon^{\perp}_{m}\omega k_{z,m}} &0 &-\frac{\left(  \varepsilon^{\parallel}_{m}k_{x,m}^{2}+\varepsilon^{\perp}_{m}k_{z,m}^{2} \right)}{\varepsilon^{\perp}_{m}\omega k_{z,m}} &0 \\
    0 &\frac{k_{x,m}}{\omega} &0 &\frac{k_{x,m}}{\omega}
    \end{pmatrix}
     \label{Leftmatrix}
\end{equation}
\end{widetext}

and
\begin{widetext}
\begin{equation}
    R_{m}=  \frac{V}{\mu_{0}\mu_{m+1}} \begin{pmatrix}
    0 &-\frac{k_{z,m+1}}{\omega} &0 &\frac{k_{z,m+1}}{\omega} \\
    \frac{\left( \varepsilon^{\parallel}_{m+1}k_{x,m+1}^{2}+\varepsilon^{\perp}_{m+1}k_{z,m+1}^{2}\right)}{\varepsilon^{\perp}_{m+1}\omega k_{z,m+1}} &0 &-\frac{\left( \varepsilon^{\parallel}_{m+1}k_{x,m+1}^{2}+\varepsilon^{\perp}_{m+1}k_{z,m+1}^{2}\right)}{\varepsilon^{\perp}_{m+1}\omega k_{z,m+1}} &0 \\
    0 &\frac{k_{x,m+1}}{\omega} &0 &\frac{k_{x,m+1}}{\omega}
    \end{pmatrix} + \begin{pmatrix}
    -\sigma_{m}^{yx} &-\sigma_{m}^{yy} &-\sigma_{m}^{yx} &-\sigma_{m}^{yy} \\
    \sigma_{m}^{xx} &\sigma_{m}^{xy} &\sigma_{m}^{xx} &\sigma_{m}^{xy}
    \end{pmatrix}
     \label{Rightmatrix}
\end{equation}
\end{widetext}
where $k_{z,m}=\sqrt{\frac{\omega^{2}}{c^{2}}\mu^{xx}_{m}\varepsilon^{\parallel}_{m}-\frac{\varepsilon^{\parallel}_{m}}{\varepsilon^{\perp}_{m}}k_{x,m}^{2}}$

We now define a scattering matrix at the $m^{th}$ interface S$_m$ such that:
\begin{equation}
    \begin{pmatrix}
    A_{x,m+1} \\
    A_{y,m+1} \\
    B_{x,m} \\
    B_{y,m}
    \end{pmatrix} = S_{m}\begin{pmatrix}
    A_{x,m} \\
    A_{y,m} \\
    B_{x,m+1} \\
    B_{y,m+1}
    \end{pmatrix}
\end{equation}
This $S_m$ is related to the interface matrix $I_{m}$ by:
\begin{equation}
    S_{m} = \begin{bmatrix}
        \left(I_{m}^{11} \right)^{-1} &-\left(I_{m}^{11} \right)^{-1}I_{m}^{12} \\
        I_{m}^{21}\left(I_{m}^{11} \right)^{-1} &I_{m}^{22}-I_{m}^{21}\left(I_{m}^{11} \right)^{-1}I_{m}^{12}
    \end{bmatrix}
    \label{scatmatrx}
\end{equation}
A global scattering matrix $S$ that describes the scattering processes of an EM wave propagating in a heterostructure composed of N-1 constituent materials is given by the super matrix form:
\begin{widetext}
    \begin{equation}
S  = \left[ \begin{matrix}
S_{1}^{-1} &-P_{0}^{\left(2,1\right)} &0 &0 &... &0 &0\\
-P_{0}^{\left(1,2\right)} &S_{2}^{-1} &-P_{0}^{\left(3,2\right)} &0 &... &0 &0\\
		0 &-P_{0}^{\left(2,3 \right)} &S_{3}^{-1} &-P_{0}^{\left(4,3 \right)} &... &0 &0 \\
		 0 &0 &-P_{0}^{\left( 3,4 \right)} &S_{4}^{-1} &... &0 &0 \\
		 \vdots &\vdots  &\vdots &\vdots &\ddots &\vdots &\vdots \\
		 0 &0 &0 &0 &... &S_{N-1}^{-1} &-P_{0}^{\left(N,N-1\right)} \\
		 0 &0 &0 &0 &... &-P_{0}^{\left(N-1,N \right)} &S_{N}^{-1}
\end{matrix}  \right]^{-1}
\label{globalscattering}
\end{equation}
\end{widetext}
Here the propagation matrices for an EM wave propagating between the $m^{th}$ and $(m+1)^{th}$ interfaces takes the form 
\begin{widetext}
\begin{equation}
    P_{0}^{m,m+1} = \begin{pmatrix}
    e^{ik_{z,m+1}d_{m+1}} &0 &0 &0 \\
    0 &e^{ik_{z,m+1}d_{m+1}} &0 &0 \\
    0 &0 &0 &0 \\
    0 &0 &0 &0
    \end{pmatrix} ,~ ~ ~ P_{0}^{m+1,m} = \begin{pmatrix}
    0 &0 &0 &0 \\
    0 &0 &0 &0 \\
    0 &0 &e^{ik_{z,m+1}d_{m+1}} &0 \\
    0 &0 &0 &e^{ik_{z,m+1}d_{m+1}}
    \end{pmatrix}
\end{equation}
\end{widetext}
and $S_{m}^{-1}$ ($m=1 \div N$) is the inversion of the matrix $S_{m}$ given in Eq. \ref{scatmatrx}. The global scattering matrix $S$ can then be written in terms of 
\begin{equation}
    S = \begin{pmatrix}
        S_{11} & S_{12} &\cdots &S_{1N} \\
        S_{21} &S_{22} &\cdots &S_{2N} \\
        \vdots &\vdots &\ddots & \vdots \\
        S_{N1} &S_{N2} &\cdots &S_{NN}
    \end{pmatrix}
\end{equation}
where $S_{ij}$ ($i,j=1\div N$) is a $4 \times 4$ block matrix element of $S$ that describes the scattering event of the EM wave that starts at the $j^{th}$ interface and ends up at the $i^{th}$ interface. In particular
\begin{equation}
    S_{ij} = \begin{pmatrix}
        S_{ij}^{11} & S_{ij}^{12} \\
        S_{ij}^{21} & S_{ij}^{22}
    \end{pmatrix} = \begin{pmatrix}
        t_{ij} & r_{ij}^{'} \\
        r_{ij} & t_{ij}^{'}
    \end{pmatrix}
\end{equation}
where $S_{ij}^{11}$ and $S_{ij}^{21}$ are the block matrices giving the transmission $t_{ij}$ and reflection $r_{ij}$ coefficients associated with the incident wave propagating along the +z direction . In contrast, $S_{ij}^{22}$ and $S_{ij}^{12}$ ($t_{ij}'$ and $r_{ij}'$) correspond to the incident wave propagating along the -z direction. For instance, the reflection coefficient of the entire system with N interfaces is derived from the $S_{11}^{21}$ element whereas the transmission coefficient of the entire system is obtained from the $S_{N1}^{11}$ element. In summary, using a global scattering matrix one can compute the optical response of the entire structure because the global scattering matrix captures what happen at each interface and within each layer of the structure. In the main text, we have calculated the imaginary part of $S_{11}^{21}$ and used it to reveal the dispersion relations for the surface plasmon-phonon-magnon polariton in a TI/AFM structure.

\bibliography{Ref}

\begin{thebibliography}{75}%
\makeatletter
\providecommand \@ifxundefined [1]{%
 \@ifx{#1\undefined}
}%
\providecommand \@ifnum [1]{%
 \ifnum #1\expandafter \@firstoftwo
 \else \expandafter \@secondoftwo
 \fi
}%
\providecommand \@ifx [1]{%
 \ifx #1\expandafter \@firstoftwo
 \else \expandafter \@secondoftwo
 \fi
}%
\providecommand \natexlab [1]{#1}%
\providecommand \enquote  [1]{``#1''}%
\providecommand \bibnamefont  [1]{#1}%
\providecommand \bibfnamefont [1]{#1}%
\providecommand \citenamefont [1]{#1}%
\providecommand \href@noop [0]{\@secondoftwo}%
\providecommand \href [0]{\begingroup \@sanitize@url \@href}%
\providecommand \@href[1]{\@@startlink{#1}\@@href}%
\providecommand \@@href[1]{\endgroup#1\@@endlink}%
\providecommand \@sanitize@url [0]{\catcode `\\12\catcode `\$12\catcode
  `\&12\catcode `\#12\catcode `\^12\catcode `\_12\catcode `\%12\relax}%
\providecommand \@@startlink[1]{}%
\providecommand \@@endlink[0]{}%
\providecommand \url  [0]{\begingroup\@sanitize@url \@url }%
\providecommand \@url [1]{\endgroup\@href {#1}{\urlprefix }}%
\providecommand \urlprefix  [0]{URL }%
\providecommand \Eprint [0]{\href }%
\providecommand \doibase [0]{https://doi.org/}%
\providecommand \selectlanguage [0]{\@gobble}%
\providecommand \bibinfo  [0]{\@secondoftwo}%
\providecommand \bibfield  [0]{\@secondoftwo}%
\providecommand \translation [1]{[#1]}%
\providecommand \BibitemOpen [0]{}%
\providecommand \bibitemStop [0]{}%
\providecommand \bibitemNoStop [0]{.\EOS\space}%
\providecommand \EOS [0]{\spacefactor3000\relax}%
\providecommand \BibitemShut  [1]{\csname bibitem#1\endcsname}%
\let\auto@bib@innerbib\@empty
\bibitem [{\citenamefont {Siegel}(2003)}]{Siegel2003}%
  \BibitemOpen
  \bibfield  {author} {\bibinfo {author} {\bibfnamefont {P.~H.}\ \bibnamefont
  {Siegel}},\ }\bibfield  {title} {\bibinfo {title} {Thz technology: An
  overview},\ }\href@noop {} {\bibfield  {journal} {\bibinfo  {journal}
  {Terahertz Sensing Technology: Volume 1: Electronic Devices and Advanced
  Systems Technology}\ ,\ \bibinfo {pages} {1}} (\bibinfo {year}
  {2003})}\BibitemShut {NoStop}%
\bibitem [{\citenamefont {Pawar}\ \emph {et~al.}(2013)\citenamefont {Pawar},
  \citenamefont {Sonawane}, \citenamefont {Erande},\ and\ \citenamefont
  {Derle}}]{Pawar2013}%
  \BibitemOpen
  \bibfield  {author} {\bibinfo {author} {\bibfnamefont {A.~Y.}\ \bibnamefont
  {Pawar}}, \bibinfo {author} {\bibfnamefont {D.~D.}\ \bibnamefont {Sonawane}},
  \bibinfo {author} {\bibfnamefont {K.~B.}\ \bibnamefont {Erande}},\ and\
  \bibinfo {author} {\bibfnamefont {D.~V.}\ \bibnamefont {Derle}},\ }\bibfield
  {title} {\bibinfo {title} {Terahertz technology and its applications},\
  }\href@noop {} {\bibfield  {journal} {\bibinfo  {journal} {Drug invention
  today}\ }\textbf {\bibinfo {volume} {5}},\ \bibinfo {pages} {157} (\bibinfo
  {year} {2013})}\BibitemShut {NoStop}%
\bibitem [{\citenamefont {Walowski}\ and\ \citenamefont
  {M{\"u}nzenberg}(2016)}]{Walowski2016}%
  \BibitemOpen
  \bibfield  {author} {\bibinfo {author} {\bibfnamefont {J.}~\bibnamefont
  {Walowski}}\ and\ \bibinfo {author} {\bibfnamefont {M.}~\bibnamefont
  {M{\"u}nzenberg}},\ }\bibfield  {title} {\bibinfo {title} {Perspective:
  Ultrafast magnetism and thz spintronics},\ }\href@noop {} {\bibfield
  {journal} {\bibinfo  {journal} {Journal of Applied Physics}\ }\textbf
  {\bibinfo {volume} {120}},\ \bibinfo {pages} {140901} (\bibinfo {year}
  {2016})}\BibitemShut {NoStop}%
\bibitem [{\citenamefont {Zaytsev}\ \emph {et~al.}(2019)\citenamefont
  {Zaytsev}, \citenamefont {Dolganova}, \citenamefont {Chernomyrdin},
  \citenamefont {Katyba}, \citenamefont {Gavdush}, \citenamefont {Cherkasova},
  \citenamefont {Komandin}, \citenamefont {Shchedrina}, \citenamefont {Khodan},
  \citenamefont {Ponomarev} \emph {et~al.}}]{Zaytsev2019}%
  \BibitemOpen
  \bibfield  {author} {\bibinfo {author} {\bibfnamefont {K.}~\bibnamefont
  {Zaytsev}}, \bibinfo {author} {\bibfnamefont {I.}~\bibnamefont {Dolganova}},
  \bibinfo {author} {\bibfnamefont {N.}~\bibnamefont {Chernomyrdin}}, \bibinfo
  {author} {\bibfnamefont {G.}~\bibnamefont {Katyba}}, \bibinfo {author}
  {\bibfnamefont {A.}~\bibnamefont {Gavdush}}, \bibinfo {author} {\bibfnamefont
  {O.}~\bibnamefont {Cherkasova}}, \bibinfo {author} {\bibfnamefont
  {G.}~\bibnamefont {Komandin}}, \bibinfo {author} {\bibfnamefont
  {M.}~\bibnamefont {Shchedrina}}, \bibinfo {author} {\bibfnamefont
  {A.}~\bibnamefont {Khodan}}, \bibinfo {author} {\bibfnamefont
  {D.}~\bibnamefont {Ponomarev}}, \emph {et~al.},\ }\bibfield  {title}
  {\bibinfo {title} {The progress and perspectives of terahertz technology for
  diagnosis of neoplasms: a review},\ }\href@noop {} {\bibfield  {journal}
  {\bibinfo  {journal} {Journal of Optics}\ }\textbf {\bibinfo {volume} {22}},\
  \bibinfo {pages} {013001} (\bibinfo {year} {2019})}\BibitemShut {NoStop}%
\bibitem [{\citenamefont {Amini}\ \emph {et~al.}(2021)\citenamefont {Amini},
  \citenamefont {Jahangiri}, \citenamefont {Ameri},\ and\ \citenamefont
  {Hemmatian}}]{Amini2021}%
  \BibitemOpen
  \bibfield  {author} {\bibinfo {author} {\bibfnamefont {T.}~\bibnamefont
  {Amini}}, \bibinfo {author} {\bibfnamefont {F.}~\bibnamefont {Jahangiri}},
  \bibinfo {author} {\bibfnamefont {Z.}~\bibnamefont {Ameri}},\ and\ \bibinfo
  {author} {\bibfnamefont {M.~A.}\ \bibnamefont {Hemmatian}},\ }\bibfield
  {title} {\bibinfo {title} {A review of feasible applications of thz waves in
  medical diagnostics and treatments},\ }\href@noop {} {\bibfield  {journal}
  {\bibinfo  {journal} {Journal of Lasers in Medical Sciences}\ }\textbf
  {\bibinfo {volume} {12}} (\bibinfo {year} {2021})}\BibitemShut {NoStop}%
\bibitem [{\citenamefont {Burford}\ and\ \citenamefont
  {El-Shenawee}(2017)}]{Burford2017}%
  \BibitemOpen
  \bibfield  {author} {\bibinfo {author} {\bibfnamefont {N.~M.}\ \bibnamefont
  {Burford}}\ and\ \bibinfo {author} {\bibfnamefont {M.~O.}\ \bibnamefont
  {El-Shenawee}},\ }\bibfield  {title} {\bibinfo {title} {Review of terahertz
  photoconductive antenna technology},\ }\href@noop {} {\bibfield  {journal}
  {\bibinfo  {journal} {Optical Engineering}\ }\textbf {\bibinfo {volume}
  {56}},\ \bibinfo {pages} {010901} (\bibinfo {year} {2017})}\BibitemShut
  {NoStop}%
\bibitem [{\citenamefont {Dang}\ \emph
  {et~al.}(2020{\natexlab{a}})\citenamefont {Dang}, \citenamefont {Hawecker},
  \citenamefont {Rongione}, \citenamefont {Baez~Flores}, \citenamefont {To},
  \citenamefont {Rojas-Sanchez}, \citenamefont {Nong}, \citenamefont
  {Mangeney}, \citenamefont {Tignon}, \citenamefont {Godel} \emph
  {et~al.}}]{Dang2020a}%
  \BibitemOpen
  \bibfield  {author} {\bibinfo {author} {\bibfnamefont {T.}~\bibnamefont
  {Dang}}, \bibinfo {author} {\bibfnamefont {J.}~\bibnamefont {Hawecker}},
  \bibinfo {author} {\bibfnamefont {E.}~\bibnamefont {Rongione}}, \bibinfo
  {author} {\bibfnamefont {G.}~\bibnamefont {Baez~Flores}}, \bibinfo {author}
  {\bibfnamefont {D.}~\bibnamefont {To}}, \bibinfo {author} {\bibfnamefont
  {J.}~\bibnamefont {Rojas-Sanchez}}, \bibinfo {author} {\bibfnamefont
  {H.}~\bibnamefont {Nong}}, \bibinfo {author} {\bibfnamefont {J.}~\bibnamefont
  {Mangeney}}, \bibinfo {author} {\bibfnamefont {J.}~\bibnamefont {Tignon}},
  \bibinfo {author} {\bibfnamefont {F.}~\bibnamefont {Godel}}, \emph {et~al.},\
  }\bibfield  {title} {\bibinfo {title} {Ultrafast spin-currents and charge
  conversion at 3 d-5 d interfaces probed by time-domain terahertz
  spectroscopy},\ }\href@noop {} {\bibfield  {journal} {\bibinfo  {journal}
  {Applied Physics Reviews}\ }\textbf {\bibinfo {volume} {7}},\ \bibinfo
  {pages} {041409} (\bibinfo {year} {2020}{\natexlab{a}})}\BibitemShut
  {NoStop}%
\bibitem [{\citenamefont {Papaioannou}\ and\ \citenamefont
  {Beigang}(2021)}]{Papaioannou2021}%
  \BibitemOpen
  \bibfield  {author} {\bibinfo {author} {\bibfnamefont {E.~T.}\ \bibnamefont
  {Papaioannou}}\ and\ \bibinfo {author} {\bibfnamefont {R.}~\bibnamefont
  {Beigang}},\ }\bibfield  {title} {\bibinfo {title} {Thz spintronic emitters:
  a review on achievements and future challenges},\ }\href@noop {} {\bibfield
  {journal} {\bibinfo  {journal} {Nanophotonics}\ }\textbf {\bibinfo {volume}
  {10}},\ \bibinfo {pages} {1243} (\bibinfo {year} {2021})}\BibitemShut
  {NoStop}%
\bibitem [{\citenamefont {Wu}\ \emph {et~al.}(2021)\citenamefont {Wu},
  \citenamefont {Yaw~Ameyaw}, \citenamefont {Doty},\ and\ \citenamefont
  {Jungfleisch}}]{Wu2021}%
  \BibitemOpen
  \bibfield  {author} {\bibinfo {author} {\bibfnamefont {W.}~\bibnamefont
  {Wu}}, \bibinfo {author} {\bibfnamefont {C.}~\bibnamefont {Yaw~Ameyaw}},
  \bibinfo {author} {\bibfnamefont {M.~F.}\ \bibnamefont {Doty}},\ and\
  \bibinfo {author} {\bibfnamefont {M.~B.}\ \bibnamefont {Jungfleisch}},\
  }\bibfield  {title} {\bibinfo {title} {Principles of spintronic thz
  emitters},\ }\href@noop {} {\bibfield  {journal} {\bibinfo  {journal}
  {Journal of Applied Physics}\ }\textbf {\bibinfo {volume} {130}},\ \bibinfo
  {pages} {091101} (\bibinfo {year} {2021})}\BibitemShut {NoStop}%
\bibitem [{\citenamefont {Seifert}\ \emph {et~al.}(2022)\citenamefont
  {Seifert}, \citenamefont {Cheng}, \citenamefont {Wei}, \citenamefont
  {Kampfrath},\ and\ \citenamefont {Qi}}]{Seifert2022}%
  \BibitemOpen
  \bibfield  {author} {\bibinfo {author} {\bibfnamefont {T.~S.}\ \bibnamefont
  {Seifert}}, \bibinfo {author} {\bibfnamefont {L.}~\bibnamefont {Cheng}},
  \bibinfo {author} {\bibfnamefont {Z.}~\bibnamefont {Wei}}, \bibinfo {author}
  {\bibfnamefont {T.}~\bibnamefont {Kampfrath}},\ and\ \bibinfo {author}
  {\bibfnamefont {J.}~\bibnamefont {Qi}},\ }\href@noop {} {\bibinfo {title}
  {Spintronic sources of ultrashort terahertz electromagnetic pulses}}
  (\bibinfo {year} {2022})\BibitemShut {NoStop}%
\bibitem [{\citenamefont {Di~Pietro}\ \emph {et~al.}(2013)\citenamefont
  {Di~Pietro}, \citenamefont {Ortolani}, \citenamefont {Limaj}, \citenamefont
  {Di~Gaspare}, \citenamefont {Giliberti}, \citenamefont {Giorgianni},
  \citenamefont {Brahlek}, \citenamefont {Bansal}, \citenamefont {Koirala},
  \citenamefont {Oh} \emph {et~al.}}]{Di2013}%
  \BibitemOpen
  \bibfield  {author} {\bibinfo {author} {\bibfnamefont {P.}~\bibnamefont
  {Di~Pietro}}, \bibinfo {author} {\bibfnamefont {M.}~\bibnamefont {Ortolani}},
  \bibinfo {author} {\bibfnamefont {O.}~\bibnamefont {Limaj}}, \bibinfo
  {author} {\bibfnamefont {A.}~\bibnamefont {Di~Gaspare}}, \bibinfo {author}
  {\bibfnamefont {V.}~\bibnamefont {Giliberti}}, \bibinfo {author}
  {\bibfnamefont {F.}~\bibnamefont {Giorgianni}}, \bibinfo {author}
  {\bibfnamefont {M.}~\bibnamefont {Brahlek}}, \bibinfo {author} {\bibfnamefont
  {N.}~\bibnamefont {Bansal}}, \bibinfo {author} {\bibfnamefont
  {N.}~\bibnamefont {Koirala}}, \bibinfo {author} {\bibfnamefont
  {S.}~\bibnamefont {Oh}}, \emph {et~al.},\ }\bibfield  {title} {\bibinfo
  {title} {Observation of dirac plasmons in a topological insulator},\
  }\href@noop {} {\bibfield  {journal} {\bibinfo  {journal} {Nature
  nanotechnology}\ }\textbf {\bibinfo {volume} {8}},\ \bibinfo {pages} {556}
  (\bibinfo {year} {2013})}\BibitemShut {NoStop}%
\bibitem [{\citenamefont {Stauber}\ \emph {et~al.}(2017)\citenamefont
  {Stauber}, \citenamefont {G{\'o}mez-Santos},\ and\ \citenamefont
  {Brey}}]{Stauber2017}%
  \BibitemOpen
  \bibfield  {author} {\bibinfo {author} {\bibfnamefont {T.}~\bibnamefont
  {Stauber}}, \bibinfo {author} {\bibfnamefont {G.}~\bibnamefont
  {G{\'o}mez-Santos}},\ and\ \bibinfo {author} {\bibfnamefont {L.}~\bibnamefont
  {Brey}},\ }\bibfield  {title} {\bibinfo {title} {Plasmonics in topological
  insulators: Spin--charge separation, the influence of the inversion layer,
  and phonon--plasmon coupling},\ }\href@noop {} {\bibfield  {journal}
  {\bibinfo  {journal} {Acs Photonics}\ }\textbf {\bibinfo {volume} {4}},\
  \bibinfo {pages} {2978} (\bibinfo {year} {2017})}\BibitemShut {NoStop}%
\bibitem [{\citenamefont {Ginley}\ \emph {et~al.}(2018)\citenamefont {Ginley},
  \citenamefont {Wang}, \citenamefont {Wang},\ and\ \citenamefont
  {Law}}]{Ginley2018}%
  \BibitemOpen
  \bibfield  {author} {\bibinfo {author} {\bibfnamefont {T.}~\bibnamefont
  {Ginley}}, \bibinfo {author} {\bibfnamefont {Y.}~\bibnamefont {Wang}},
  \bibinfo {author} {\bibfnamefont {Z.}~\bibnamefont {Wang}},\ and\ \bibinfo
  {author} {\bibfnamefont {S.}~\bibnamefont {Law}},\ }\bibfield  {title}
  {\bibinfo {title} {Dirac plasmons and beyond: the past, present, and future
  of plasmonics in 3d topological insulators},\ }\href@noop {} {\bibfield
  {journal} {\bibinfo  {journal} {MRS Communications}\ }\textbf {\bibinfo
  {volume} {8}},\ \bibinfo {pages} {782} (\bibinfo {year} {2018})}\BibitemShut
  {NoStop}%
\bibitem [{\citenamefont {Di~Pietro}\ \emph {et~al.}(2020)\citenamefont
  {Di~Pietro}, \citenamefont {Adhlakha}, \citenamefont {Piccirilli},
  \citenamefont {Di~Gaspare}, \citenamefont {Moon}, \citenamefont {Oh},
  \citenamefont {Di~Mitri}, \citenamefont {Spampinati}, \citenamefont
  {Perucchi},\ and\ \citenamefont {Lupi}}]{Di2020}%
  \BibitemOpen
  \bibfield  {author} {\bibinfo {author} {\bibfnamefont {P.}~\bibnamefont
  {Di~Pietro}}, \bibinfo {author} {\bibfnamefont {N.}~\bibnamefont {Adhlakha}},
  \bibinfo {author} {\bibfnamefont {F.}~\bibnamefont {Piccirilli}}, \bibinfo
  {author} {\bibfnamefont {A.}~\bibnamefont {Di~Gaspare}}, \bibinfo {author}
  {\bibfnamefont {J.}~\bibnamefont {Moon}}, \bibinfo {author} {\bibfnamefont
  {S.}~\bibnamefont {Oh}}, \bibinfo {author} {\bibfnamefont {S.}~\bibnamefont
  {Di~Mitri}}, \bibinfo {author} {\bibfnamefont {S.}~\bibnamefont
  {Spampinati}}, \bibinfo {author} {\bibfnamefont {A.}~\bibnamefont
  {Perucchi}},\ and\ \bibinfo {author} {\bibfnamefont {S.}~\bibnamefont
  {Lupi}},\ }\bibfield  {title} {\bibinfo {title} {Terahertz tuning of dirac
  plasmons in ${\mathrm{bi}}_{2}{\mathrm{se}}_{3}$ topological insulator},\
  }\href {https://doi.org/10.1103/PhysRevLett.124.226403} {\bibfield  {journal}
  {\bibinfo  {journal} {Phys. Rev. Lett.}\ }\textbf {\bibinfo {volume} {124}},\
  \bibinfo {pages} {226403} (\bibinfo {year} {2020})}\BibitemShut {NoStop}%
\bibitem [{\citenamefont {Jungfleisch}\ \emph {et~al.}(2018)\citenamefont
  {Jungfleisch}, \citenamefont {Zhang},\ and\ \citenamefont
  {Hoffmann}}]{Jungfleisch2018}%
  \BibitemOpen
  \bibfield  {author} {\bibinfo {author} {\bibfnamefont {M.~B.}\ \bibnamefont
  {Jungfleisch}}, \bibinfo {author} {\bibfnamefont {W.}~\bibnamefont {Zhang}},\
  and\ \bibinfo {author} {\bibfnamefont {A.}~\bibnamefont {Hoffmann}},\
  }\bibfield  {title} {\bibinfo {title} {Perspectives of antiferromagnetic
  spintronics},\ }\href@noop {} {\bibfield  {journal} {\bibinfo  {journal}
  {Physics Letters A}\ }\textbf {\bibinfo {volume} {382}},\ \bibinfo {pages}
  {865} (\bibinfo {year} {2018})}\BibitemShut {NoStop}%
\bibitem [{\citenamefont {Gibertini}\ \emph {et~al.}(2019)\citenamefont
  {Gibertini}, \citenamefont {Koperski}, \citenamefont {Morpurgo},\ and\
  \citenamefont {Novoselov}}]{Gibertini2019}%
  \BibitemOpen
  \bibfield  {author} {\bibinfo {author} {\bibfnamefont {M.}~\bibnamefont
  {Gibertini}}, \bibinfo {author} {\bibfnamefont {M.}~\bibnamefont {Koperski}},
  \bibinfo {author} {\bibfnamefont {A.~F.}\ \bibnamefont {Morpurgo}},\ and\
  \bibinfo {author} {\bibfnamefont {K.~S.}\ \bibnamefont {Novoselov}},\
  }\bibfield  {title} {\bibinfo {title} {Magnetic 2d materials and
  heterostructures},\ }\href@noop {} {\bibfield  {journal} {\bibinfo  {journal}
  {Nature nanotechnology}\ }\textbf {\bibinfo {volume} {14}},\ \bibinfo {pages}
  {408} (\bibinfo {year} {2019})}\BibitemShut {NoStop}%
\bibitem [{\citenamefont {Huang}\ \emph {et~al.}(2020)\citenamefont {Huang},
  \citenamefont {McGuire}, \citenamefont {May}, \citenamefont {Xiao},
  \citenamefont {Jarillo-Herrero},\ and\ \citenamefont {Xu}}]{Huang2020}%
  \BibitemOpen
  \bibfield  {author} {\bibinfo {author} {\bibfnamefont {B.}~\bibnamefont
  {Huang}}, \bibinfo {author} {\bibfnamefont {M.~A.}\ \bibnamefont {McGuire}},
  \bibinfo {author} {\bibfnamefont {A.~F.}\ \bibnamefont {May}}, \bibinfo
  {author} {\bibfnamefont {D.}~\bibnamefont {Xiao}}, \bibinfo {author}
  {\bibfnamefont {P.}~\bibnamefont {Jarillo-Herrero}},\ and\ \bibinfo {author}
  {\bibfnamefont {X.}~\bibnamefont {Xu}},\ }\bibfield  {title} {\bibinfo
  {title} {Emergent phenomena and proximity effects in two-dimensional magnets
  and heterostructures},\ }\href@noop {} {\bibfield  {journal} {\bibinfo
  {journal} {Nature Materials}\ }\textbf {\bibinfo {volume} {19}},\ \bibinfo
  {pages} {1276} (\bibinfo {year} {2020})}\BibitemShut {NoStop}%
\bibitem [{\citenamefont {Zhang}\ \emph {et~al.}(2020)\citenamefont {Zhang},
  \citenamefont {Li}, \citenamefont {Weber}, \citenamefont {Goldberger},
  \citenamefont {Mak},\ and\ \citenamefont {Shan}}]{Zhang2020}%
  \BibitemOpen
  \bibfield  {author} {\bibinfo {author} {\bibfnamefont {X.-X.}\ \bibnamefont
  {Zhang}}, \bibinfo {author} {\bibfnamefont {L.}~\bibnamefont {Li}}, \bibinfo
  {author} {\bibfnamefont {D.}~\bibnamefont {Weber}}, \bibinfo {author}
  {\bibfnamefont {J.}~\bibnamefont {Goldberger}}, \bibinfo {author}
  {\bibfnamefont {K.~F.}\ \bibnamefont {Mak}},\ and\ \bibinfo {author}
  {\bibfnamefont {J.}~\bibnamefont {Shan}},\ }\bibfield  {title} {\bibinfo
  {title} {Gate-tunable spin waves in antiferromagnetic atomic bilayers},\
  }\href@noop {} {\bibfield  {journal} {\bibinfo  {journal} {Nature materials}\
  }\textbf {\bibinfo {volume} {19}},\ \bibinfo {pages} {838} (\bibinfo {year}
  {2020})}\BibitemShut {NoStop}%
\bibitem [{\citenamefont {Yang}\ \emph {et~al.}(2021)\citenamefont {Yang},
  \citenamefont {Zhang},\ and\ \citenamefont {Jiang}}]{Yang2021}%
  \BibitemOpen
  \bibfield  {author} {\bibinfo {author} {\bibfnamefont {S.}~\bibnamefont
  {Yang}}, \bibinfo {author} {\bibfnamefont {T.}~\bibnamefont {Zhang}},\ and\
  \bibinfo {author} {\bibfnamefont {C.}~\bibnamefont {Jiang}},\ }\bibfield
  {title} {\bibinfo {title} {van der waals magnets: material family, detection
  and modulation of magnetism, and perspective in spintronics},\ }\href@noop {}
  {\bibfield  {journal} {\bibinfo  {journal} {Advanced Science}\ }\textbf
  {\bibinfo {volume} {8}},\ \bibinfo {pages} {2002488} (\bibinfo {year}
  {2021})}\BibitemShut {NoStop}%
\bibitem [{\citenamefont {Jiang}\ \emph {et~al.}(2021)\citenamefont {Jiang},
  \citenamefont {Liu}, \citenamefont {Xing}, \citenamefont {Liu}, \citenamefont
  {Guo}, \citenamefont {Liu},\ and\ \citenamefont {Zhao}}]{Jiang2021}%
  \BibitemOpen
  \bibfield  {author} {\bibinfo {author} {\bibfnamefont {X.}~\bibnamefont
  {Jiang}}, \bibinfo {author} {\bibfnamefont {Q.}~\bibnamefont {Liu}}, \bibinfo
  {author} {\bibfnamefont {J.}~\bibnamefont {Xing}}, \bibinfo {author}
  {\bibfnamefont {N.}~\bibnamefont {Liu}}, \bibinfo {author} {\bibfnamefont
  {Y.}~\bibnamefont {Guo}}, \bibinfo {author} {\bibfnamefont {Z.}~\bibnamefont
  {Liu}},\ and\ \bibinfo {author} {\bibfnamefont {J.}~\bibnamefont {Zhao}},\
  }\bibfield  {title} {\bibinfo {title} {Recent progress on 2d magnets:
  Fundamental mechanism, structural design and modification},\ }\href@noop {}
  {\bibfield  {journal} {\bibinfo  {journal} {Applied Physics Reviews}\
  }\textbf {\bibinfo {volume} {8}},\ \bibinfo {pages} {031305} (\bibinfo {year}
  {2021})}\BibitemShut {NoStop}%
\bibitem [{\citenamefont {Zhang}\ \emph {et~al.}(2021)\citenamefont {Zhang},
  \citenamefont {Ozerov}, \citenamefont {Bostr{\"o}m}, \citenamefont {Cui},
  \citenamefont {Suri}, \citenamefont {Jiang}, \citenamefont {Wang},
  \citenamefont {Wu}, \citenamefont {Hwangbo}, \citenamefont {Chu} \emph
  {et~al.}}]{Zhang2021}%
  \BibitemOpen
  \bibfield  {author} {\bibinfo {author} {\bibfnamefont {Q.}~\bibnamefont
  {Zhang}}, \bibinfo {author} {\bibfnamefont {M.}~\bibnamefont {Ozerov}},
  \bibinfo {author} {\bibfnamefont {E.~V.}\ \bibnamefont {Bostr{\"o}m}},
  \bibinfo {author} {\bibfnamefont {J.}~\bibnamefont {Cui}}, \bibinfo {author}
  {\bibfnamefont {N.}~\bibnamefont {Suri}}, \bibinfo {author} {\bibfnamefont
  {Q.}~\bibnamefont {Jiang}}, \bibinfo {author} {\bibfnamefont
  {C.}~\bibnamefont {Wang}}, \bibinfo {author} {\bibfnamefont {F.}~\bibnamefont
  {Wu}}, \bibinfo {author} {\bibfnamefont {K.}~\bibnamefont {Hwangbo}},
  \bibinfo {author} {\bibfnamefont {J.-H.}\ \bibnamefont {Chu}}, \emph
  {et~al.},\ }\bibfield  {title} {\bibinfo {title} {Coherent strong-coupling of
  terahertz magnons and phonons in a van der waals antiferromagnetic
  insulator},\ }\href@noop {} {\bibfield  {journal} {\bibinfo  {journal} {arXiv
  preprint arXiv:2108.11619}\ } (\bibinfo {year} {2021})}\BibitemShut {NoStop}%
\bibitem [{\citenamefont {Belvin}\ \emph {et~al.}(2021)\citenamefont {Belvin},
  \citenamefont {Baldini}, \citenamefont {Ozel}, \citenamefont {Mao},
  \citenamefont {Po}, \citenamefont {Allington}, \citenamefont {Son},
  \citenamefont {Kim}, \citenamefont {Kim}, \citenamefont {Hwang} \emph
  {et~al.}}]{Belvin2021}%
  \BibitemOpen
  \bibfield  {author} {\bibinfo {author} {\bibfnamefont {C.~A.}\ \bibnamefont
  {Belvin}}, \bibinfo {author} {\bibfnamefont {E.}~\bibnamefont {Baldini}},
  \bibinfo {author} {\bibfnamefont {I.~O.}\ \bibnamefont {Ozel}}, \bibinfo
  {author} {\bibfnamefont {D.}~\bibnamefont {Mao}}, \bibinfo {author}
  {\bibfnamefont {H.~C.}\ \bibnamefont {Po}}, \bibinfo {author} {\bibfnamefont
  {C.~J.}\ \bibnamefont {Allington}}, \bibinfo {author} {\bibfnamefont
  {S.}~\bibnamefont {Son}}, \bibinfo {author} {\bibfnamefont {B.~H.}\
  \bibnamefont {Kim}}, \bibinfo {author} {\bibfnamefont {J.}~\bibnamefont
  {Kim}}, \bibinfo {author} {\bibfnamefont {I.}~\bibnamefont {Hwang}}, \emph
  {et~al.},\ }\bibfield  {title} {\bibinfo {title} {Exciton-driven
  antiferromagnetic metal in a correlated van der waals insulator},\
  }\href@noop {} {\bibfield  {journal} {\bibinfo  {journal} {Nature
  communications}\ }\textbf {\bibinfo {volume} {12}},\ \bibinfo {pages} {1}
  (\bibinfo {year} {2021})}\BibitemShut {NoStop}%
\bibitem [{\citenamefont {Han}\ \emph {et~al.}(2019)\citenamefont {Han},
  \citenamefont {Zhang}, \citenamefont {Hou}, \citenamefont {Siddiqui},\ and\
  \citenamefont {Liu}}]{Han2019}%
  \BibitemOpen
  \bibfield  {author} {\bibinfo {author} {\bibfnamefont {J.}~\bibnamefont
  {Han}}, \bibinfo {author} {\bibfnamefont {P.}~\bibnamefont {Zhang}}, \bibinfo
  {author} {\bibfnamefont {J.~T.}\ \bibnamefont {Hou}}, \bibinfo {author}
  {\bibfnamefont {S.~A.}\ \bibnamefont {Siddiqui}},\ and\ \bibinfo {author}
  {\bibfnamefont {L.}~\bibnamefont {Liu}},\ }\bibfield  {title} {\bibinfo
  {title} {Mutual control of coherent spin waves and magnetic domain walls in a
  magnonic device},\ }\href@noop {} {\bibfield  {journal} {\bibinfo  {journal}
  {Science}\ }\textbf {\bibinfo {volume} {366}},\ \bibinfo {pages} {1121}
  (\bibinfo {year} {2019})}\BibitemShut {NoStop}%
\bibitem [{\citenamefont {Fulara}\ \emph {et~al.}(2019)\citenamefont {Fulara},
  \citenamefont {Zahedinejad}, \citenamefont {Khymyn}, \citenamefont {Awad},
  \citenamefont {Muralidhar}, \citenamefont {Dvornik},\ and\ \citenamefont
  {{\AA}kerman}}]{Fulara2019}%
  \BibitemOpen
  \bibfield  {author} {\bibinfo {author} {\bibfnamefont {H.}~\bibnamefont
  {Fulara}}, \bibinfo {author} {\bibfnamefont {M.}~\bibnamefont {Zahedinejad}},
  \bibinfo {author} {\bibfnamefont {R.}~\bibnamefont {Khymyn}}, \bibinfo
  {author} {\bibfnamefont {A.}~\bibnamefont {Awad}}, \bibinfo {author}
  {\bibfnamefont {S.}~\bibnamefont {Muralidhar}}, \bibinfo {author}
  {\bibfnamefont {M.}~\bibnamefont {Dvornik}},\ and\ \bibinfo {author}
  {\bibfnamefont {J.}~\bibnamefont {{\AA}kerman}},\ }\bibfield  {title}
  {\bibinfo {title} {Spin-orbit torque--driven propagating spin waves},\
  }\href@noop {} {\bibfield  {journal} {\bibinfo  {journal} {Science advances}\
  }\textbf {\bibinfo {volume} {5}},\ \bibinfo {pages} {eaax8467} (\bibinfo
  {year} {2019})}\BibitemShut {NoStop}%
\bibitem [{\citenamefont {Wang}\ \emph {et~al.}(2019)\citenamefont {Wang},
  \citenamefont {Zhu}, \citenamefont {Yang}, \citenamefont {Lee}, \citenamefont
  {Mishra}, \citenamefont {Go}, \citenamefont {Oh}, \citenamefont {Kim},
  \citenamefont {Cai}, \citenamefont {Liu} \emph {et~al.}}]{Wang2019}%
  \BibitemOpen
  \bibfield  {author} {\bibinfo {author} {\bibfnamefont {Y.}~\bibnamefont
  {Wang}}, \bibinfo {author} {\bibfnamefont {D.}~\bibnamefont {Zhu}}, \bibinfo
  {author} {\bibfnamefont {Y.}~\bibnamefont {Yang}}, \bibinfo {author}
  {\bibfnamefont {K.}~\bibnamefont {Lee}}, \bibinfo {author} {\bibfnamefont
  {R.}~\bibnamefont {Mishra}}, \bibinfo {author} {\bibfnamefont
  {G.}~\bibnamefont {Go}}, \bibinfo {author} {\bibfnamefont {S.-H.}\
  \bibnamefont {Oh}}, \bibinfo {author} {\bibfnamefont {D.-H.}\ \bibnamefont
  {Kim}}, \bibinfo {author} {\bibfnamefont {K.}~\bibnamefont {Cai}}, \bibinfo
  {author} {\bibfnamefont {E.}~\bibnamefont {Liu}}, \emph {et~al.},\ }\bibfield
   {title} {\bibinfo {title} {Magnetization switching by magnon-mediated spin
  torque through an antiferromagnetic insulator},\ }\href@noop {} {\bibfield
  {journal} {\bibinfo  {journal} {Science}\ }\textbf {\bibinfo {volume}
  {366}},\ \bibinfo {pages} {1125} (\bibinfo {year} {2019})}\BibitemShut
  {NoStop}%
\bibitem [{\citenamefont {Liu}\ \emph {et~al.}(2021)\citenamefont {Liu},
  \citenamefont {Granados~del \'Aguila}, \citenamefont {Bhowmick},
  \citenamefont {Gan}, \citenamefont {Thu Ha~Do}, \citenamefont {Prosnikov},
  \citenamefont {Sedmidubsk\'y}, \citenamefont {Sofer}, \citenamefont
  {Christianen}, \citenamefont {Sengupta},\ and\ \citenamefont
  {Xiong}}]{Liu2021}%
  \BibitemOpen
  \bibfield  {author} {\bibinfo {author} {\bibfnamefont {S.}~\bibnamefont
  {Liu}}, \bibinfo {author} {\bibfnamefont {A.}~\bibnamefont {Granados~del
  \'Aguila}}, \bibinfo {author} {\bibfnamefont {D.}~\bibnamefont {Bhowmick}},
  \bibinfo {author} {\bibfnamefont {C.~K.}\ \bibnamefont {Gan}}, \bibinfo
  {author} {\bibfnamefont {T.}~\bibnamefont {Thu Ha~Do}}, \bibinfo {author}
  {\bibfnamefont {M.~A.}\ \bibnamefont {Prosnikov}}, \bibinfo {author}
  {\bibfnamefont {D.}~\bibnamefont {Sedmidubsk\'y}}, \bibinfo {author}
  {\bibfnamefont {Z.}~\bibnamefont {Sofer}}, \bibinfo {author} {\bibfnamefont
  {P.~C.~M.}\ \bibnamefont {Christianen}}, \bibinfo {author} {\bibfnamefont
  {P.}~\bibnamefont {Sengupta}},\ and\ \bibinfo {author} {\bibfnamefont
  {Q.}~\bibnamefont {Xiong}},\ }\bibfield  {title} {\bibinfo {title} {Direct
  observation of magnon-phonon strong coupling in two-dimensional
  antiferromagnet at high magnetic fields},\ }\href
  {https://doi.org/10.1103/PhysRevLett.127.097401} {\bibfield  {journal}
  {\bibinfo  {journal} {Phys. Rev. Lett.}\ }\textbf {\bibinfo {volume} {127}},\
  \bibinfo {pages} {097401} (\bibinfo {year} {2021})}\BibitemShut {NoStop}%
\bibitem [{\citenamefont {Yuan}\ and\ \citenamefont {Wang}(2017)}]{Yuan2017}%
  \BibitemOpen
  \bibfield  {author} {\bibinfo {author} {\bibfnamefont {H.}~\bibnamefont
  {Yuan}}\ and\ \bibinfo {author} {\bibfnamefont {X.}~\bibnamefont {Wang}},\
  }\bibfield  {title} {\bibinfo {title} {Magnon-photon coupling in
  antiferromagnets},\ }\href@noop {} {\bibfield  {journal} {\bibinfo  {journal}
  {Applied Physics Letters}\ }\textbf {\bibinfo {volume} {110}},\ \bibinfo
  {pages} {082403} (\bibinfo {year} {2017})}\BibitemShut {NoStop}%
\bibitem [{\citenamefont {Golovchanskiy}\ \emph {et~al.}(2021)\citenamefont
  {Golovchanskiy}, \citenamefont {Abramov}, \citenamefont {Stolyarov},
  \citenamefont {Weides}, \citenamefont {Ryazanov}, \citenamefont {Golubov},
  \citenamefont {Ustinov},\ and\ \citenamefont
  {Kupriyanov}}]{Golovchanskiy2021}%
  \BibitemOpen
  \bibfield  {author} {\bibinfo {author} {\bibfnamefont {I.~A.}\ \bibnamefont
  {Golovchanskiy}}, \bibinfo {author} {\bibfnamefont {N.~N.}\ \bibnamefont
  {Abramov}}, \bibinfo {author} {\bibfnamefont {V.~S.}\ \bibnamefont
  {Stolyarov}}, \bibinfo {author} {\bibfnamefont {M.}~\bibnamefont {Weides}},
  \bibinfo {author} {\bibfnamefont {V.~V.}\ \bibnamefont {Ryazanov}}, \bibinfo
  {author} {\bibfnamefont {A.~A.}\ \bibnamefont {Golubov}}, \bibinfo {author}
  {\bibfnamefont {A.~V.}\ \bibnamefont {Ustinov}},\ and\ \bibinfo {author}
  {\bibfnamefont {M.~Y.}\ \bibnamefont {Kupriyanov}},\ }\bibfield  {title}
  {\bibinfo {title} {Ultrastrong photon-to-magnon coupling in multilayered
  heterostructures involving superconducting coherence via ferromagnetic
  layers},\ }\href@noop {} {\bibfield  {journal} {\bibinfo  {journal} {Science
  advances}\ }\textbf {\bibinfo {volume} {7}},\ \bibinfo {pages} {eabe8638}
  (\bibinfo {year} {2021})}\BibitemShut {NoStop}%
\bibitem [{\citenamefont {Xiao}\ \emph {et~al.}(2021)\citenamefont {Xiao},
  \citenamefont {Yan}, \citenamefont {Bai}, \citenamefont {Guo}, \citenamefont
  {Hu},\ and\ \citenamefont {Xia}}]{Xiao2021}%
  \BibitemOpen
  \bibfield  {author} {\bibinfo {author} {\bibfnamefont {Y.}~\bibnamefont
  {Xiao}}, \bibinfo {author} {\bibfnamefont {X.~H.}\ \bibnamefont {Yan}},
  \bibinfo {author} {\bibfnamefont {L.~H.}\ \bibnamefont {Bai}}, \bibinfo
  {author} {\bibfnamefont {H.}~\bibnamefont {Guo}}, \bibinfo {author}
  {\bibfnamefont {C.~M.}\ \bibnamefont {Hu}},\ and\ \bibinfo {author}
  {\bibfnamefont {K.}~\bibnamefont {Xia}},\ }\bibfield  {title} {\bibinfo
  {title} {Magnon photon coupling for magnetization antiparallel to the
  magnetic field},\ }\href {https://doi.org/10.1103/PhysRevB.103.104432}
  {\bibfield  {journal} {\bibinfo  {journal} {Phys. Rev. B}\ }\textbf {\bibinfo
  {volume} {103}},\ \bibinfo {pages} {104432} (\bibinfo {year}
  {2021})}\BibitemShut {NoStop}%
\bibitem [{\citenamefont {Henriques}\ \emph {et~al.}(2022)\citenamefont
  {Henriques}, \citenamefont {Ant{\~a}o},\ and\ \citenamefont
  {Peres}}]{Henriques2022}%
  \BibitemOpen
  \bibfield  {author} {\bibinfo {author} {\bibfnamefont {J.}~\bibnamefont
  {Henriques}}, \bibinfo {author} {\bibfnamefont {T.}~\bibnamefont
  {Ant{\~a}o}},\ and\ \bibinfo {author} {\bibfnamefont {N.}~\bibnamefont
  {Peres}},\ }\bibfield  {title} {\bibinfo {title} {Laser induced enhanced
  coupling between photons and squeezed magnons in antiferromagnets},\
  }\href@noop {} {\bibfield  {journal} {\bibinfo  {journal} {Journal of
  Physics: Condensed Matter}\ }\textbf {\bibinfo {volume} {34}},\ \bibinfo
  {pages} {245802} (\bibinfo {year} {2022})}\BibitemShut {NoStop}%
\bibitem [{\citenamefont {T{\"o}rm{\"a}}\ and\ \citenamefont
  {Barnes}(2014)}]{Torma2014}%
  \BibitemOpen
  \bibfield  {author} {\bibinfo {author} {\bibfnamefont {P.}~\bibnamefont
  {T{\"o}rm{\"a}}}\ and\ \bibinfo {author} {\bibfnamefont {W.~L.}\ \bibnamefont
  {Barnes}},\ }\bibfield  {title} {\bibinfo {title} {Strong coupling between
  surface plasmon polaritons and emitters: a review},\ }\href@noop {}
  {\bibfield  {journal} {\bibinfo  {journal} {Reports on Progress in Physics}\
  }\textbf {\bibinfo {volume} {78}},\ \bibinfo {pages} {013901} (\bibinfo
  {year} {2014})}\BibitemShut {NoStop}%
\bibitem [{\citenamefont {Forn-D\'{\i}az}\ \emph {et~al.}(2019)\citenamefont
  {Forn-D\'{\i}az}, \citenamefont {Lamata}, \citenamefont {Rico}, \citenamefont
  {Kono},\ and\ \citenamefont {Solano}}]{FornDiaz2019}%
  \BibitemOpen
  \bibfield  {author} {\bibinfo {author} {\bibfnamefont {P.}~\bibnamefont
  {Forn-D\'{\i}az}}, \bibinfo {author} {\bibfnamefont {L.}~\bibnamefont
  {Lamata}}, \bibinfo {author} {\bibfnamefont {E.}~\bibnamefont {Rico}},
  \bibinfo {author} {\bibfnamefont {J.}~\bibnamefont {Kono}},\ and\ \bibinfo
  {author} {\bibfnamefont {E.}~\bibnamefont {Solano}},\ }\bibfield  {title}
  {\bibinfo {title} {Ultrastrong coupling regimes of light-matter
  interaction},\ }\href {https://doi.org/10.1103/RevModPhys.91.025005}
  {\bibfield  {journal} {\bibinfo  {journal} {Rev. Mod. Phys.}\ }\textbf
  {\bibinfo {volume} {91}},\ \bibinfo {pages} {025005} (\bibinfo {year}
  {2019})}\BibitemShut {NoStop}%
\bibitem [{\citenamefont {Frisk~Kockum}\ \emph {et~al.}(2019)\citenamefont
  {Frisk~Kockum}, \citenamefont {Miranowicz}, \citenamefont {De~Liberato},
  \citenamefont {Savasta},\ and\ \citenamefont {Nori}}]{Frisk2019}%
  \BibitemOpen
  \bibfield  {author} {\bibinfo {author} {\bibfnamefont {A.}~\bibnamefont
  {Frisk~Kockum}}, \bibinfo {author} {\bibfnamefont {A.}~\bibnamefont
  {Miranowicz}}, \bibinfo {author} {\bibfnamefont {S.}~\bibnamefont
  {De~Liberato}}, \bibinfo {author} {\bibfnamefont {S.}~\bibnamefont
  {Savasta}},\ and\ \bibinfo {author} {\bibfnamefont {F.}~\bibnamefont
  {Nori}},\ }\bibfield  {title} {\bibinfo {title} {Ultrastrong coupling between
  light and matter},\ }\href@noop {} {\bibfield  {journal} {\bibinfo  {journal}
  {Nature Reviews Physics}\ }\textbf {\bibinfo {volume} {1}},\ \bibinfo {pages}
  {19} (\bibinfo {year} {2019})}\BibitemShut {NoStop}%
\bibitem [{\citenamefont {Sivarajah}\ \emph {et~al.}(2019)\citenamefont
  {Sivarajah}, \citenamefont {Steinbacher}, \citenamefont {Dastrup},
  \citenamefont {Lu}, \citenamefont {Xiang}, \citenamefont {Ren}, \citenamefont
  {Kamba}, \citenamefont {Cao},\ and\ \citenamefont {Nelson}}]{Sivarajah2019}%
  \BibitemOpen
  \bibfield  {author} {\bibinfo {author} {\bibfnamefont {P.}~\bibnamefont
  {Sivarajah}}, \bibinfo {author} {\bibfnamefont {A.}~\bibnamefont
  {Steinbacher}}, \bibinfo {author} {\bibfnamefont {B.}~\bibnamefont
  {Dastrup}}, \bibinfo {author} {\bibfnamefont {J.}~\bibnamefont {Lu}},
  \bibinfo {author} {\bibfnamefont {M.}~\bibnamefont {Xiang}}, \bibinfo
  {author} {\bibfnamefont {W.}~\bibnamefont {Ren}}, \bibinfo {author}
  {\bibfnamefont {S.}~\bibnamefont {Kamba}}, \bibinfo {author} {\bibfnamefont
  {S.}~\bibnamefont {Cao}},\ and\ \bibinfo {author} {\bibfnamefont {K.~A.}\
  \bibnamefont {Nelson}},\ }\bibfield  {title} {\bibinfo {title} {Thz-frequency
  magnon-phonon-polaritons in the collective strong-coupling regime},\
  }\href@noop {} {\bibfield  {journal} {\bibinfo  {journal} {Journal of Applied
  Physics}\ }\textbf {\bibinfo {volume} {125}},\ \bibinfo {pages} {213103}
  (\bibinfo {year} {2019})}\BibitemShut {NoStop}%
\bibitem [{\citenamefont {Pitarke}\ \emph {et~al.}(2006)\citenamefont
  {Pitarke}, \citenamefont {Silkin}, \citenamefont {Chulkov},\ and\
  \citenamefont {Echenique}}]{Pitarke2006}%
  \BibitemOpen
  \bibfield  {author} {\bibinfo {author} {\bibfnamefont {J.}~\bibnamefont
  {Pitarke}}, \bibinfo {author} {\bibfnamefont {V.}~\bibnamefont {Silkin}},
  \bibinfo {author} {\bibfnamefont {E.}~\bibnamefont {Chulkov}},\ and\ \bibinfo
  {author} {\bibfnamefont {P.}~\bibnamefont {Echenique}},\ }\bibfield  {title}
  {\bibinfo {title} {Theory of surface plasmons and surface-plasmon
  polaritons},\ }\href@noop {} {\bibfield  {journal} {\bibinfo  {journal}
  {Reports on progress in physics}\ }\textbf {\bibinfo {volume} {70}},\
  \bibinfo {pages} {1} (\bibinfo {year} {2006})}\BibitemShut {NoStop}%
\bibitem [{\citenamefont {Zhang}\ \emph {et~al.}(2012)\citenamefont {Zhang},
  \citenamefont {Zhang},\ and\ \citenamefont {Xu}}]{Zhang2012}%
  \BibitemOpen
  \bibfield  {author} {\bibinfo {author} {\bibfnamefont {J.}~\bibnamefont
  {Zhang}}, \bibinfo {author} {\bibfnamefont {L.}~\bibnamefont {Zhang}},\ and\
  \bibinfo {author} {\bibfnamefont {W.}~\bibnamefont {Xu}},\ }\bibfield
  {title} {\bibinfo {title} {Surface plasmon polaritons: physics and
  applications},\ }\href@noop {} {\bibfield  {journal} {\bibinfo  {journal}
  {Journal of Physics D: Applied Physics}\ }\textbf {\bibinfo {volume} {45}},\
  \bibinfo {pages} {113001} (\bibinfo {year} {2012})}\BibitemShut {NoStop}%
\bibitem [{\citenamefont {Stauber}\ \emph {et~al.}(2013)\citenamefont
  {Stauber}, \citenamefont {G\'omez-Santos},\ and\ \citenamefont
  {Brey}}]{Stauber2013}%
  \BibitemOpen
  \bibfield  {author} {\bibinfo {author} {\bibfnamefont {T.}~\bibnamefont
  {Stauber}}, \bibinfo {author} {\bibfnamefont {G.}~\bibnamefont
  {G\'omez-Santos}},\ and\ \bibinfo {author} {\bibfnamefont {L.}~\bibnamefont
  {Brey}},\ }\bibfield  {title} {\bibinfo {title} {Spin-charge separation of
  plasmonic excitations in thin topological insulators},\ }\href
  {https://doi.org/10.1103/PhysRevB.88.205427} {\bibfield  {journal} {\bibinfo
  {journal} {Phys. Rev. B}\ }\textbf {\bibinfo {volume} {88}},\ \bibinfo
  {pages} {205427} (\bibinfo {year} {2013})}\BibitemShut {NoStop}%
\bibitem [{\citenamefont {Qi}\ \emph {et~al.}(2014)\citenamefont {Qi},
  \citenamefont {Liu},\ and\ \citenamefont {Xie}}]{Qi2014}%
  \BibitemOpen
  \bibfield  {author} {\bibinfo {author} {\bibfnamefont {J.}~\bibnamefont
  {Qi}}, \bibinfo {author} {\bibfnamefont {H.}~\bibnamefont {Liu}},\ and\
  \bibinfo {author} {\bibfnamefont {X.~C.}\ \bibnamefont {Xie}},\ }\bibfield
  {title} {\bibinfo {title} {Surface plasmon polaritons in topological
  insulators},\ }\href {https://doi.org/10.1103/PhysRevB.89.155420} {\bibfield
  {journal} {\bibinfo  {journal} {Phys. Rev. B}\ }\textbf {\bibinfo {volume}
  {89}},\ \bibinfo {pages} {155420} (\bibinfo {year} {2014})}\BibitemShut
  {NoStop}%
\bibitem [{\citenamefont {Deshko}\ \emph {et~al.}(2016)\citenamefont {Deshko},
  \citenamefont {Krusin-Elbaum}, \citenamefont {Menon}, \citenamefont
  {Khanikaev},\ and\ \citenamefont {Trevino}}]{Deshko2016}%
  \BibitemOpen
  \bibfield  {author} {\bibinfo {author} {\bibfnamefont {Y.}~\bibnamefont
  {Deshko}}, \bibinfo {author} {\bibfnamefont {L.}~\bibnamefont
  {Krusin-Elbaum}}, \bibinfo {author} {\bibfnamefont {V.}~\bibnamefont
  {Menon}}, \bibinfo {author} {\bibfnamefont {A.}~\bibnamefont {Khanikaev}},\
  and\ \bibinfo {author} {\bibfnamefont {J.}~\bibnamefont {Trevino}},\
  }\bibfield  {title} {\bibinfo {title} {Surface plasmon polaritons in
  topological insulator nano-films and superlattices},\ }\href@noop {}
  {\bibfield  {journal} {\bibinfo  {journal} {Optics Express}\ }\textbf
  {\bibinfo {volume} {24}},\ \bibinfo {pages} {7398} (\bibinfo {year}
  {2016})}\BibitemShut {NoStop}%
\bibitem [{\citenamefont {Wang}\ \emph {et~al.}(2020)\citenamefont {Wang},
  \citenamefont {Ginley}, \citenamefont {Mambakkam}, \citenamefont {Chandan},
  \citenamefont {Zhang}, \citenamefont {Ni},\ and\ \citenamefont
  {Law}}]{Wang2020}%
  \BibitemOpen
  \bibfield  {author} {\bibinfo {author} {\bibfnamefont {Z.}~\bibnamefont
  {Wang}}, \bibinfo {author} {\bibfnamefont {T.~P.}\ \bibnamefont {Ginley}},
  \bibinfo {author} {\bibfnamefont {S.~V.}\ \bibnamefont {Mambakkam}}, \bibinfo
  {author} {\bibfnamefont {G.}~\bibnamefont {Chandan}}, \bibinfo {author}
  {\bibfnamefont {Y.}~\bibnamefont {Zhang}}, \bibinfo {author} {\bibfnamefont
  {C.}~\bibnamefont {Ni}},\ and\ \bibinfo {author} {\bibfnamefont
  {S.}~\bibnamefont {Law}},\ }\bibfield  {title} {\bibinfo {title} {Plasmon
  coupling in topological insulator multilayers},\ }\href
  {https://doi.org/10.1103/PhysRevMaterials.4.115202} {\bibfield  {journal}
  {\bibinfo  {journal} {Phys. Rev. Materials}\ }\textbf {\bibinfo {volume}
  {4}},\ \bibinfo {pages} {115202} (\bibinfo {year} {2020})}\BibitemShut
  {NoStop}%
\bibitem [{\citenamefont {Almeida}\ and\ \citenamefont
  {Mills}(1988)}]{Almeida1988}%
  \BibitemOpen
  \bibfield  {author} {\bibinfo {author} {\bibfnamefont {N.~S.}\ \bibnamefont
  {Almeida}}\ and\ \bibinfo {author} {\bibfnamefont {D.~L.}\ \bibnamefont
  {Mills}},\ }\bibfield  {title} {\bibinfo {title} {Dynamical response of
  antiferromagnets in an oblique magnetic field: Application to surface
  magnons},\ }\href {https://doi.org/10.1103/PhysRevB.37.3400} {\bibfield
  {journal} {\bibinfo  {journal} {Phys. Rev. B}\ }\textbf {\bibinfo {volume}
  {37}},\ \bibinfo {pages} {3400} (\bibinfo {year} {1988})}\BibitemShut
  {NoStop}%
\bibitem [{\citenamefont {Dumelow}\ and\ \citenamefont
  {Oliveros}(1997)}]{Dumelow1997}%
  \BibitemOpen
  \bibfield  {author} {\bibinfo {author} {\bibfnamefont {T.}~\bibnamefont
  {Dumelow}}\ and\ \bibinfo {author} {\bibfnamefont {M.~C.}\ \bibnamefont
  {Oliveros}},\ }\bibfield  {title} {\bibinfo {title} {Continuum model of
  confined magnon polaritons in superlattices of antiferromagnets},\ }\href
  {https://doi.org/10.1103/PhysRevB.55.994} {\bibfield  {journal} {\bibinfo
  {journal} {Phys. Rev. B}\ }\textbf {\bibinfo {volume} {55}},\ \bibinfo
  {pages} {994} (\bibinfo {year} {1997})}\BibitemShut {NoStop}%
\bibitem [{\citenamefont {Sloan}\ \emph {et~al.}(2019)\citenamefont {Sloan},
  \citenamefont {Rivera}, \citenamefont {Joannopoulos}, \citenamefont
  {Kaminer},\ and\ \citenamefont {Solja\ifmmode \check{c}\else
  \v{c}\fi{}i\ifmmode~\acute{c}\else \'{c}\fi{}}}]{Sloan2019}%
  \BibitemOpen
  \bibfield  {author} {\bibinfo {author} {\bibfnamefont {J.}~\bibnamefont
  {Sloan}}, \bibinfo {author} {\bibfnamefont {N.}~\bibnamefont {Rivera}},
  \bibinfo {author} {\bibfnamefont {J.~D.}\ \bibnamefont {Joannopoulos}},
  \bibinfo {author} {\bibfnamefont {I.}~\bibnamefont {Kaminer}},\ and\ \bibinfo
  {author} {\bibfnamefont {M.}~\bibnamefont {Solja\ifmmode \check{c}\else
  \v{c}\fi{}i\ifmmode~\acute{c}\else \'{c}\fi{}}},\ }\bibfield  {title}
  {\bibinfo {title} {Controlling spins with surface magnon polaritons},\ }\href
  {https://doi.org/10.1103/PhysRevB.100.235453} {\bibfield  {journal} {\bibinfo
   {journal} {Phys. Rev. B}\ }\textbf {\bibinfo {volume} {100}},\ \bibinfo
  {pages} {235453} (\bibinfo {year} {2019})}\BibitemShut {NoStop}%
\bibitem [{\citenamefont {Mac\^edo}\ and\ \citenamefont
  {Camley}(2019)}]{Macedo2019}%
  \BibitemOpen
  \bibfield  {author} {\bibinfo {author} {\bibfnamefont {R.}~\bibnamefont
  {Mac\^edo}}\ and\ \bibinfo {author} {\bibfnamefont {R.~E.}\ \bibnamefont
  {Camley}},\ }\bibfield  {title} {\bibinfo {title} {Engineering terahertz
  surface magnon-polaritons in hyperbolic antiferromagnets},\ }\href
  {https://doi.org/10.1103/PhysRevB.99.014437} {\bibfield  {journal} {\bibinfo
  {journal} {Phys. Rev. B}\ }\textbf {\bibinfo {volume} {99}},\ \bibinfo
  {pages} {014437} (\bibinfo {year} {2019})}\BibitemShut {NoStop}%
\bibitem [{\citenamefont {Vasconcelos}\ \emph {et~al.}(2020)\citenamefont
  {Vasconcelos}, \citenamefont {Cottam},\ and\ \citenamefont
  {Anselmo}}]{Vasconcelos2020}%
  \BibitemOpen
  \bibfield  {author} {\bibinfo {author} {\bibfnamefont {M.}~\bibnamefont
  {Vasconcelos}}, \bibinfo {author} {\bibfnamefont {M.}~\bibnamefont
  {Cottam}},\ and\ \bibinfo {author} {\bibfnamefont {D.}~\bibnamefont
  {Anselmo}},\ }\bibfield  {title} {\bibinfo {title} {Magnon-polaritons in
  graphene/gyromagnetic slab heterostructures},\ }\href@noop {} {\bibfield
  {journal} {\bibinfo  {journal} {Journal of Physics: Condensed Matter}\
  }\textbf {\bibinfo {volume} {33}},\ \bibinfo {pages} {055801} (\bibinfo
  {year} {2020})}\BibitemShut {NoStop}%
\bibitem [{\citenamefont {Hao}\ \emph {et~al.}(2021)\citenamefont {Hao},
  \citenamefont {Fu}, \citenamefont {Zhou},\ and\ \citenamefont
  {Wang}}]{Hao2021}%
  \BibitemOpen
  \bibfield  {author} {\bibinfo {author} {\bibfnamefont {S.}~\bibnamefont
  {Hao}}, \bibinfo {author} {\bibfnamefont {S.}~\bibnamefont {Fu}}, \bibinfo
  {author} {\bibfnamefont {S.}~\bibnamefont {Zhou}},\ and\ \bibinfo {author}
  {\bibfnamefont {X.-Z.}\ \bibnamefont {Wang}},\ }\bibfield  {title} {\bibinfo
  {title} {Dyakonov surface magnons and magnon polaritons},\ }\href@noop {}
  {\bibfield  {journal} {\bibinfo  {journal} {Physical Review B}\ }\textbf
  {\bibinfo {volume} {104}},\ \bibinfo {pages} {045407} (\bibinfo {year}
  {2021})}\BibitemShut {NoStop}%
\bibitem [{\citenamefont {Bludov}\ \emph {et~al.}(2019)\citenamefont {Bludov},
  \citenamefont {Gomes}, \citenamefont {Farias}, \citenamefont
  {Fern{\'a}ndez-Rossier}, \citenamefont {Vasilevskiy},\ and\ \citenamefont
  {Peres}}]{Bludov2019}%
  \BibitemOpen
  \bibfield  {author} {\bibinfo {author} {\bibfnamefont {Y.~V.}\ \bibnamefont
  {Bludov}}, \bibinfo {author} {\bibfnamefont {J.~N.}\ \bibnamefont {Gomes}},
  \bibinfo {author} {\bibfnamefont {G.~d.~A.}\ \bibnamefont {Farias}}, \bibinfo
  {author} {\bibfnamefont {J.}~\bibnamefont {Fern{\'a}ndez-Rossier}}, \bibinfo
  {author} {\bibfnamefont {M.}~\bibnamefont {Vasilevskiy}},\ and\ \bibinfo
  {author} {\bibfnamefont {N.~M.}\ \bibnamefont {Peres}},\ }\bibfield  {title}
  {\bibinfo {title} {Hybrid plasmon-magnon polaritons in
  graphene-antiferromagnet heterostructures},\ }\href@noop {} {\bibfield
  {journal} {\bibinfo  {journal} {2D Materials}\ }\textbf {\bibinfo {volume}
  {6}},\ \bibinfo {pages} {045003} (\bibinfo {year} {2019})}\BibitemShut
  {NoStop}%
\bibitem [{\citenamefont {Pikalov}\ \emph {et~al.}(2021)\citenamefont
  {Pikalov}, \citenamefont {Dorofeenko},\ and\ \citenamefont
  {Granovsky}}]{Pikalov2021}%
  \BibitemOpen
  \bibfield  {author} {\bibinfo {author} {\bibfnamefont {A.~M.}\ \bibnamefont
  {Pikalov}}, \bibinfo {author} {\bibfnamefont {A.~V.}\ \bibnamefont
  {Dorofeenko}},\ and\ \bibinfo {author} {\bibfnamefont {A.}~\bibnamefont
  {Granovsky}},\ }\bibfield  {title} {\bibinfo {title} {Plasmon--magnon
  interaction in the (graphene--antiferromagnetic insulator) system},\
  }\href@noop {} {\bibfield  {journal} {\bibinfo  {journal} {JETP Letters}\
  }\textbf {\bibinfo {volume} {113}},\ \bibinfo {pages} {521} (\bibinfo {year}
  {2021})}\BibitemShut {NoStop}%
\bibitem [{\citenamefont {To}\ \emph {et~al.}(2022{\natexlab{a}})\citenamefont
  {To}, \citenamefont {Wang}, \citenamefont {Liu}, \citenamefont {Wu},
  \citenamefont {Jungfleisch}, \citenamefont {Xiao}, \citenamefont {Zide},
  \citenamefont {Law},\ and\ \citenamefont {Doty}}]{To2022b}%
  \BibitemOpen
  \bibfield  {author} {\bibinfo {author} {\bibfnamefont {D.~Q.}\ \bibnamefont
  {To}}, \bibinfo {author} {\bibfnamefont {Z.}~\bibnamefont {Wang}}, \bibinfo
  {author} {\bibfnamefont {Y.}~\bibnamefont {Liu}}, \bibinfo {author}
  {\bibfnamefont {W.}~\bibnamefont {Wu}}, \bibinfo {author} {\bibfnamefont
  {M.~B.}\ \bibnamefont {Jungfleisch}}, \bibinfo {author} {\bibfnamefont
  {J.~Q.}\ \bibnamefont {Xiao}}, \bibinfo {author} {\bibfnamefont {J.~M.~O.}\
  \bibnamefont {Zide}}, \bibinfo {author} {\bibfnamefont {S.}~\bibnamefont
  {Law}},\ and\ \bibinfo {author} {\bibfnamefont {M.~F.}\ \bibnamefont
  {Doty}},\ }\bibfield  {title} {\bibinfo {title} {Surface
  plasmon-phonon-magnon polariton in a topological insulator-antiferromagnetic
  bilayer structure},\ }\href
  {https://doi.org/10.1103/PhysRevMaterials.6.085201} {\bibfield  {journal}
  {\bibinfo  {journal} {Phys. Rev. Materials}\ }\textbf {\bibinfo {volume}
  {6}},\ \bibinfo {pages} {085201} (\bibinfo {year}
  {2022}{\natexlab{a}})}\BibitemShut {NoStop}%
\bibitem [{\citenamefont {Wildes}\ \emph {et~al.}(2012)\citenamefont {Wildes},
  \citenamefont {Rule}, \citenamefont {Bewley}, \citenamefont {Enderle},\ and\
  \citenamefont {Hicks}}]{Wildes2012}%
  \BibitemOpen
  \bibfield  {author} {\bibinfo {author} {\bibfnamefont {A.}~\bibnamefont
  {Wildes}}, \bibinfo {author} {\bibfnamefont {K.~C.}\ \bibnamefont {Rule}},
  \bibinfo {author} {\bibfnamefont {R.}~\bibnamefont {Bewley}}, \bibinfo
  {author} {\bibfnamefont {M.}~\bibnamefont {Enderle}},\ and\ \bibinfo {author}
  {\bibfnamefont {T.~J.}\ \bibnamefont {Hicks}},\ }\bibfield  {title} {\bibinfo
  {title} {The magnon dynamics and spin exchange parameters of feps3},\
  }\href@noop {} {\bibfield  {journal} {\bibinfo  {journal} {Journal of
  Physics: Condensed Matter}\ }\textbf {\bibinfo {volume} {24}},\ \bibinfo
  {pages} {416004} (\bibinfo {year} {2012})}\BibitemShut {NoStop}%
\bibitem [{\citenamefont {Lan\ifmmode~\mbox{\c{c}}\else \c{c}\fi{}on}\ \emph
  {et~al.}(2016)\citenamefont {Lan\ifmmode~\mbox{\c{c}}\else \c{c}\fi{}on},
  \citenamefont {Walker}, \citenamefont {Ressouche}, \citenamefont {Ouladdiaf},
  \citenamefont {Rule}, \citenamefont {McIntyre}, \citenamefont {Hicks},
  \citenamefont {R\o{}nnow},\ and\ \citenamefont {Wildes}}]{Lan2016}%
  \BibitemOpen
  \bibfield  {author} {\bibinfo {author} {\bibfnamefont {D.}~\bibnamefont
  {Lan\ifmmode~\mbox{\c{c}}\else \c{c}\fi{}on}}, \bibinfo {author}
  {\bibfnamefont {H.~C.}\ \bibnamefont {Walker}}, \bibinfo {author}
  {\bibfnamefont {E.}~\bibnamefont {Ressouche}}, \bibinfo {author}
  {\bibfnamefont {B.}~\bibnamefont {Ouladdiaf}}, \bibinfo {author}
  {\bibfnamefont {K.~C.}\ \bibnamefont {Rule}}, \bibinfo {author}
  {\bibfnamefont {G.~J.}\ \bibnamefont {McIntyre}}, \bibinfo {author}
  {\bibfnamefont {T.~J.}\ \bibnamefont {Hicks}}, \bibinfo {author}
  {\bibfnamefont {H.~M.}\ \bibnamefont {R\o{}nnow}},\ and\ \bibinfo {author}
  {\bibfnamefont {A.~R.}\ \bibnamefont {Wildes}},\ }\bibfield  {title}
  {\bibinfo {title} {Magnetic structure and magnon dynamics of the
  quasi-two-dimensional antiferromagnet ${\mathrm{feps}}_{3}$},\ }\href
  {https://doi.org/10.1103/PhysRevB.94.214407} {\bibfield  {journal} {\bibinfo
  {journal} {Phys. Rev. B}\ }\textbf {\bibinfo {volume} {94}},\ \bibinfo
  {pages} {214407} (\bibinfo {year} {2016})}\BibitemShut {NoStop}%
\bibitem [{\citenamefont {Olsen}(2021)}]{Olsen2021}%
  \BibitemOpen
  \bibfield  {author} {\bibinfo {author} {\bibfnamefont {T.}~\bibnamefont
  {Olsen}},\ }\bibfield  {title} {\bibinfo {title} {Magnetic anisotropy and
  exchange interactions of two-dimensional feps3, nips3 and mnps3 from first
  principles calculations},\ }\href@noop {} {\bibfield  {journal} {\bibinfo
  {journal} {Journal of Physics D: Applied Physics}\ }\textbf {\bibinfo
  {volume} {54}},\ \bibinfo {pages} {314001} (\bibinfo {year}
  {2021})}\BibitemShut {NoStop}%
\bibitem [{\citenamefont {McCreary}\ \emph {et~al.}(2020)\citenamefont
  {McCreary}, \citenamefont {Simpson}, \citenamefont {Mai}, \citenamefont
  {McMichael}, \citenamefont {Douglas}, \citenamefont {Butch}, \citenamefont
  {Dennis}, \citenamefont {Vald\'es~Aguilar},\ and\ \citenamefont
  {Hight~Walker}}]{McCreary2020}%
  \BibitemOpen
  \bibfield  {author} {\bibinfo {author} {\bibfnamefont {A.}~\bibnamefont
  {McCreary}}, \bibinfo {author} {\bibfnamefont {J.~R.}\ \bibnamefont
  {Simpson}}, \bibinfo {author} {\bibfnamefont {T.~T.}\ \bibnamefont {Mai}},
  \bibinfo {author} {\bibfnamefont {R.~D.}\ \bibnamefont {McMichael}}, \bibinfo
  {author} {\bibfnamefont {J.~E.}\ \bibnamefont {Douglas}}, \bibinfo {author}
  {\bibfnamefont {N.}~\bibnamefont {Butch}}, \bibinfo {author} {\bibfnamefont
  {C.}~\bibnamefont {Dennis}}, \bibinfo {author} {\bibfnamefont
  {R.}~\bibnamefont {Vald\'es~Aguilar}},\ and\ \bibinfo {author} {\bibfnamefont
  {A.~R.}\ \bibnamefont {Hight~Walker}},\ }\bibfield  {title} {\bibinfo {title}
  {Quasi-two-dimensional magnon identification in antiferromagnetic
  $\mathrm{FeP}{\mathrm{s}}_{3}$ via magneto-raman spectroscopy},\ }\href
  {https://doi.org/10.1103/PhysRevB.101.064416} {\bibfield  {journal} {\bibinfo
   {journal} {Phys. Rev. B}\ }\textbf {\bibinfo {volume} {101}},\ \bibinfo
  {pages} {064416} (\bibinfo {year} {2020})}\BibitemShut {NoStop}%
\bibitem [{\citenamefont {Zhu}\ \emph {et~al.}(2018)\citenamefont {Zhu},
  \citenamefont {Ralph},\ and\ \citenamefont {Buhrman}}]{Zhu2018}%
  \BibitemOpen
  \bibfield  {author} {\bibinfo {author} {\bibfnamefont {L.~J.}\ \bibnamefont
  {Zhu}}, \bibinfo {author} {\bibfnamefont {D.~C.}\ \bibnamefont {Ralph}},\
  and\ \bibinfo {author} {\bibfnamefont {R.~A.}\ \bibnamefont {Buhrman}},\
  }\bibfield  {title} {\bibinfo {title} {Irrelevance of magnetic proximity
  effect to spin-orbit torques in heavy-metal/ferromagnet bilayers},\ }\href
  {https://doi.org/10.1103/PhysRevB.98.134406} {\bibfield  {journal} {\bibinfo
  {journal} {Phys. Rev. B}\ }\textbf {\bibinfo {volume} {98}},\ \bibinfo
  {pages} {134406} (\bibinfo {year} {2018})}\BibitemShut {NoStop}%
\bibitem [{\citenamefont {To}\ \emph {et~al.}(2022{\natexlab{b}})\citenamefont
  {To}, \citenamefont {Wang}, \citenamefont {Ho}, \citenamefont {Hu},
  \citenamefont {Acuna}, \citenamefont {Liu}, \citenamefont {Bryant},
  \citenamefont {Janotti}, \citenamefont {Zide}, \citenamefont {Law},\ and\
  \citenamefont {Doty}}]{To2022a}%
  \BibitemOpen
  \bibfield  {author} {\bibinfo {author} {\bibfnamefont {D.~Q.}\ \bibnamefont
  {To}}, \bibinfo {author} {\bibfnamefont {Z.}~\bibnamefont {Wang}}, \bibinfo
  {author} {\bibfnamefont {D.~Q.}\ \bibnamefont {Ho}}, \bibinfo {author}
  {\bibfnamefont {R.}~\bibnamefont {Hu}}, \bibinfo {author} {\bibfnamefont
  {W.}~\bibnamefont {Acuna}}, \bibinfo {author} {\bibfnamefont
  {Y.}~\bibnamefont {Liu}}, \bibinfo {author} {\bibfnamefont {G.~W.}\
  \bibnamefont {Bryant}}, \bibinfo {author} {\bibfnamefont {A.}~\bibnamefont
  {Janotti}}, \bibinfo {author} {\bibfnamefont {J.~M.~O.}\ \bibnamefont
  {Zide}}, \bibinfo {author} {\bibfnamefont {S.}~\bibnamefont {Law}},\ and\
  \bibinfo {author} {\bibfnamefont {M.~F.}\ \bibnamefont {Doty}},\ }\bibfield
  {title} {\bibinfo {title} {Strong coupling between a topological insulator
  and a iii-v heterostructure at terahertz frequency},\ }\href
  {https://doi.org/10.1103/PhysRevMaterials.6.035201} {\bibfield  {journal}
  {\bibinfo  {journal} {Phys. Rev. Materials}\ }\textbf {\bibinfo {volume}
  {6}},\ \bibinfo {pages} {035201} (\bibinfo {year}
  {2022}{\natexlab{b}})}\BibitemShut {NoStop}%
\bibitem [{\citenamefont {Richter}\ and\ \citenamefont
  {Becker}(1977)}]{Richter1977}%
  \BibitemOpen
  \bibfield  {author} {\bibinfo {author} {\bibfnamefont {W.}~\bibnamefont
  {Richter}}\ and\ \bibinfo {author} {\bibfnamefont {C.}~\bibnamefont
  {Becker}},\ }\bibfield  {title} {\bibinfo {title} {A raman and far-infrared
  investigation of phonons in the rhombohedral v2--vi3 compounds bi2te3,
  bi2se3, sb2te3 and bi2 (te1- xsex) 3 (0< x< 1),(bi1- ysby) 2te3 (0< y< 1)},\
  }\href@noop {} {\bibfield  {journal} {\bibinfo  {journal} {physica status
  solidi (b)}\ }\textbf {\bibinfo {volume} {84}},\ \bibinfo {pages} {619}
  (\bibinfo {year} {1977})}\BibitemShut {NoStop}%
\bibitem [{\citenamefont {Ghosh}\ \emph {et~al.}(2022)\citenamefont {Ghosh},
  \citenamefont {Birowska}, \citenamefont {Ghose}, \citenamefont {Rybak},
  \citenamefont {Maity}, \citenamefont {Ghosh}, \citenamefont {Das},
  \citenamefont {Bera}, \citenamefont {Bhardwaj}, \citenamefont {Nandi} \emph
  {et~al.}}]{Ghosh2022}%
  \BibitemOpen
  \bibfield  {author} {\bibinfo {author} {\bibfnamefont {A.}~\bibnamefont
  {Ghosh}}, \bibinfo {author} {\bibfnamefont {M.}~\bibnamefont {Birowska}},
  \bibinfo {author} {\bibfnamefont {P.~K.}\ \bibnamefont {Ghose}}, \bibinfo
  {author} {\bibfnamefont {M.}~\bibnamefont {Rybak}}, \bibinfo {author}
  {\bibfnamefont {S.}~\bibnamefont {Maity}}, \bibinfo {author} {\bibfnamefont
  {S.}~\bibnamefont {Ghosh}}, \bibinfo {author} {\bibfnamefont
  {B.}~\bibnamefont {Das}}, \bibinfo {author} {\bibfnamefont {S.}~\bibnamefont
  {Bera}}, \bibinfo {author} {\bibfnamefont {S.}~\bibnamefont {Bhardwaj}},
  \bibinfo {author} {\bibfnamefont {S.}~\bibnamefont {Nandi}}, \emph {et~al.},\
  }\bibfield  {title} {\bibinfo {title} {Anisotropic magnetodielectric coupling
  in layered antiferromagnetic feps $ \_3$},\ }\href@noop {} {\bibfield
  {journal} {\bibinfo  {journal} {arXiv preprint arXiv:2208.02729}\ } (\bibinfo
  {year} {2022})}\BibitemShut {NoStop}%
\bibitem [{\citenamefont {Joy}\ and\ \citenamefont
  {Vasudevan}(1992)}]{Joy1992}%
  \BibitemOpen
  \bibfield  {author} {\bibinfo {author} {\bibfnamefont {P.~A.}\ \bibnamefont
  {Joy}}\ and\ \bibinfo {author} {\bibfnamefont {S.}~\bibnamefont
  {Vasudevan}},\ }\bibfield  {title} {\bibinfo {title} {Magnetism in the
  layered transition-metal thiophosphates m${\mathrm{ps}}_{3}$ (m=mn, fe, and
  ni)},\ }\href {https://doi.org/10.1103/PhysRevB.46.5425} {\bibfield
  {journal} {\bibinfo  {journal} {Phys. Rev. B}\ }\textbf {\bibinfo {volume}
  {46}},\ \bibinfo {pages} {5425} (\bibinfo {year} {1992})}\BibitemShut
  {NoStop}%
\bibitem [{\citenamefont {Subramanian}\ \emph {et~al.}(1989)\citenamefont
  {Subramanian}, \citenamefont {Shannon}, \citenamefont {Chai}, \citenamefont
  {Abraham},\ and\ \citenamefont {Wintersgill}}]{Subramanian1989}%
  \BibitemOpen
  \bibfield  {author} {\bibinfo {author} {\bibfnamefont {M.}~\bibnamefont
  {Subramanian}}, \bibinfo {author} {\bibfnamefont {R.}~\bibnamefont
  {Shannon}}, \bibinfo {author} {\bibfnamefont {B.}~\bibnamefont {Chai}},
  \bibinfo {author} {\bibfnamefont {M.}~\bibnamefont {Abraham}},\ and\ \bibinfo
  {author} {\bibfnamefont {M.}~\bibnamefont {Wintersgill}},\ }\bibfield
  {title} {\bibinfo {title} {Dielectric constants of beo, mgo, and cao using
  the two-terminal method},\ }\href@noop {} {\bibfield  {journal} {\bibinfo
  {journal} {Physics and chemistry of minerals}\ }\textbf {\bibinfo {volume}
  {16}},\ \bibinfo {pages} {741} (\bibinfo {year} {1989})}\BibitemShut
  {NoStop}%
\bibitem [{\citenamefont {To}(2019)}]{To2019}%
  \BibitemOpen
  \bibfield  {author} {\bibinfo {author} {\bibfnamefont {D.-Q.}\ \bibnamefont
  {To}},\ }\emph {\bibinfo {title} {Advanced kp multiband methods for
  semiconductor-based spinorbitronics}},\ \href@noop {} {Ph.D. thesis},\
  \bibinfo  {school} {Institut polytechnique de Paris} (\bibinfo {year}
  {2019})\BibitemShut {NoStop}%
\bibitem [{\citenamefont {Dang}\ \emph
  {et~al.}(2020{\natexlab{b}})\citenamefont {Dang}, \citenamefont
  {Barbedienne}, \citenamefont {To}, \citenamefont {Rongione}, \citenamefont
  {Reyren}, \citenamefont {Godel}, \citenamefont {Collin}, \citenamefont
  {George},\ and\ \citenamefont {Jaffr\`es}}]{Dang2020}%
  \BibitemOpen
  \bibfield  {author} {\bibinfo {author} {\bibfnamefont {T.~H.}\ \bibnamefont
  {Dang}}, \bibinfo {author} {\bibfnamefont {Q.}~\bibnamefont {Barbedienne}},
  \bibinfo {author} {\bibfnamefont {D.~Q.}\ \bibnamefont {To}}, \bibinfo
  {author} {\bibfnamefont {E.}~\bibnamefont {Rongione}}, \bibinfo {author}
  {\bibfnamefont {N.}~\bibnamefont {Reyren}}, \bibinfo {author} {\bibfnamefont
  {F.}~\bibnamefont {Godel}}, \bibinfo {author} {\bibfnamefont
  {S.}~\bibnamefont {Collin}}, \bibinfo {author} {\bibfnamefont {J.~M.}\
  \bibnamefont {George}},\ and\ \bibinfo {author} {\bibfnamefont
  {H.}~\bibnamefont {Jaffr\`es}},\ }\bibfield  {title} {\bibinfo {title}
  {Anomalous hall effect in $3d/5d$ multilayers mediated by interface
  scattering and nonlocal spin conductivity},\ }\href
  {https://doi.org/10.1103/PhysRevB.102.144405} {\bibfield  {journal} {\bibinfo
   {journal} {Phys. Rev. B}\ }\textbf {\bibinfo {volume} {102}},\ \bibinfo
  {pages} {144405} (\bibinfo {year} {2020}{\natexlab{b}})}\BibitemShut
  {NoStop}%
\bibitem [{\citenamefont {To}\ \emph {et~al.}(2021)\citenamefont {To},
  \citenamefont {Dang}, \citenamefont {Vila}, \citenamefont {Attan\'e},
  \citenamefont {Bibes},\ and\ \citenamefont {Jaffr\`es}}]{To2021}%
  \BibitemOpen
  \bibfield  {author} {\bibinfo {author} {\bibfnamefont {D.~Q.}\ \bibnamefont
  {To}}, \bibinfo {author} {\bibfnamefont {T.~H.}\ \bibnamefont {Dang}},
  \bibinfo {author} {\bibfnamefont {L.}~\bibnamefont {Vila}}, \bibinfo {author}
  {\bibfnamefont {J.~P.}\ \bibnamefont {Attan\'e}}, \bibinfo {author}
  {\bibfnamefont {M.}~\bibnamefont {Bibes}},\ and\ \bibinfo {author}
  {\bibfnamefont {H.}~\bibnamefont {Jaffr\`es}},\ }\bibfield  {title} {\bibinfo
  {title} {Spin to charge conversion at rashba-split ${\mathbf{srtio}}_{3}$
  interfaces from resonant tunneling},\ }\href
  {https://doi.org/10.1103/PhysRevResearch.3.043170} {\bibfield  {journal}
  {\bibinfo  {journal} {Phys. Rev. Research}\ }\textbf {\bibinfo {volume}
  {3}},\ \bibinfo {pages} {043170} (\bibinfo {year} {2021})}\BibitemShut
  {NoStop}%
\bibitem [{\citenamefont {Woessner}\ \emph {et~al.}(2015)\citenamefont
  {Woessner}, \citenamefont {Lundeberg}, \citenamefont {Gao}, \citenamefont
  {Principi}, \citenamefont {Alonso-Gonz{\'a}lez}, \citenamefont {Carrega},
  \citenamefont {Watanabe}, \citenamefont {Taniguchi}, \citenamefont {Vignale},
  \citenamefont {Polini} \emph {et~al.}}]{Woessner2015}%
  \BibitemOpen
  \bibfield  {author} {\bibinfo {author} {\bibfnamefont {A.}~\bibnamefont
  {Woessner}}, \bibinfo {author} {\bibfnamefont {M.~B.}\ \bibnamefont
  {Lundeberg}}, \bibinfo {author} {\bibfnamefont {Y.}~\bibnamefont {Gao}},
  \bibinfo {author} {\bibfnamefont {A.}~\bibnamefont {Principi}}, \bibinfo
  {author} {\bibfnamefont {P.}~\bibnamefont {Alonso-Gonz{\'a}lez}}, \bibinfo
  {author} {\bibfnamefont {M.}~\bibnamefont {Carrega}}, \bibinfo {author}
  {\bibfnamefont {K.}~\bibnamefont {Watanabe}}, \bibinfo {author}
  {\bibfnamefont {T.}~\bibnamefont {Taniguchi}}, \bibinfo {author}
  {\bibfnamefont {G.}~\bibnamefont {Vignale}}, \bibinfo {author} {\bibfnamefont
  {M.}~\bibnamefont {Polini}}, \emph {et~al.},\ }\bibfield  {title} {\bibinfo
  {title} {Highly confined low-loss plasmons in graphene--boron nitride
  heterostructures},\ }\href@noop {} {\bibfield  {journal} {\bibinfo  {journal}
  {Nature materials}\ }\textbf {\bibinfo {volume} {14}},\ \bibinfo {pages}
  {421} (\bibinfo {year} {2015})}\BibitemShut {NoStop}%
\bibitem [{\citenamefont {Kumar}\ \emph {et~al.}(2015)\citenamefont {Kumar},
  \citenamefont {Low}, \citenamefont {Fung}, \citenamefont {Avouris},\ and\
  \citenamefont {Fang}}]{Kumar2015}%
  \BibitemOpen
  \bibfield  {author} {\bibinfo {author} {\bibfnamefont {A.}~\bibnamefont
  {Kumar}}, \bibinfo {author} {\bibfnamefont {T.}~\bibnamefont {Low}}, \bibinfo
  {author} {\bibfnamefont {K.~H.}\ \bibnamefont {Fung}}, \bibinfo {author}
  {\bibfnamefont {P.}~\bibnamefont {Avouris}},\ and\ \bibinfo {author}
  {\bibfnamefont {N.~X.}\ \bibnamefont {Fang}},\ }\bibfield  {title} {\bibinfo
  {title} {Tunable light--matter interaction and the role of hyperbolicity in
  graphene--hbn system},\ }\href@noop {} {\bibfield  {journal} {\bibinfo
  {journal} {Nano letters}\ }\textbf {\bibinfo {volume} {15}},\ \bibinfo
  {pages} {3172} (\bibinfo {year} {2015})}\BibitemShut {NoStop}%
\bibitem [{\citenamefont {Bezares}\ \emph {et~al.}(2017)\citenamefont
  {Bezares}, \citenamefont {Sanctis}, \citenamefont {Saavedra}, \citenamefont
  {Woessner}, \citenamefont {Alonso-Gonzalez}, \citenamefont {Amenabar},
  \citenamefont {Chen}, \citenamefont {Bointon}, \citenamefont {Dai},
  \citenamefont {Fogler} \emph {et~al.}}]{Bezares2017}%
  \BibitemOpen
  \bibfield  {author} {\bibinfo {author} {\bibfnamefont {F.~J.}\ \bibnamefont
  {Bezares}}, \bibinfo {author} {\bibfnamefont {A.~D.}\ \bibnamefont
  {Sanctis}}, \bibinfo {author} {\bibfnamefont {J.}~\bibnamefont {Saavedra}},
  \bibinfo {author} {\bibfnamefont {A.}~\bibnamefont {Woessner}}, \bibinfo
  {author} {\bibfnamefont {P.}~\bibnamefont {Alonso-Gonzalez}}, \bibinfo
  {author} {\bibfnamefont {I.}~\bibnamefont {Amenabar}}, \bibinfo {author}
  {\bibfnamefont {J.}~\bibnamefont {Chen}}, \bibinfo {author} {\bibfnamefont
  {T.~H.}\ \bibnamefont {Bointon}}, \bibinfo {author} {\bibfnamefont
  {S.}~\bibnamefont {Dai}}, \bibinfo {author} {\bibfnamefont {M.~M.}\
  \bibnamefont {Fogler}}, \emph {et~al.},\ }\bibfield  {title} {\bibinfo
  {title} {Intrinsic plasmon--phonon interactions in highly doped graphene: A
  near-field imaging study},\ }\href@noop {} {\bibfield  {journal} {\bibinfo
  {journal} {Nano letters}\ }\textbf {\bibinfo {volume} {17}},\ \bibinfo
  {pages} {5908} (\bibinfo {year} {2017})}\BibitemShut {NoStop}%
\bibitem [{\citenamefont {Epstein}\ \emph {et~al.}(2020)\citenamefont
  {Epstein}, \citenamefont {Alcaraz}, \citenamefont {Huang}, \citenamefont
  {Pusapati}, \citenamefont {Hugonin}, \citenamefont {Kumar}, \citenamefont
  {Deputy}, \citenamefont {Khodkov}, \citenamefont {Rappoport}, \citenamefont
  {Hong} \emph {et~al.}}]{Epstein2020}%
  \BibitemOpen
  \bibfield  {author} {\bibinfo {author} {\bibfnamefont {I.}~\bibnamefont
  {Epstein}}, \bibinfo {author} {\bibfnamefont {D.}~\bibnamefont {Alcaraz}},
  \bibinfo {author} {\bibfnamefont {Z.}~\bibnamefont {Huang}}, \bibinfo
  {author} {\bibfnamefont {V.-V.}\ \bibnamefont {Pusapati}}, \bibinfo {author}
  {\bibfnamefont {J.-P.}\ \bibnamefont {Hugonin}}, \bibinfo {author}
  {\bibfnamefont {A.}~\bibnamefont {Kumar}}, \bibinfo {author} {\bibfnamefont
  {X.~M.}\ \bibnamefont {Deputy}}, \bibinfo {author} {\bibfnamefont
  {T.}~\bibnamefont {Khodkov}}, \bibinfo {author} {\bibfnamefont {T.~G.}\
  \bibnamefont {Rappoport}}, \bibinfo {author} {\bibfnamefont {J.-Y.}\
  \bibnamefont {Hong}}, \emph {et~al.},\ }\bibfield  {title} {\bibinfo {title}
  {Far-field excitation of single graphene plasmon cavities with
  ultracompressed mode volumes},\ }\href@noop {} {\bibfield  {journal}
  {\bibinfo  {journal} {Science}\ }\textbf {\bibinfo {volume} {368}},\ \bibinfo
  {pages} {1219} (\bibinfo {year} {2020})}\BibitemShut {NoStop}%
\bibitem [{\citenamefont {Szunyogh}\ \emph {et~al.}(2009)\citenamefont
  {Szunyogh}, \citenamefont {Lazarovits}, \citenamefont {Udvardi},
  \citenamefont {Jackson},\ and\ \citenamefont {Nowak}}]{Szunyogh2009}%
  \BibitemOpen
  \bibfield  {author} {\bibinfo {author} {\bibfnamefont {L.}~\bibnamefont
  {Szunyogh}}, \bibinfo {author} {\bibfnamefont {B.}~\bibnamefont
  {Lazarovits}}, \bibinfo {author} {\bibfnamefont {L.}~\bibnamefont {Udvardi}},
  \bibinfo {author} {\bibfnamefont {J.}~\bibnamefont {Jackson}},\ and\ \bibinfo
  {author} {\bibfnamefont {U.}~\bibnamefont {Nowak}},\ }\bibfield  {title}
  {\bibinfo {title} {Giant magnetic anisotropy of the bulk antiferromagnets
  irmn and ${\text{irmn}}_{3}$ from first principles},\ }\href
  {https://doi.org/10.1103/PhysRevB.79.020403} {\bibfield  {journal} {\bibinfo
  {journal} {Phys. Rev. B}\ }\textbf {\bibinfo {volume} {79}},\ \bibinfo
  {pages} {020403} (\bibinfo {year} {2009})}\BibitemShut {NoStop}%
\bibitem [{\citenamefont {Wang}\ \emph {et~al.}(2017)\citenamefont {Wang},
  \citenamefont {Tang}, \citenamefont {Du},\ and\ \citenamefont
  {Wan}}]{Wang2017}%
  \BibitemOpen
  \bibfield  {author} {\bibinfo {author} {\bibfnamefont {D.}~\bibnamefont
  {Wang}}, \bibinfo {author} {\bibfnamefont {F.}~\bibnamefont {Tang}}, \bibinfo
  {author} {\bibfnamefont {Y.}~\bibnamefont {Du}},\ and\ \bibinfo {author}
  {\bibfnamefont {X.}~\bibnamefont {Wan}},\ }\bibfield  {title} {\bibinfo
  {title} {First-principles study of the giant magnetic anisotropy energy in
  bulk ${\mathrm{na}}_{4}{\mathrm{iro}}_{4}$},\ }\href
  {https://doi.org/10.1103/PhysRevB.96.205159} {\bibfield  {journal} {\bibinfo
  {journal} {Phys. Rev. B}\ }\textbf {\bibinfo {volume} {96}},\ \bibinfo
  {pages} {205159} (\bibinfo {year} {2017})}\BibitemShut {NoStop}%
\bibitem [{\citenamefont {Albaridy}\ \emph {et~al.}(2020)\citenamefont
  {Albaridy}, \citenamefont {Manchon},\ and\ \citenamefont
  {Schwingenschl{\"o}gl}}]{Albaridy2020}%
  \BibitemOpen
  \bibfield  {author} {\bibinfo {author} {\bibfnamefont {R.}~\bibnamefont
  {Albaridy}}, \bibinfo {author} {\bibfnamefont {A.}~\bibnamefont {Manchon}},\
  and\ \bibinfo {author} {\bibfnamefont {U.}~\bibnamefont
  {Schwingenschl{\"o}gl}},\ }\bibfield  {title} {\bibinfo {title} {Tunable
  magnetic anisotropy in cr--trihalide janus monolayers},\ }\href@noop {}
  {\bibfield  {journal} {\bibinfo  {journal} {Journal of Physics: Condensed
  Matter}\ }\textbf {\bibinfo {volume} {32}},\ \bibinfo {pages} {355702}
  (\bibinfo {year} {2020})}\BibitemShut {NoStop}%
\bibitem [{\citenamefont {Momma}\ and\ \citenamefont
  {Izumi}(2008)}]{Momma2008}%
  \BibitemOpen
  \bibfield  {author} {\bibinfo {author} {\bibfnamefont {K.}~\bibnamefont
  {Momma}}\ and\ \bibinfo {author} {\bibfnamefont {F.}~\bibnamefont {Izumi}},\
  }\bibfield  {title} {\bibinfo {title} {Vesta: a three-dimensional
  visualization system for electronic and structural analysis},\ }\href@noop {}
  {\bibfield  {journal} {\bibinfo  {journal} {Journal of Applied
  crystallography}\ }\textbf {\bibinfo {volume} {41}},\ \bibinfo {pages} {653}
  (\bibinfo {year} {2008})}\BibitemShut {NoStop}%
\bibitem [{\citenamefont {Li}\ \emph {et~al.}(2019)\citenamefont {Li},
  \citenamefont {Jiang}, \citenamefont {Li}, \citenamefont {Xu},\ and\
  \citenamefont {Duan}}]{Li2109}%
  \BibitemOpen
  \bibfield  {author} {\bibinfo {author} {\bibfnamefont {Y.}~\bibnamefont
  {Li}}, \bibinfo {author} {\bibfnamefont {Z.}~\bibnamefont {Jiang}}, \bibinfo
  {author} {\bibfnamefont {J.}~\bibnamefont {Li}}, \bibinfo {author}
  {\bibfnamefont {S.}~\bibnamefont {Xu}},\ and\ \bibinfo {author}
  {\bibfnamefont {W.}~\bibnamefont {Duan}},\ }\bibfield  {title} {\bibinfo
  {title} {Magnetic anisotropy of the two-dimensional ferromagnetic insulator
  ${\mathrm{mnbi}}_{2}{\mathrm{te}}_{4}$},\ }\href
  {https://doi.org/10.1103/PhysRevB.100.134438} {\bibfield  {journal} {\bibinfo
   {journal} {Phys. Rev. B}\ }\textbf {\bibinfo {volume} {100}},\ \bibinfo
  {pages} {134438} (\bibinfo {year} {2019})}\BibitemShut {NoStop}%
\bibitem [{\citenamefont {Wildes}\ \emph {et~al.}(2020)\citenamefont {Wildes},
  \citenamefont {Lan\ifmmode~\mbox{\c{c}}\else \c{c}\fi{}on}, \citenamefont
  {Chan}, \citenamefont {Weickert}, \citenamefont {Harrison}, \citenamefont
  {Simonet}, \citenamefont {Zhitomirsky}, \citenamefont {Gvozdikova},
  \citenamefont {Ziman},\ and\ \citenamefont {R\o{}nnow}}]{Wildes2020a}%
  \BibitemOpen
  \bibfield  {author} {\bibinfo {author} {\bibfnamefont {A.~R.}\ \bibnamefont
  {Wildes}}, \bibinfo {author} {\bibfnamefont {D.}~\bibnamefont
  {Lan\ifmmode~\mbox{\c{c}}\else \c{c}\fi{}on}}, \bibinfo {author}
  {\bibfnamefont {M.~K.}\ \bibnamefont {Chan}}, \bibinfo {author}
  {\bibfnamefont {F.}~\bibnamefont {Weickert}}, \bibinfo {author}
  {\bibfnamefont {N.}~\bibnamefont {Harrison}}, \bibinfo {author}
  {\bibfnamefont {V.}~\bibnamefont {Simonet}}, \bibinfo {author} {\bibfnamefont
  {M.~E.}\ \bibnamefont {Zhitomirsky}}, \bibinfo {author} {\bibfnamefont
  {M.~V.}\ \bibnamefont {Gvozdikova}}, \bibinfo {author} {\bibfnamefont
  {T.}~\bibnamefont {Ziman}},\ and\ \bibinfo {author} {\bibfnamefont {H.~M.}\
  \bibnamefont {R\o{}nnow}},\ }\bibfield  {title} {\bibinfo {title} {High field
  magnetization of ${\mathrm{feps}}_{3}$},\ }\href
  {https://doi.org/10.1103/PhysRevB.101.024415} {\bibfield  {journal} {\bibinfo
   {journal} {Phys. Rev. B}\ }\textbf {\bibinfo {volume} {101}},\ \bibinfo
  {pages} {024415} (\bibinfo {year} {2020})}\BibitemShut {NoStop}%
\bibitem [{\citenamefont {Rezende}\ \emph {et~al.}(2019)\citenamefont
  {Rezende}, \citenamefont {Azevedo},\ and\ \citenamefont
  {Rodr{\'\i}guez-Su{\'a}rez}}]{Rezende2019}%
  \BibitemOpen
  \bibfield  {author} {\bibinfo {author} {\bibfnamefont {S.~M.}\ \bibnamefont
  {Rezende}}, \bibinfo {author} {\bibfnamefont {A.}~\bibnamefont {Azevedo}},\
  and\ \bibinfo {author} {\bibfnamefont {R.~L.}\ \bibnamefont
  {Rodr{\'\i}guez-Su{\'a}rez}},\ }\bibfield  {title} {\bibinfo {title}
  {Introduction to antiferromagnetic magnons},\ }\href@noop {} {\bibfield
  {journal} {\bibinfo  {journal} {Journal of Applied Physics}\ }\textbf
  {\bibinfo {volume} {126}},\ \bibinfo {pages} {151101} (\bibinfo {year}
  {2019})}\BibitemShut {NoStop}%
\bibitem [{\citenamefont {Jackson}(1999)}]{Jackson1999}%
  \BibitemOpen
  \bibfield  {author} {\bibinfo {author} {\bibfnamefont {J.~D.}\ \bibnamefont
  {Jackson}},\ }\href@noop {} {\emph {\bibinfo {title} {Classical
  electrodynamics, 3rd ed.}}}\ (\bibinfo  {publisher} {Willey, New Jersey},\
  \bibinfo {year} {1999})\BibitemShut {NoStop}%
\bibitem [{\citenamefont {Zangwill}(2012)}]{Zangwill_2012}%
  \BibitemOpen
  \bibfield  {author} {\bibinfo {author} {\bibfnamefont {A.}~\bibnamefont
  {Zangwill}},\ }\href {https://doi.org/10.1017/CBO9781139034777} {\emph
  {\bibinfo {title} {Modern Electrodynamics}}}\ (\bibinfo  {publisher}
  {Cambridge University Press, Cambridge},\ \bibinfo {year} {2012})\BibitemShut
  {NoStop}%
\end{thebibliography}%



\end{document}